\documentclass[aps, prd, twocolumn, amsmath, floats,floatfix, superscriptaddress, nofootinbib]{revtex4-1}

\usepackage{float}
\usepackage{amssymb}
\usepackage{amsmath}
\usepackage{verbatim}
\usepackage{mathrsfs}
\usepackage{amsfonts}
\usepackage{latexsym}
\usepackage{epsfig}
\usepackage{color}
\usepackage{graphicx,subfigure}
\usepackage{units}
\usepackage{envmath}
\usepackage{natbib}
\usepackage[dvipsnames, table,usenames,svgnames]{xcolor}
\usepackage[normalem]{ulem}
\usepackage[linktocpage]{hyperref}

\begin{document}

%%%%%%%%%%%%%%%%%%
%%%   MACROS   %%%
%%%%%%%%%%%%%%%%%%

\definecolor{orange}{rgb}{0.9,0.45,0}

\newcommand{\re}{\mbox{Re}}
\newcommand{\im}{\mbox{Im}}

\newcommand{\tf}[1]{\textcolor{red}{TF: #1}}
\newcommand{\nsg}[1]{\textcolor{cyan}{NSG: #1}}
\newcommand{\fdg}[1]{\textcolor{magenta}{FDG: #1}}
\newcommand{\mmt}[1]{\textcolor{purple}{ #1}}
\newcommand{\pcd}[1]{\textcolor{Green}{#1}}

\def\CovDev{D}
\def\Res{{\mathcal R}}
\def\Gammaflat{\hat \Gamma}
\def\metricflat{\hat \gamma}
\def\Dflat{\hat {\mathcal D}}
\def\part_n{\partial_\perp}

%=== Definition for abbreviations ===
\def\Lie{\mathcal{L}}
\def\A{\mathcal{X}}
\def\Aphi{\A_{\phi}}
\def\hAphi{\hat{\A}_{\phi}}
\def\E{\mathcal{E}}
\def\Ham{\mathcal{H}}
\def\M{\mathcal{M}}
\def\R{\mathcal{R}}
\def\p{\partial}

\def\hg{\hat{\gamma}}
\def\hA{\hat{A}}
\def\hD{\hat{D}}
\def\hE{\hat{E}}
\def\hR{\hat{R}}
\def\hcA{\hat{\mathcal{A}}}
\def\hDelt{\hat{\triangle}}

\def\na{\nabla}
\def\dif{{\rm{d}}}
\def\non{\nonumber}
\newcommand{\erf}{\textrm{erf}}
%====================================

\renewcommand{\t}{\times}

% \long\def\symbolfootnote[#1]#2{\begingroup%
% \def\thefootnote{\fnsymbol{footnote}}\footnote[#1]{#2}\endgroup}

%%%%%%%%%%%%%%%%%
%%%   TITLE   %%%
%%%%%%%%%%%%%%%%%

\title{Impact of ultralight bosonic dark matter on the dynamical bar-mode instability of rotating neutron stars}
 
\author{Fabrizio Di Giovanni}
\affiliation{Departamento de
  Astronom\'{\i}a y Astrof\'{\i}sica, Universitat de Val\`encia,
  Dr. Moliner 50, 46100, Burjassot (Val\`encia), Spain}

\author{Nicolas Sanchis-Gual}
\affiliation{Departamento de
  Astronom\'{\i}a y Astrof\'{\i}sica, Universitat de Val\`encia,
  Dr. Moliner 50, 46100, Burjassot (Val\`encia), Spain}
  \affiliation{Departamento  de  Matem\'{a}tica  da  Universidade  de  Aveiro  and  Centre  for  Research  and  Development in  Mathematics  and  Applications  (CIDMA),  Campus  de  Santiago,  3810-183  Aveiro,  Portugal}

\author{Davide Guerra}
\affiliation{Departamento de
  Astronom\'{\i}a y Astrof\'{\i}sica, Universitat de Val\`encia,
  Dr. Moliner 50, 46100, Burjassot (Val\`encia), Spain}

\author{Miquel Miravet-Ten\'es}
\affiliation{Departamento de
  Astronom\'{\i}a y Astrof\'{\i}sica, Universitat de Val\`encia,
  Dr. Moliner 50, 46100, Burjassot (Val\`encia), Spain}
  
 \author{Pablo Cerd\'a-Dur\'an}
\affiliation{Departamento de
  Astronom\'{\i}a y Astrof\'{\i}sica, Universitat de Val\`encia,
  Dr. Moliner 50, 46100, Burjassot (Val\`encia), Spain}  

\author{Jos\'e A. Font}
\affiliation{Departamento de
  Astronom\'{\i}a y Astrof\'{\i}sica, Universitat de Val\`encia,
  Dr. Moliner 50, 46100, Burjassot (Val\`encia), Spain}
\affiliation{Observatori Astron\`omic, Universitat de Val\`encia, C/ Catedr\'atico 
  Jos\'e Beltr\'an 2, 46980, Paterna (Val\`encia), Spain}

%%%%%%%%%%%%%%%%
%%%   DATE   %%%
%%%%%%%%%%%%%%%%

\date{2022}

%%%%%%%%%%%%%%%%%%%%
%%%   ABSTRACT   %%%
%%%%%%%%%%%%%%%%%%%%

\begin{abstract} 
We investigate the effects ultralight bosonic field dark matter may have on the dynamics of unstable differentially-rotating neutron stars prone to the bar-mode instability. To this aim we perform numerical simulations in general relativity of rotating neutron stars accreting an initial spherically symmetric bosonic field cloud, solving the Einstein-(complex, massive) Klein-Gordon-Euler and the Einstein-(complex) Proca-Euler systems. We find that the presence of the bosonic field can critically modify the development of the bar-mode instability of neutron stars, depending on the total mass of the bosonic field and on the boson particle mass. In some cases, the accreting bosonic field can even quench the dominant $\ell=m=2$ mode of the bar-deformation by dynamically forming a mixed (fermion-boson) star that retains part of the angular momentum of the original neutron star. However, the mixed star undergoes the development of a mixed bar that leads to significant gravitational-wave emission, substantially different to that of the isolated neutron star. Our results indicate that dark-matter accretion in neutron stars could change the frequency of the expected emission of the bar-mode instability, which would have an important impact on ongoing searches for continuous gravitational waves.
\end{abstract}

%%%%%%%%%%%%%%%%
%%%   PACS   %%%
%%%%%%%%%%%%%%%%

% \pacs{
% 95.30.Sf, % relativity and gravitation
% 04.70.Bw, 
% 04.40.Nr, 
% 04.25.dg
% }

%%%%%%%%%%%%%%%%%%%%%%
%%%   MAKE TITLE   %%%
%%%%%%%%%%%%%%%%%%%%%%
\preprint{ET-0105A-22}

\maketitle

%{\bf \large Charged spherical boson stars in a cavity}
\vspace{0.8cm}

%%%%%%%%%%%%%%%%%%%%%%
\section{Introduction}

Differential rotation is expected to occur in neutron stars. It can be present in proto-neutron stars (PNS) formed in core-collapse supernova (CCSN) explosions, in the transient post-merger remnants that form after binary neutron star (BNS) mergers, and in X-ray binary systems where accretion can trigger high-amplitude oscillation (axial fluid) r-modes that might impact the neutron star rotation. In addition, rotating neutron stars are also expected to be subject to various types of non-axisymmetric instabilities (for reviews see~\cite{Glampedakis2017,Paschalidis2017} and references therein). For sufficiently high values of the ratio of the rotational kinetic energy $T$ and the gravitational potential energy $W$, namely $\beta\equiv T/|W|\gtrsim0.27$, neutron stars are subject to the dynamical bar-mode instability. Through this instability the star is deformed into a bar by virtue of the nonlinear growth of the  $\ell=2$ oscillation mode ($\ell$ being the spherical harmonic index) which leads to the emission of high-frequency (kHz) gravitational waves~\cite{Shibata2000,Baiotti2007,Loeffler2015}. As the degree of differential rotation increases, rotating stars are dynamically unstable against bar-mode deformation even for values of $\beta$ of order 0.01~\cite{Shibata2002,Watts:2005,Cerda-Duran2007,Corvino2010,Yoshida2017}. Moreover, highly differentially rotating neutron stars can also become unstable to a dynamical one-arm ($m=1$, spiral) instability~\cite{Centrella2001,Saijo2006}. At lower rotation rates secular nonaxisymmetric instabilities can also appear, driven by gravitational radiation (through the Chandrasekhar-Friedman-Schutz mechanism) or by viscosity (the latter, however, not being a generic instability in rotating neutron stars). 

Interestingly, this phenomenology might not be exclusive of rotating compact bodies composed only of fermionic matter. Recently we have shown through numerical-relativity simulations of spinning bosonic stars~\cite{sanchis2019nonlinear,DiGiovanni:2020ror} that those hypothetical objects can also be affected by the same type of dynamical bar-mode instabilities that operate in rapidly-rotating neutron stars. Bosonic stars are self-gravitating compact objects that can be constructed by minimally coupling a complex, massive bosonic field, either scalar or vector, to Einstein's gravity~\cite{Kaup:1968zz,Ruffini:1969qy,brito2016proca}. They can form dynamically from incomplete gravitational collapse through the gravitational cooling mechanism~\cite{Seidel:1993zk,di2018dynamical} and are composed of ultralight bosonic fields that could account for (part of) dark matter. The fields' particles have masses that range from $10^{-10}$ to $10^{-22}$ eV and have been motivated by String Theory~\cite{Arvanitaki:2009fg,Arvanitaki:2010sy} and by simple extensions of the Standard Model of particles~\cite{Freitas:2021cfi}. Such stars could be detected through their gravitational-wave emission in mergers~\cite{CalderonBustillo:2020srq} or through their effective shadow~\cite{olivares,Herdeiro:2021lwl}. Both, linear analysis and numerical simulations have shown that spherical bosonic stars are dynamically robust~\cite{Gleiser:1988ih,Lee:1988av,Seidel:1990jh,jetzer:1992,escorihuela2017quasistationary,sanchis2017numerical} (see~\cite{liebling2017dynamical} for a review). However, spinning bosonic stars can undergo bar-mode deformation~\cite{sanchis2019nonlinear,DiGiovanni:2020ror}, during which the angular momentum of the star is emitted and the star decays into a non-spinning configuration. In particular, spinning scalar mini-boson stars without self-interaction terms in the potential and some spinning vector boson star models are bar-mode unstable. Mechanisms to stabilize unstable bosonic stars, either in spherical symmetry or in the rotating case, have been studied recently. Those include combinations of independent bosonic fields only interacting through gravity, such as $\ell-$boson stars~\cite{Alcubierre:2018ahf,Alcubierre:2019qnh,jaramillo2020dynamical} and multistate, multifield boson stars~\cite{Bernal:2009zy,Guzman:2019gqc,guzman2021stability,sanchis2021multifield} as well as the addition of self-interaction terms in the potential~\cite{Siemonsen:2020hcg,Sanchis-Gual:2021phr}. In the former two cases the combination of a stable bosonic star with an unstable one stabilizes the mixture, even in the spinning case. 

These recent findings provide a theoretical motivation to study what could be the possible impact of adding a bosonic field to a rapidly rotating neutron star, particularly regarding the development of the bar-mode instability of the star. In addition to neutron stars and boson stars, macroscopic composites of fermions and bosons, dubbed fermion-boson stars, have also been proposed~\cite{HENRIQUES1990511,jetzer1990,valdez2013dynamical,brito2015accretion,brito2016interaction,valdez2020fermion,DiGiovanni:2020frc,Roque:2021,kain2021fermion,sanchis2022ultralight}. Such mixed configurations could form from the condensation of some primordial gas containing both types of particles or through episodes of accretion. These mixed configurations conform an extended parameter space that depends on the combination of the number of fermions and (ultralight) bosons. While hypothetical there have been proposals to endow these compact objects with potential astrophysical relevance. For example, in~\cite{DiGiovanni:2021ejn} spherically symmetric fermion-boson stars have been proposed to help explain the tension in the measurements of neutron star masses and radii reported in recent multi-messenger observations and nuclear-physics experiments. 

In this work we perform numerical-relativity simulations of three unstable differentially rotating neutron stars with an initial bosonic field distribution surrounding the star (the field can be both scalar and vector). We explore the effects of the field on the dynamics of the neutron stars by varying the initial energy of the cloud, from a small fraction to a mass comparable to that of the neutron star. In addition, we also consider three different values of the bosonic particle mass $\mu$. Our simulations show that, in all cases, the bosonic field is quickly accreted by the neutron star and condensates into a non-spinning bosonic star within its rotating fermion counterpart - a dark matter core. The impact of this core on the development of the bar-mode instability is noticeable. We find that the larger the bosonic total mass and the lower $\mu$, the instability takes longer to set in. However, within the range of parameters of our study, the bar-mode deformation of the neutron star seems an inavoidable outcome. On the other hand, the modification in the dynamics of the composite star affects significantly the associated gravitational-wave emission as compared to the case of a bar-mode unstable neutron star without a bosonic core.

This paper is organized as follows: in Section~\ref{sec2} we introduce the matter model we employ and set up the basic equations of motion to solve. Section~\ref{sec3} addresses the issue of initial data. The numerical framework for our simulations is described in Section~\ref{numerics} while the results and analysis of those simulations are presented in Section~\ref{results}. Finally, we outline our conclusions and final remarks in Section~\ref{conclusions}. Throughout this work we use units such that the relevant fundamental constants are equal to one ($G=c=M_{\odot}=1$).

\section{Formalism}
\label{sec2}
%%%%%%%%%%%%%%%%%%%%%%

\subsection{Equations of motion}

We assume that bosonic and fermionic matter are both minimally coupled to Einstein's gravity,
\begin{equation}
 R_{\alpha\beta}-\frac{1}{2}g_{\alpha\beta}R=8\pi T_{\alpha\beta} \,.
\label{eq:Einstein}
\end{equation}
Therefore, the total stress-energy tensor describing the matter content is given by the superposition of both contributions, one coming from a perfect fluid and the other from a scalar/vector complex field:
\begin{eqnarray}
T_{\mu\nu}&=& T_{\mu\nu}^{\rm{fluid}} + T_{\mu\nu}^{({\rm s})} ,
\end{eqnarray}
where superscript $({\rm s})$ stands for the spin of the bosonic particle, i.e.~$0$ for the case of a scalar field and $1$ for a vector (Proca) field. The contribution for the perfect fluid reads
\begin{equation}
T_{\mu\nu}^{\rm{fluid}} = [\rho(1+\epsilon) + P] u_{\mu}u_{\nu} + P g_{\mu\nu}\,,
\end{equation}
where $P$ is the pressure of the perfect fluid, $\rho$ its rest-mass density, $\epsilon$ its specific internal energy, and $u^{\mu} =(W/\alpha,W(v^{i}-\beta^{i}/\alpha)) $ is the fluid's 4-velocity, $W$ being the Lorentz factor and $v^{i}$ the fluid 3-velocity as seen by Eulerian observers. The contributions from the bosonic field are specified in the subsection~\ref{RNS-Scalar-eqs} and~\ref{RNS-Proca-eqs}.

The evolution equations are given by Einstein's equations $(\ref{eq:Einstein})$, by the conservation laws of the fluid stress-energy tensor and baryonic particles
\begin{eqnarray} 
\nabla_{\mu}T^{\mu\nu}_{\rm{fluid}} = 0\,, \label{conservation_laws1} \\
\nabla_{\mu}(\rho u^{\mu}) = 0\,, \label{conservation_laws2} 
\end{eqnarray}
together with a choice of an equation of state (EoS) for the fluid, and by the equations of motion for the bosonic field. For the construction of the initial data we consider a polytropic EoS,
\begin{equation}
P = K \rho^{\Gamma},
\end{equation}
with $K=100$ and $\Gamma=2$. The equations of motion of the bosonic field are the Klein-Gordon equation for a complex scalar field $\phi$,
\begin{equation}
\nabla_{\mu}\nabla^{\mu} \phi= \mu_{(0)}^2 \phi \,, \label{Klein-Gordon}
\end{equation}
 and the Proca equations for a complex vector field $\mathcal{A}^{\mu}$,
\begin{equation}
\nabla_{\mu} \mathcal{F}^{\mu\nu} + \mu_{(1)}^2 \mathcal{A}^{\nu} = 0\,. \label{Proca-equations}
\end{equation}
In the previous equations $\nabla_{\mu}$ is the covariant derivative with respect to the metric $g_{\mu\nu}$ and $\mu_{({\rm s})}$ is the mass of the particle for the scalar field (${\rm s}=0$) or the vector field (${\rm s}=1$). We consider the spacetime line element
\begin{eqnarray}
ds^2&=&g_{\mu\nu}dx^{\mu}dx^{\nu} \nonumber \\
    &=&-(\alpha^2-\beta_{i}\beta^{i})dt^2 + 2\gamma_{ij}\beta^idtdx^j + \gamma_{ij}dx^idx^j,
\end{eqnarray}
where $\alpha$ is the lapse function, $\beta^i$ is the shift vector, and $\gamma_{ij}$ is the spatial metric. We employ the Baumgarte-Shapiro-Shibata-Nakamura (BSSN) formulation of Einstein’s equations~\cite{nakamura1987general,Shibata95,baumgarte1998numerical}, references to which the reader is addressed for details. The BSSN equations involve energy-momentum source terms, namely the energy density $\E$, the momentum density $j_{i}$ measured by a normal observer $n^{\mu}$, and the spatial projection of the stress-energy tensor $S_{ij}$, which read
\begin{align}
\E&= n^{\mu}n^{\nu}T_{\mu \nu}, \\
j_i&=-\gamma_{i}^{\mu}n^{\nu}T_{\mu \nu}, \\
S_{ij}&= \gamma_{i}^{\mu} \gamma_{j}^{\nu} T_{\mu \nu},
\end{align}
 where the unit normal vector is $n^{\mu} = \frac{1}{\alpha}(1,-\beta^{i})$ and $\gamma_{i}^{\mu}$ is the spatial projection operator. The source terms for the perfect fluid read
\begin{eqnarray}\label{matter-NS}
\E^{\rm{fluid}} &=& (\rho(1+\epsilon)+P)W^2 -P\,, \label{rhoFS}\\
j^{\rm{fluid}}_i&=& (\rho(1+\epsilon)+P) W^2 v_{i}\,,\\
S^{\rm{fluid}}_{ij}&=& (\rho(1+\epsilon)+P) W^2 v_{i}v_{j} + \gamma_{ij} P\,.
\end{eqnarray}
% and the ones from the bosonic field are described in the following subsections.

%%%%%%%%%%%%%%%%%%%%%%

%%%%%%%%%%%%%%%%%%%%%%
\subsection{Einstein-Klein-Gordon-Euler system} \label{RNS-Scalar-eqs}
%%%%%%%%%%%%%%%%%%%%%%
The stress-energy tensor associated with the scalar field $\phi$ is
\begin{eqnarray}
T_{\mu\nu}^{(0)}= - \frac{1}{2}g_{\mu\nu}\partial_{\alpha}\bar{\phi}\partial^{\alpha}\phi - V(\phi) 
%\nonumber \\
 + \frac{1}{2}(\partial_{\mu}\bar{\phi}\partial_{\nu}\phi+\partial_{\mu}\phi\partial_{\nu}\bar{\phi})\,, 
 \nonumber \\
 \label{Tmunu-scalar} 
\end{eqnarray}
where for the potential of the scalar field we consider that of a mini-boson star~\cite{liebling2017dynamical},
\begin{eqnarray}
V(\phi) =  \frac{1}{2} \mu_{(0)}^2\bar{\phi}\phi\,.
\end{eqnarray}
In the previous two equations the bar symbol denotes complex conjugation. As customary, in order to write the Klein-Gordon equation~(\ref{Klein-Gordon}) as a first-order system we introduce the scalar-field conjugate momentum
\begin{equation}\label{scalarPi}
\Pi = -\frac{1}{\alpha}(\partial_t - \mathcal{L}_{\beta})\,\phi\,.
\end{equation}
The source terms for this system read
\begin{align}\label{matter-scalar}
\E^{(0)}&= \frac{1}{2} \left(\bar{\Pi}\,\Pi + \mu_{(0)}^{2} \bar{\phi}\phi + \frac{1}{2} \lambda (\bar{\phi}\phi)^{2} + D^{i}\bar{\phi}D_i\phi\right)\,, \\
j^{(0)}_i&=\frac{1}{2}(\bar{\Pi}\nabla_{i}\phi + \Pi \nabla_{i}\bar{\phi})\,, \\
S^{(0)}_{ij}&= \frac{1}{2} (\nabla_{i}\bar{\phi}\nabla_j\phi + \nabla_{j}\bar{\phi}\nabla_i\phi) + \frac{1}{2}\gamma_{ij}(\bar{\Pi}\,\Pi \nonumber \\
	& -\mu_{(0)}^{2} \bar{\phi}\,\phi - \frac{1}{2} \lambda (\bar{\phi}\phi)^{2} - D^k\bar{\phi}\nabla_{k}\phi)\,.
\end{align}
The set of evolution equations for the scalar field are described in~\cite{Okawa:2014nda}.

%%%%%%%%%%%%%%%%%%
\subsection{Einstein-Proca-Euler system} \label{RNS-Proca-eqs}
%%%%%%%%%%%%%%%%%
The stress-energy tensor for a vector field $\mathcal{A}_{\mu}$ is
\begin{eqnarray}
T_{\mu\nu}^{(1)}&=& -\mathcal{F}_{\lambda(\mu}  \bar {\mathcal{F}}_{\nu)}^{\,\,\lambda}-\frac{1}{4}g_{\mu\nu}\mathcal{F}_{\lambda\alpha}\bar{\mathcal{F}}^{\lambda\alpha} 
\nonumber \\
&+& \mu^2_{(1)} \left[
\mathcal{A}_{(\mu}\bar{\mathcal{A}}_{\nu)}-\frac{1}{2}g_{\mu\nu}\mathcal{A}_{\lambda}\bar{\mathcal{A}}^{\lambda}
\right]\,, \label{Tmunu-Proca}
\end{eqnarray}
 where $\mathcal{F}_{\mu\nu} = \nabla_{\mu}\mathcal{A}_{\nu} - \nabla_{\nu}\mathcal{A}_{\mu}$ is the field strength, and the index notation $(\mu, \nu)$ indicates, as usual, index symmetrization. We cast the splitting of the Proca $1-$form $\mathcal{A}^{\mu}$ into its scalar potential $\mathcal{X}_{\phi}$, its $3$-vector potential $\mathcal{X}_{i}$, and the 3-dimensional electric $E_{i}$ and magnetic $B_i$ field, defined by
\begin{eqnarray}\label{matter-vector}
\mathcal{X}_{\phi}&=&-n^{\mu}\mathcal{A}_{\mu}\  , \label{propot}\\
\mathcal{X}_{i}&=&\gamma^{\mu}_{i}\mathcal{A}_{\mu}\ ,\\
E^{i}&=&-i\,\frac{\gamma^{ij}}{\alpha}\,\biggl(D_{j} (\alpha\mathcal{X}_{\phi})+\partial_{t}\mathcal{X}_{j}\biggl) ,\\ 
B^{i}&=& \epsilon^{ijk} D_{j}\mathcal{X}_k,
\end{eqnarray}
where $\epsilon^{ijk}$ is the Levi-Civita tensor. The  energy-momentum source terms for this system read
\begin{align}
\E^{(1)}&=\frac{1}{2} \gamma_{ij}(\bar{E}^iE^j + \bar{B}^iB^j) + \frac{1}{2}\mu_{(1)}^2(\bar{\mathcal{X}}_{\phi}\mathcal{X}_{\phi} + \gamma^{ij}\bar{\mathcal{X}}_{i}\mathcal{X}_{j}) , \\
j^{(1)}_i&=\frac{1}{2}\mu_{(1)}^2 (\bar{\mathcal{X}_{\phi}}\mathcal{X}_{i} + \mathcal{X}_{\phi}\bar{\mathcal{X}}_{i}), \\
S^{(1)}_{ij}&= -\gamma_{ik}\gamma_{jl}(\bar{E}^k E^l + \bar{B}^k B^l) + \frac{1}{2}\gamma_{ij}(\bar{E}^k E_{k}  \nonumber \\
&+ \bar{B}^k B_{k} + \mu_{(1)}^2 \bar{\mathcal{X}}_{\phi}\mathcal{X}_{\phi} - \mu_{(1)}^2 \bar{\mathcal{X}}^k \mathcal{X}_{k}) + \mu_{(1)}^{2}\bar{\mathcal{X}}_{i} \mathcal{X}_{j}.
\end{align}
The set of evolution equations for the Proca field are described in~\cite{Zilhao:2015tya}.

%%%%%%%%%%%%%%%%%%%%%%
\section{Initial Data}
\label{sec3}
%%%%%%%%%%%%%%%%%%%%%%

We construct configurations describing a cloud of bosonic matter surrounding a rotating neutron star (RNS) model. As scalar spinning mini-boson stars may develop non-axisymmetric instabilities, as shown in~\cite{sanchis2019nonlinear,DiGiovanni:2020ror,Siemonsen:2020hcg}, we consider purely spherically symmetric scalar field clouds with zero angular momentum. However, for the vector field clouds we construct also models with non-zero angular momentum. To obtain physical initial data it is mandatory to solve the Einstein Hamiltonian and  momentum constraint equations. Moreover, for the case of a vector field, an additional constraint comes into play, the Gauss constraint $D_{i} E^{i} = \mu_{(1)}^2 \mathcal{X}_{\phi}$, where $D_i$ stands for the covariant derivative with respect to the 3-metric $\gamma_{ij}$. In this section we schematically describe the procedure to construct constraint-satisfying initial data for the physical situation we are considering. 

We begin by building highly differentially RNS models, which we choose among the bar-mode unstable models considered in~\cite{Baiotti2007}. We employ the \textsc{RNS} numerical code~\cite{Stergioulas:1995b} to construct such configurations. We then add a bosonic cloud assuming a harmonic time dependence and a particular cloud ``shape". For spherically symmetric scalar clouds we consider a Gaussian radial profile for the scalar field, yielding
\begin{equation}
\phi(r,t) = A_{0}\,e^{-\frac{r^2}{\sigma^2}} e^{i\omega t}, \label{scalarfield}
\end{equation}
where parameters $A_{0}$ and $\sigma$ are the amplitude and the width of the Gaussian shell, respectively, and $\omega$ is the initial frequency of the field. The ansatz for the vector field is more involved as there are several field component involved and we must also solve the Gauss constraint. We address the interested reader to the appendix of~\cite{DiGiovanni:2020ror} and to~\cite{Zilhao:2015tya} for specific details about the initial data for the vector field case.

Once we have constructed a RNS model and the surrounding bosonic cloud, we can evaluate the source terms entering in the Hamiltonian and momentum constraints. Those are going to be simply the sum of the terms from the fermionic matter~\ref{matter-NS} and the ones from either the scalar field~\ref{matter-scalar} or the vector field~\ref{matter-vector}. Finally as initial guess for the spacetime variables we consider those take the values of the isolated RNS and we solve the constraint equations with the updated matter source terms iteratively until  convergence is reached. This procedure is described in more detail in the appendix of~\cite{DiGiovanni:2020ror} and it relies on the so-called conformally flat approximation (CFC) of the full Einstein equations, as described in~\cite{CorderoCarrion:2008nf}. 

To characterize our initial models we compute several physical quantities: the angular velocity of the fluid $\Omega$, the baryonic mass $M_{0}$, the gravitational mass $M_{\rm{grav}}$ , the internal energy $E_{\rm{int}}$, the angular momentum $J_{\rm{NS}}$, the rotational kinetic energy $T$, and the gravitational binding energy $W$ of the RNS, respectively defined as
\begin{eqnarray}
\Omega &=& \frac{u^{\phi}}{u_{t}}\,, \\
M_{0}&=& \int d^3x D \sqrt{\gamma}\,,\\
M_{\rm{grav}}&=& \int d^3x (-2T^{0}_{0} + T^{\mu}_{\mu}) \alpha \sqrt{\gamma}\,,\\
E_{\rm{int}}&=& \int d^3x D\epsilon \sqrt{\gamma}\,,\\
J_{\rm{NS}}&=& \int d^3x T^{0}_{\varphi} \alpha \sqrt{\gamma}\,,\\
T&=& \frac{1}{2} \int d^3x \Omega T^{0}_{\varphi} \alpha \sqrt{\gamma}\,,\\
W &=& M_{0} + E_{\rm{int}} + T - M_{\rm{grav}}\,. \label{eq:calcW}
\end{eqnarray}
We note that in the previous equations we employ only the stress-energy contribution from the perfect fluid but we omit the subscript ``fluid'' to simplify the notation. We also recall the notation for the ratio between the rotational and binding energy $\beta\equiv T/|W|$.

The bosonic cloud is instead characterized by the total mass and angular momentum stored in it, which are evaluated as:
\begin{eqnarray}
M_{\rm{cloud}}&=& \int d^3x (-2T^{0}_{0} + T^{\mu}_{\mu}) \alpha \sqrt{\gamma}\,,\\
J_{\rm{cloud}}&=& \int d^3x T^{0}_{\varphi} \alpha \sqrt{\gamma}\,,
\end{eqnarray}
where again we omit the subscript $({\rm s})$ which identifies the scalar/vector field.

In Table~\ref{table:models1} we summarize the parameters and the main properties of the models we consider for this study. We choose three RNS models, labelled D2, U7, and U13 (see~\cite{Baiotti2007} and references therein for details), and different cloud parameters for the scalar and Proca fields that surround them. Since the models used in this work are constructed using the CFC formalism, which is an approximation of the full Einstein equations, the global quantities of our RNS models show a small discrepancy with respect to the original models generated by the \textsc{RNS} numerical code. For instance, comparing the physical quantities ($M_{0}$, $M_{\rm{grav}}$, $T$, $E_{\rm{int}}$) in Table~\ref{table:models1} for model U13 with respect to those shown in~\cite{Baiotti2007}, a discrepancy of order $0.3\%$ is observed, which is expected for the 2PN error resulting from the CFC approximation. The discrepancy becomes higher for more compact neutron stars like model D2 where it is of order $1\%$. We note that the relative error in the evaluation of the binding energy $W$ is higher because it is a 1PN correction to the energy of the system. In practice this means that while the error is small for  $M_{0}$ and $M_{\rm{grav}}$ (e.g. $ 4\times 10^{-3}$ for model U13, for values of $M_{0} \sim M_{\rm{grav}} \sim1.5$), this same error, contributing to the error of $W$ through Eq.~\eqref{eq:calcW}, is larger for $W$ itself (which has a value of $W=7.452\times 10^{-2}$ for U13) resulting in a  $5\%$ error, still consistent with its expected post-Newtonian order.  A similar effect is observed for $\beta$, which is also a 1PN quantity.

\begin{table*}
\caption{Models of RNS with an accreting scalar/Proca cloud. From left to right the columns report: the name of the RNS model~\cite{Baiotti2007}, its central rest-mass density $\rho_{c}$, its baryon mass $M_{0}$, its gravitational mass $M_{\rm{grav}}$, its angular momentum $J_{\rm{NS}}$, its kinetic energy $T$, its binding energy $W$, the ratio between rotational and binding energy $\beta$, the type of bosonic cloud, the mass parameter of the scalar/vector boson $\mu$, the amplitude of the Gaussian profile $A_0$, and the total mass stored in the cloud $M_{\rm{cloud}}$. Most models have an $\ell=m=0$ bosonic cloud and the amplitude of the Gaussian profile $\sigma=60$. Only the last model in the Table corresponds to a spinning $\ell=1$, $m=\pm 1$ Proca cloud with angular momentum $J^{(1)}_{\rm{cloud}}=\pm1.315$. The width of the Gaussian cloud $\sigma$ is also indicated.}
\centering
\begin{tabular}{c | c c c c c c c | c c c c}
\hline
\multicolumn{12}{c}{\cellcolor[gray]{0.8}{\textbf{$l=m=0$, $\sigma=60$}}} \\
\hline

RNS model & $\rho_{c} (10^{-4})$ & $M_{0}$ & $M_{\rm{grav}}$ & $J_{\rm{NS}}$ & $T(10^{-2})$ & $W (10^{-2})$ & $\beta$ & Cloud & $\mu$ & $A_{0} (10^{-3})$ & $M_{\rm{cloud}}$  \\
\hline
U13 & 0.599  & 1.506 & 1.466 & 3.757 &2.188 & 7.452 & 0.294 & None& -  & - & - \\
U13-a & 0.599  & 1.600 & 1.521 & 3.980 & 2.319 & 11.53 & 0.201 & Scalar & 1.0  & 1.1  & 0.628 \\
U13-b & 0.599  & 1.600 & 1.521 & 3.980 & 2.319  & 11.53 & 0.201 & Scalar & 0.5  & 2.2  & 0.629 \\
U13-c & 0.599 & 1.592 & 1.516 & 3.961  & 2.308 & 11.17 & 0.206 & Scalar & 0.33  & 3.2  & 0.578 \\
U13-d & 0.599  & 1.556 & 1.496 & 3.874 & 2.256 & 9.595 & 0.235 & Scalar & 0.33  & 2.5  & 0.346 \\
U13-e & 0.599  & 1.531 & 1.481 & 3.811 & 2.220 & 8.499 & 0.261 & Scalar & 0.33  & 1.8  & 0.176 \\
U7 & 1.406  & 1.512 & 1.462 & 3.406 & 2.337 &  8.366 & 0.279 & None & - & - & - \\
U7-a & 1.406  & 1.563 & 1.495 & 3.523 & 2.418 &  10.79 & 0.224 & Scalar & 0.5 & 1.7 & 0.368	 \\
D2 & 3.154  & 2.752 & 2.614 & 7.583 & 9.211 & 30.50& 0.302 & None & -  & - & - \\
D2-a & 3.154  & 2.862 & 2.678 & 7.870 & 9.560 & 35.75 & 0.267 & Scalar & 0.33  & 1.9  & 0.222 \\
D2-b & 3.154 & 2.956 & 2.733 & 8.119  & 9.867 & 40.24 & 0.245 & Scalar & 0.5  & 1.6  & 0.372 \\
D2-c & 3.154 & 2.850 & 2.671 & 7.838 &9.520 & 35.18 & 0.270  & Scalar & 1.0  & 0.6  & 0.205 \\
U13-f & 0.599  & 1.591 & 1.507 & 3.960 & 2.306 & 12.00 & 0.192 & Proca & 0.5  & 23 & 1.182 \\
U13-g & 0.599  &  1.541 & 1.488 & 3.836 & 2.234 & 8.766 & 0.255 & Proca & 1.0 & 3.0 & 0.505 \\
U13-h & 0.599  & 1.590 &  1.520 & 3.958 & 2.306 & 10.65 & 0.216 & Proca & 1.0  & 4.5 & 1.172 \\
\hline
\multicolumn{12}{c}{\cellcolor[gray]{0.8}{\textbf{ $l=1$, $m=\pm 1$, $\sigma=40$}}} \\
\hline
U13-i & 0.599  & 1.658 &  1.564 & 4.122 & 1.382 & 13.17 & 0.182 & Proca & 1.0  & 3.0 $\times 10^{-4}$ & 1.302 \\
\hline
\end{tabular}
\label{table:models1}
\end{table*}

\begin{figure*}[t!]
\includegraphics[width=0.24\linewidth]{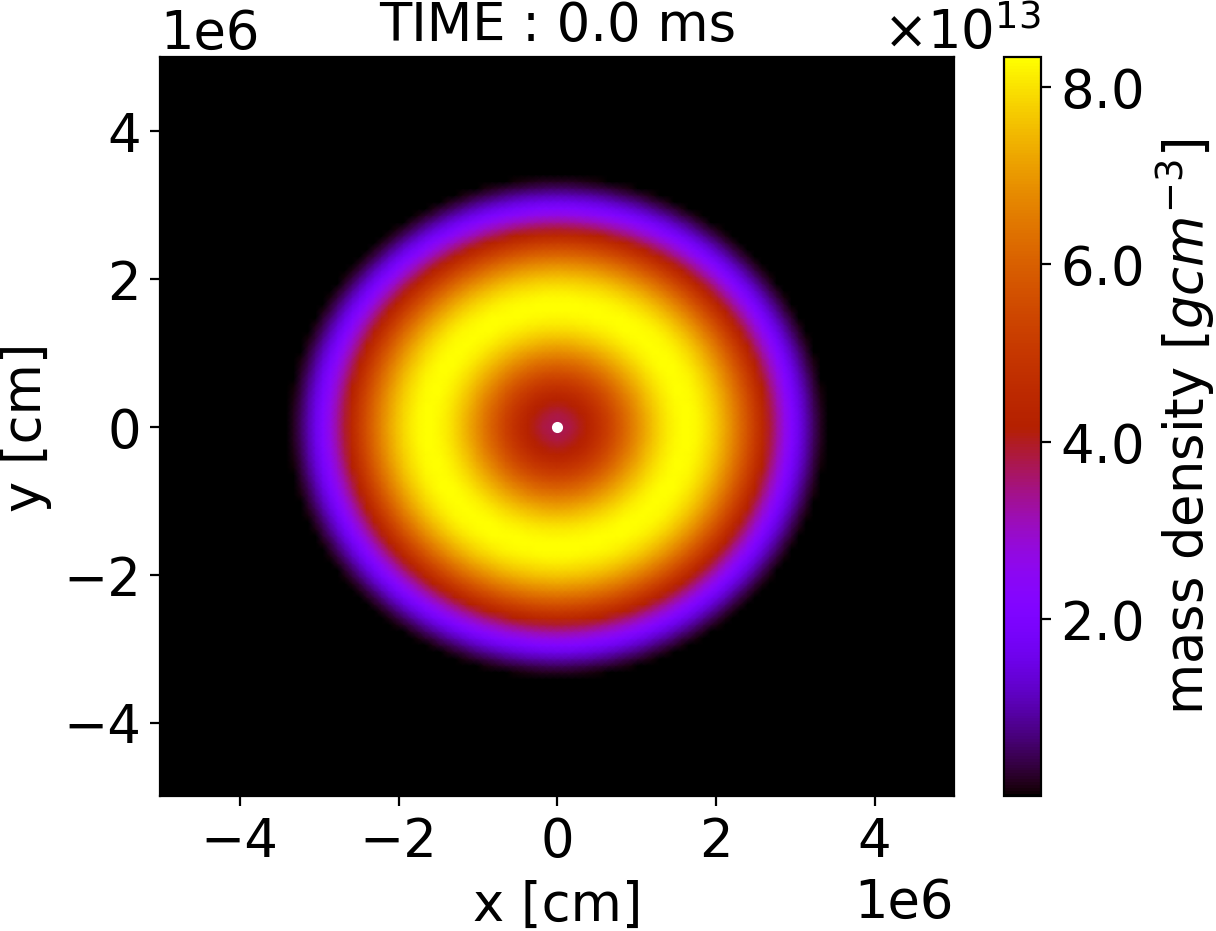}\hspace{-0.0\linewidth}
\includegraphics[width=0.24\linewidth]{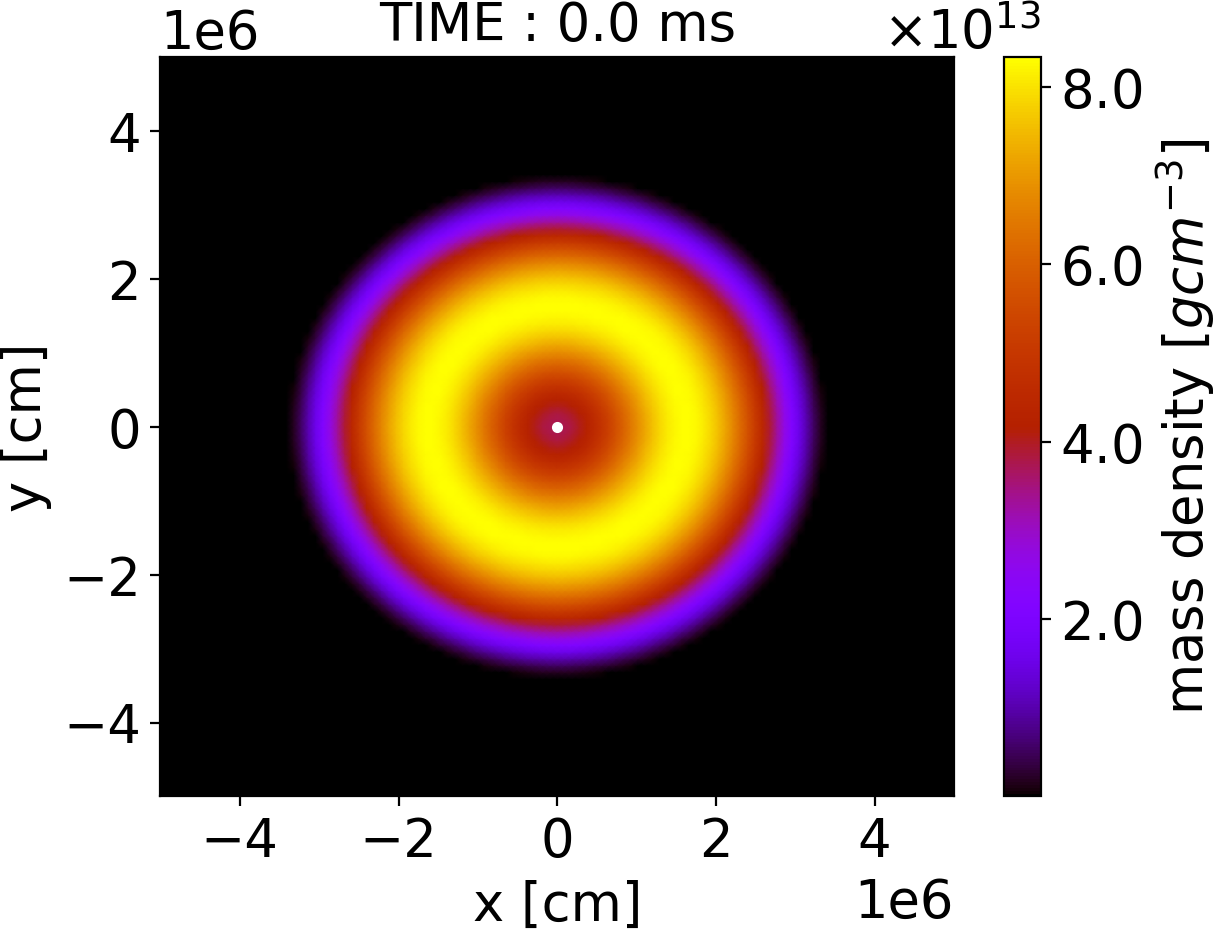}\hspace{-0.0\linewidth}
\includegraphics[width=0.24\linewidth]{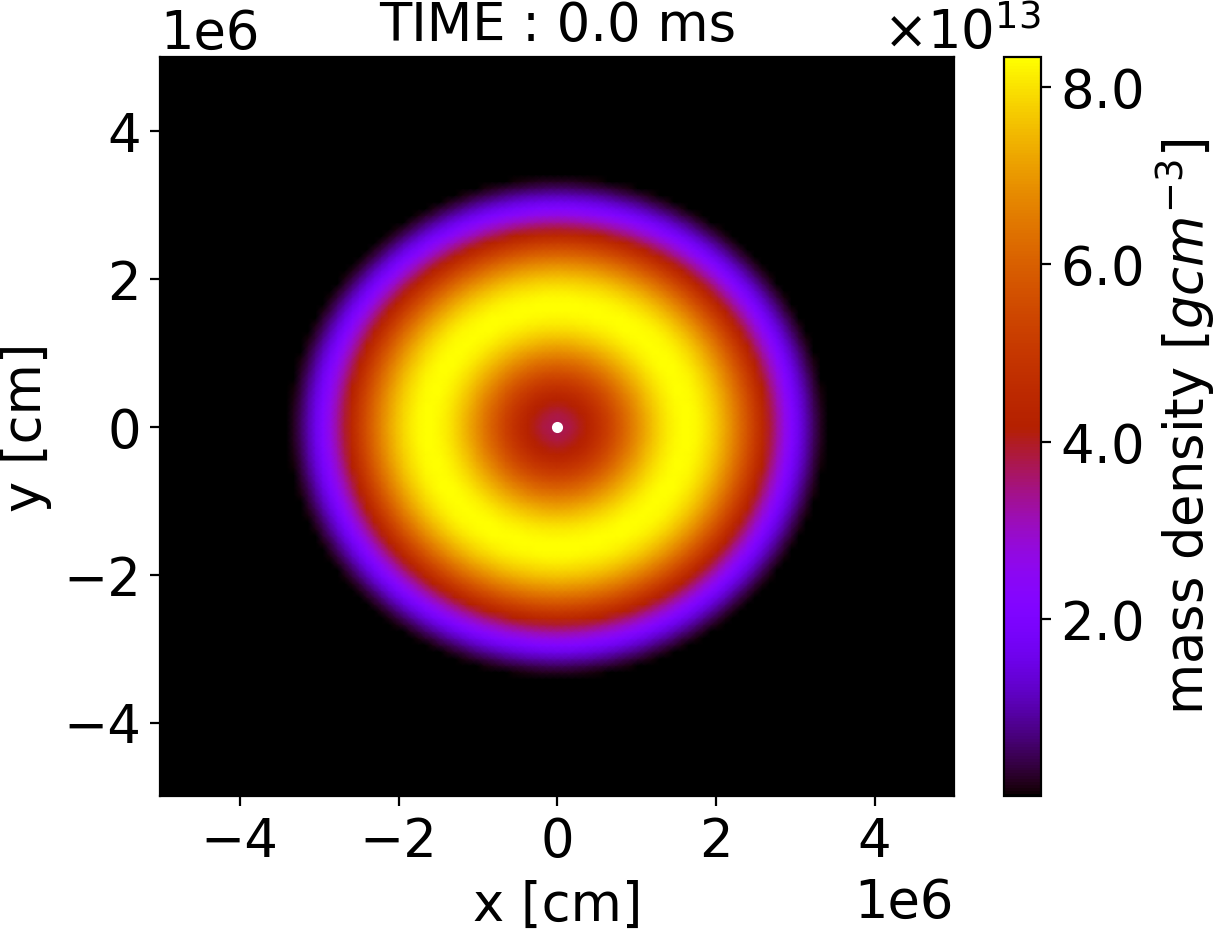}\hspace{-0.0\linewidth}
\includegraphics[width=0.24\linewidth]{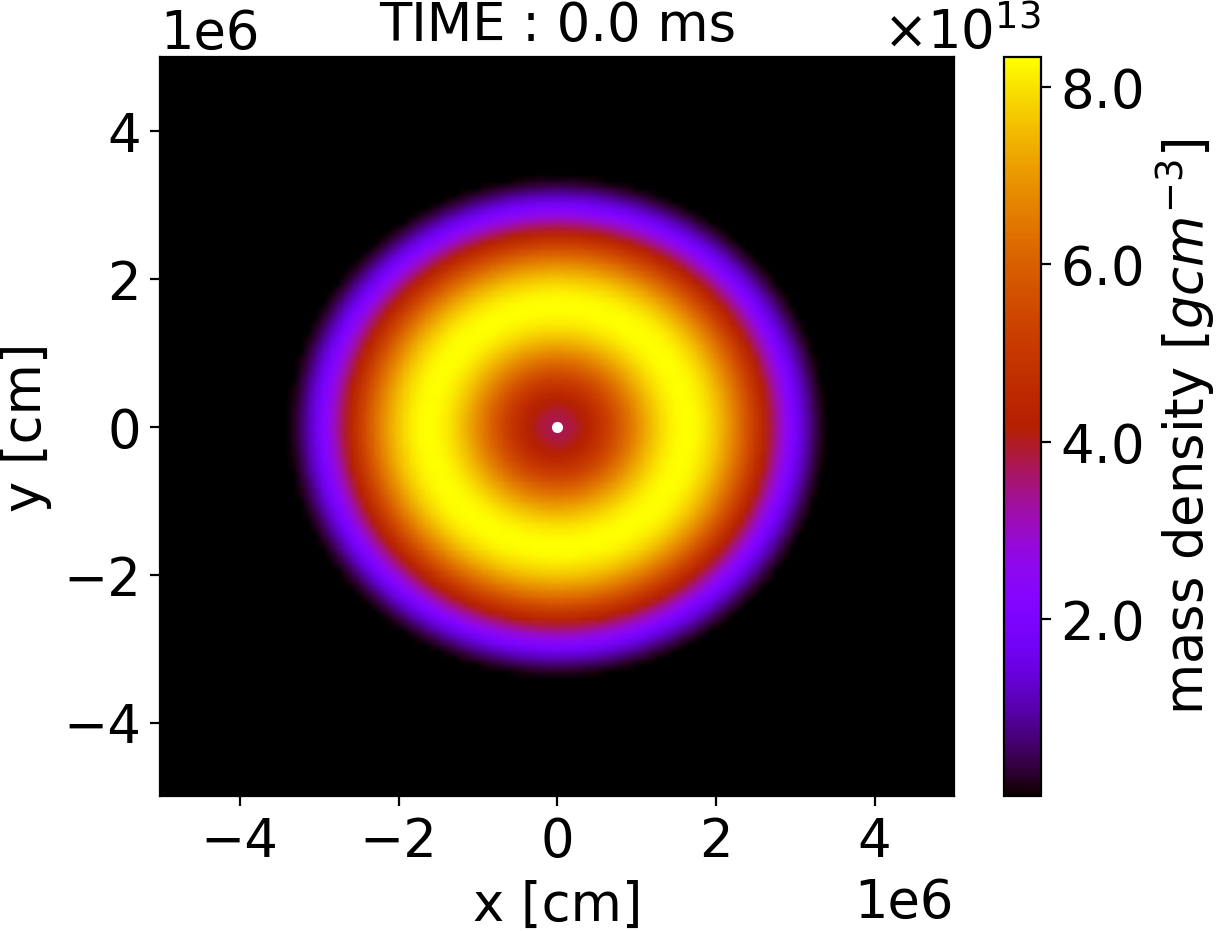}\\
\includegraphics[width=0.24\linewidth]{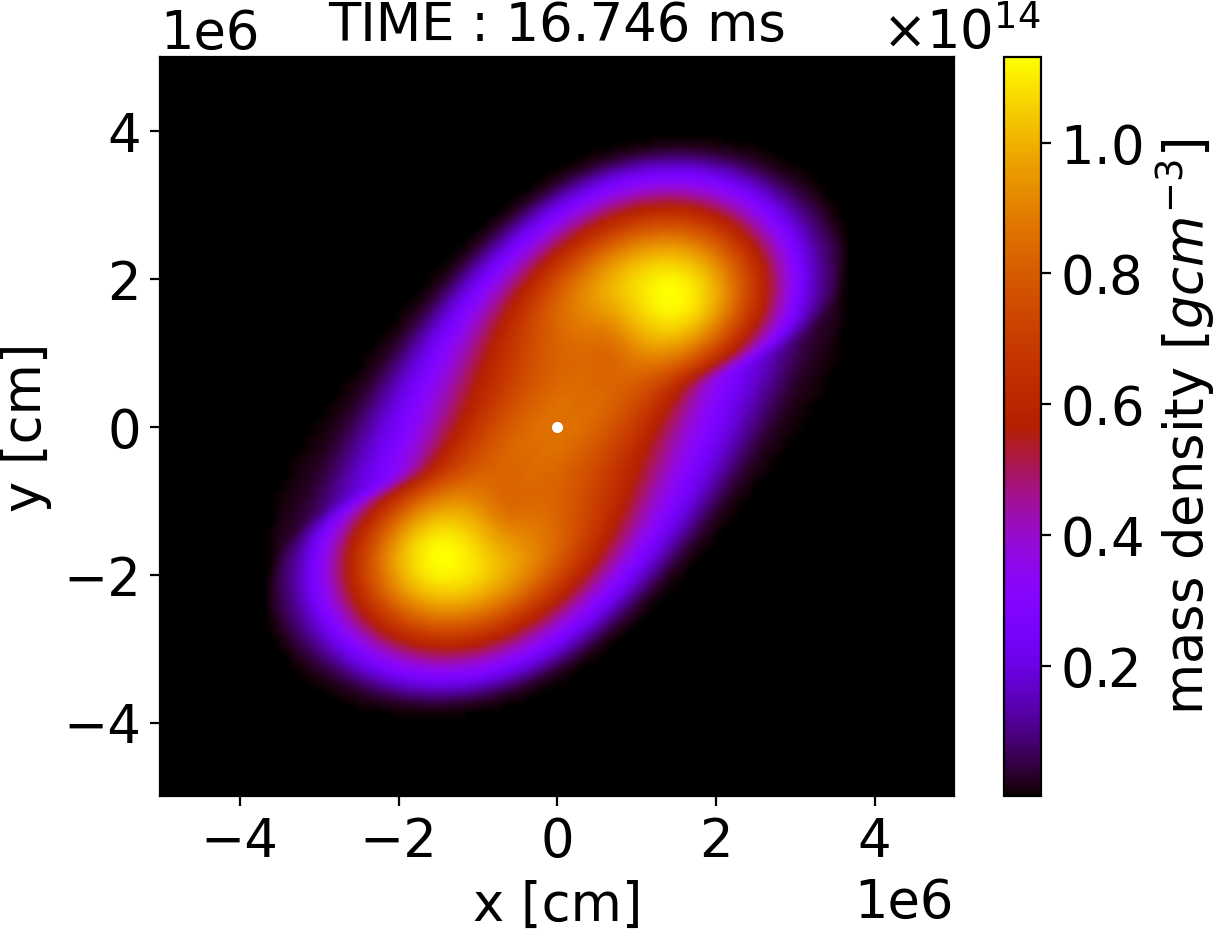}\hspace{-0.0\linewidth}
\includegraphics[width=0.24\linewidth]{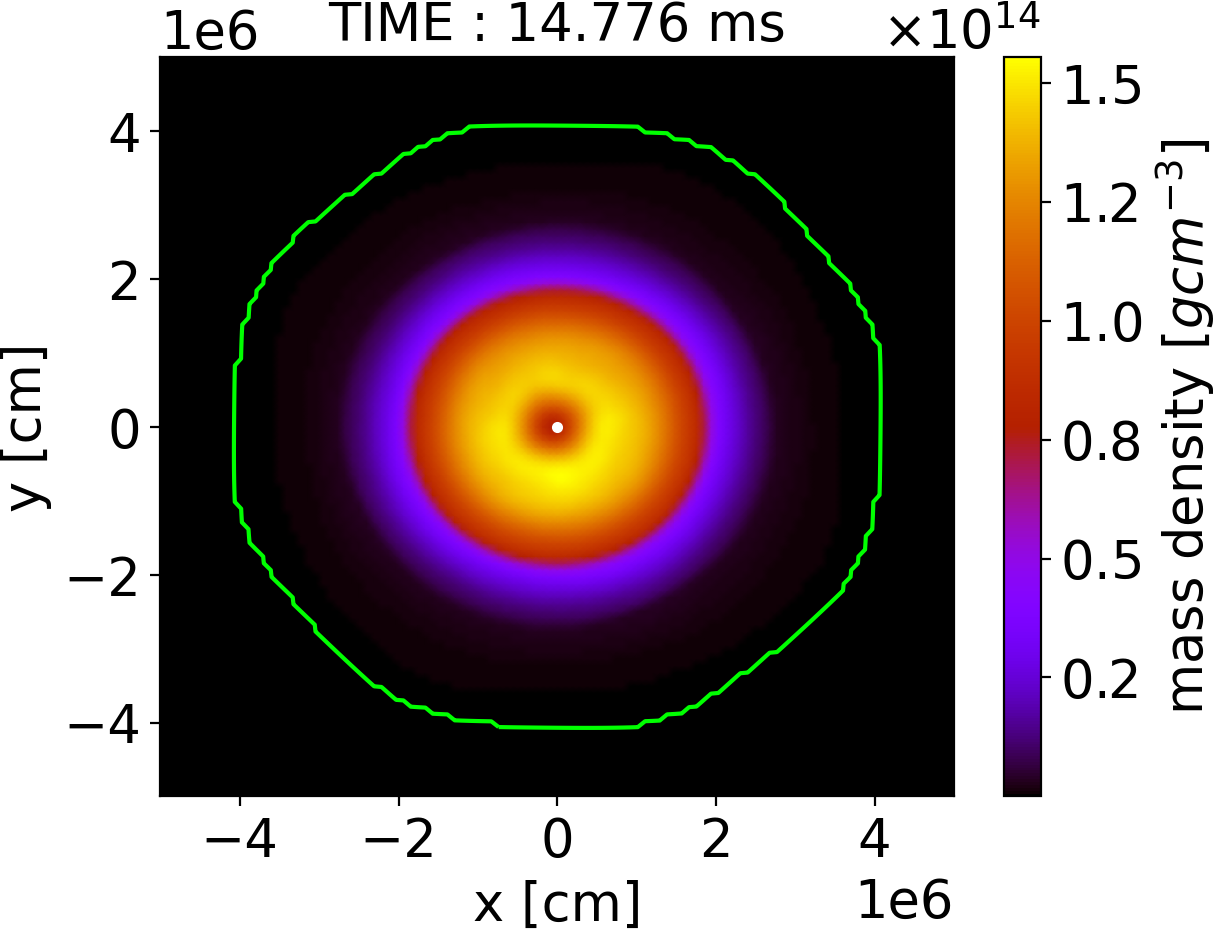}\hspace{-0.0\linewidth}
\includegraphics[width=0.24\linewidth]{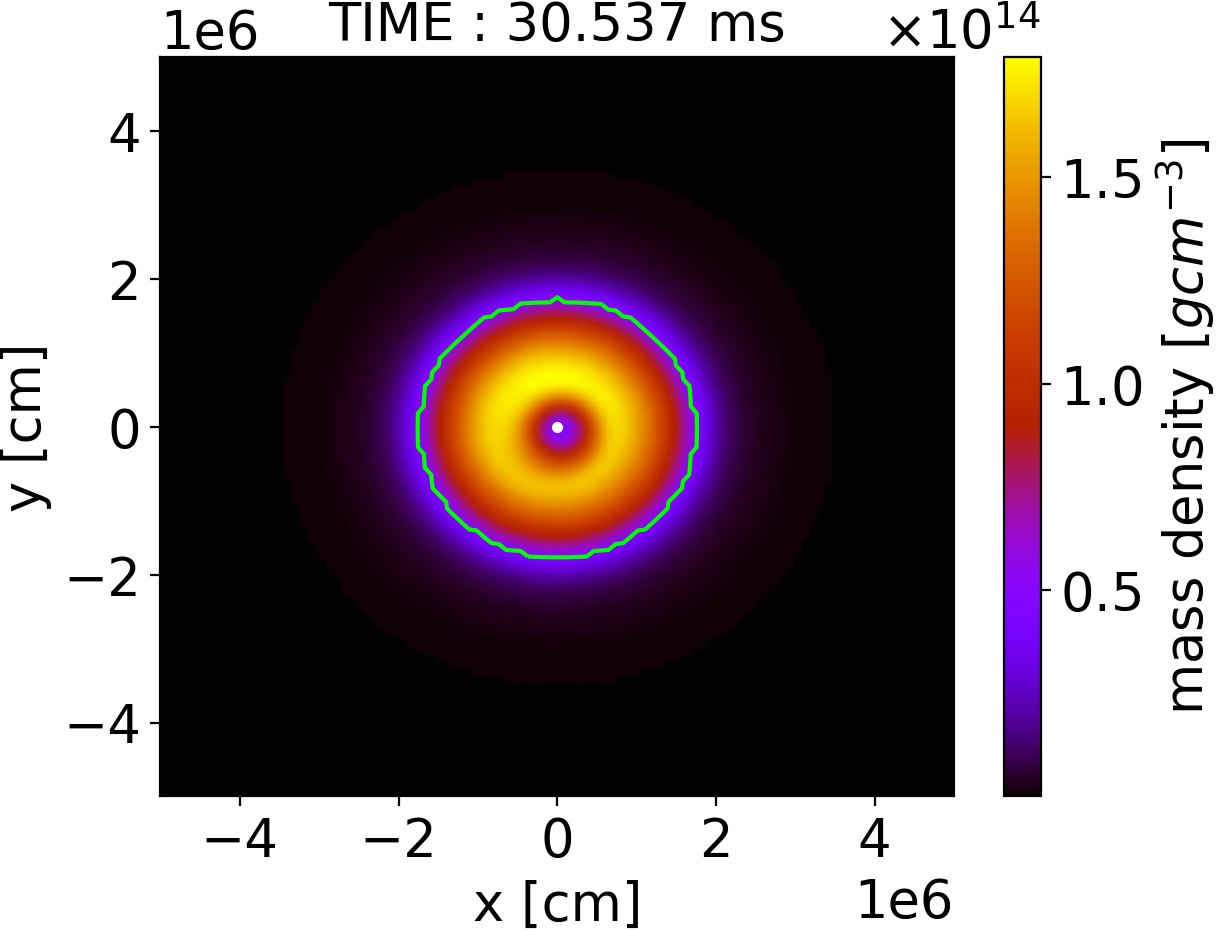}\hspace{-0.0\linewidth}
\includegraphics[width=0.24\linewidth]{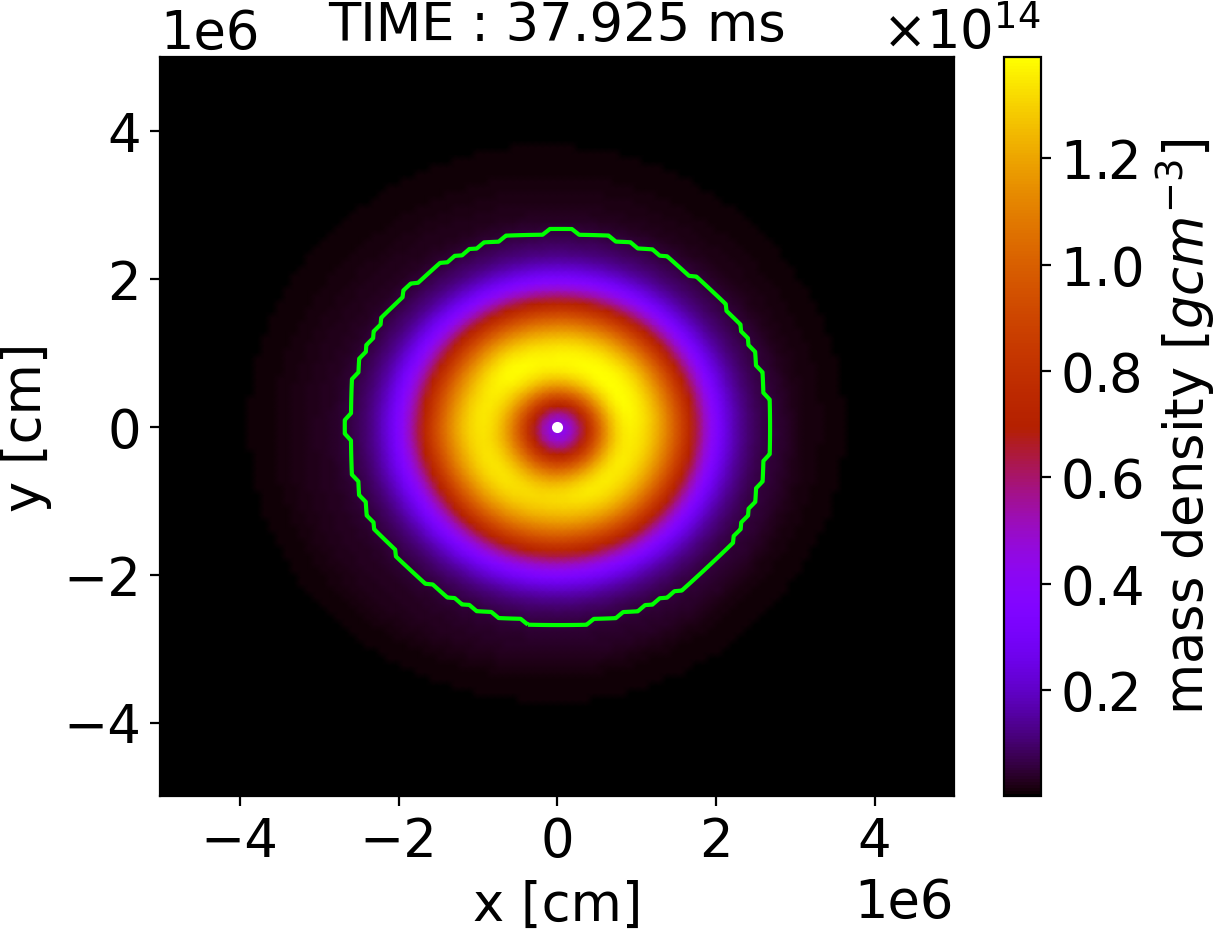}\\
\includegraphics[width=0.24\linewidth]{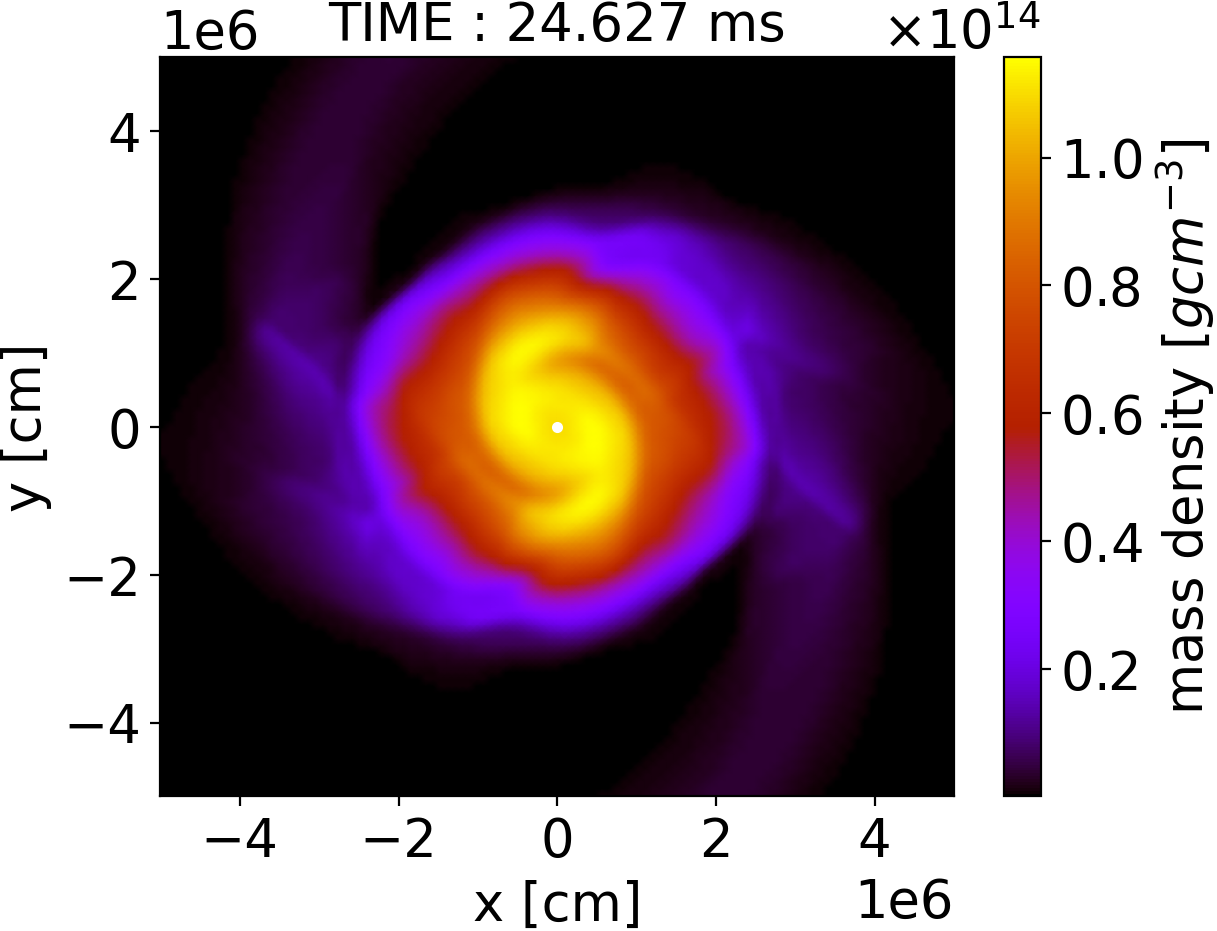}\hspace{-0.0\linewidth}
\includegraphics[width=0.24\linewidth]{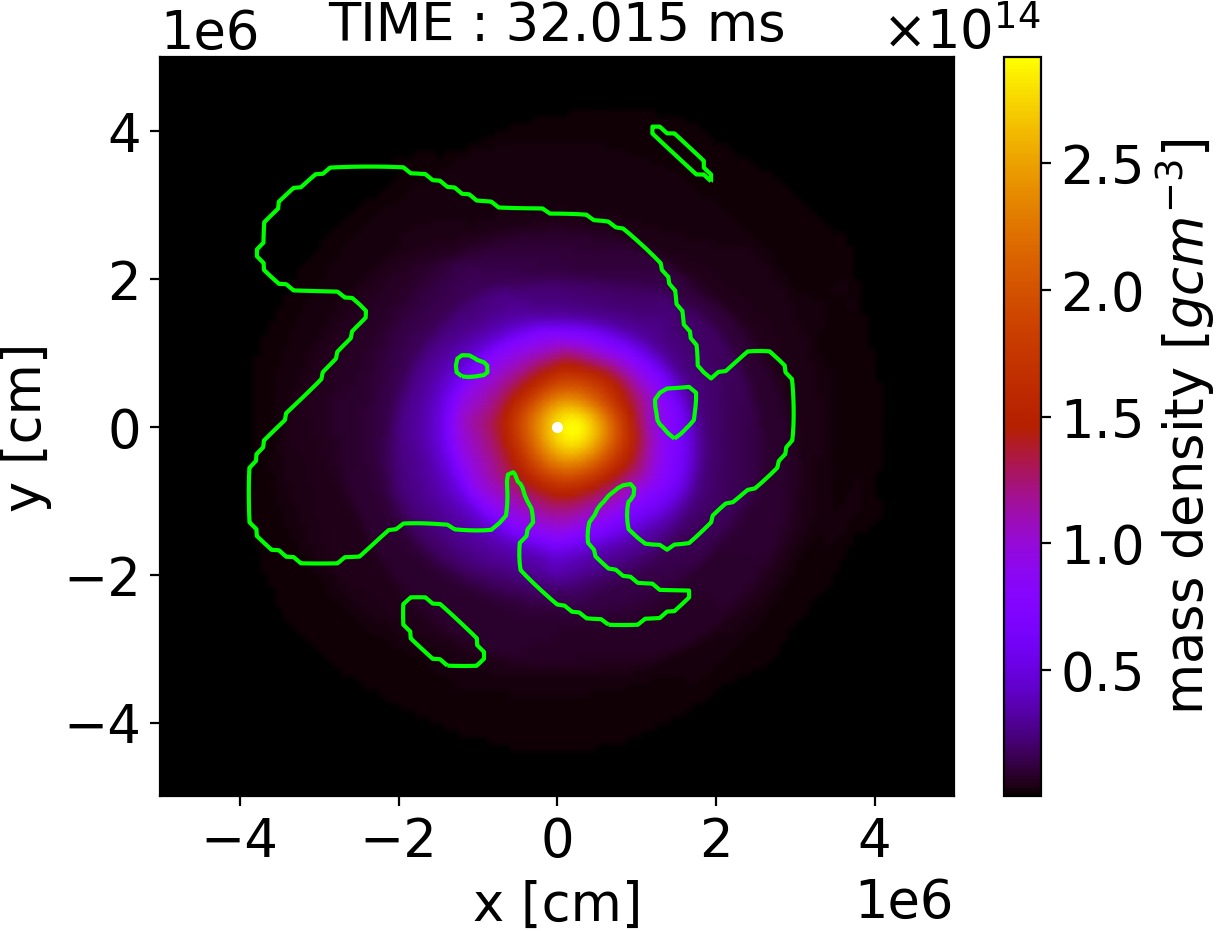}\hspace{-0.0\linewidth}
\includegraphics[width=0.24\linewidth]{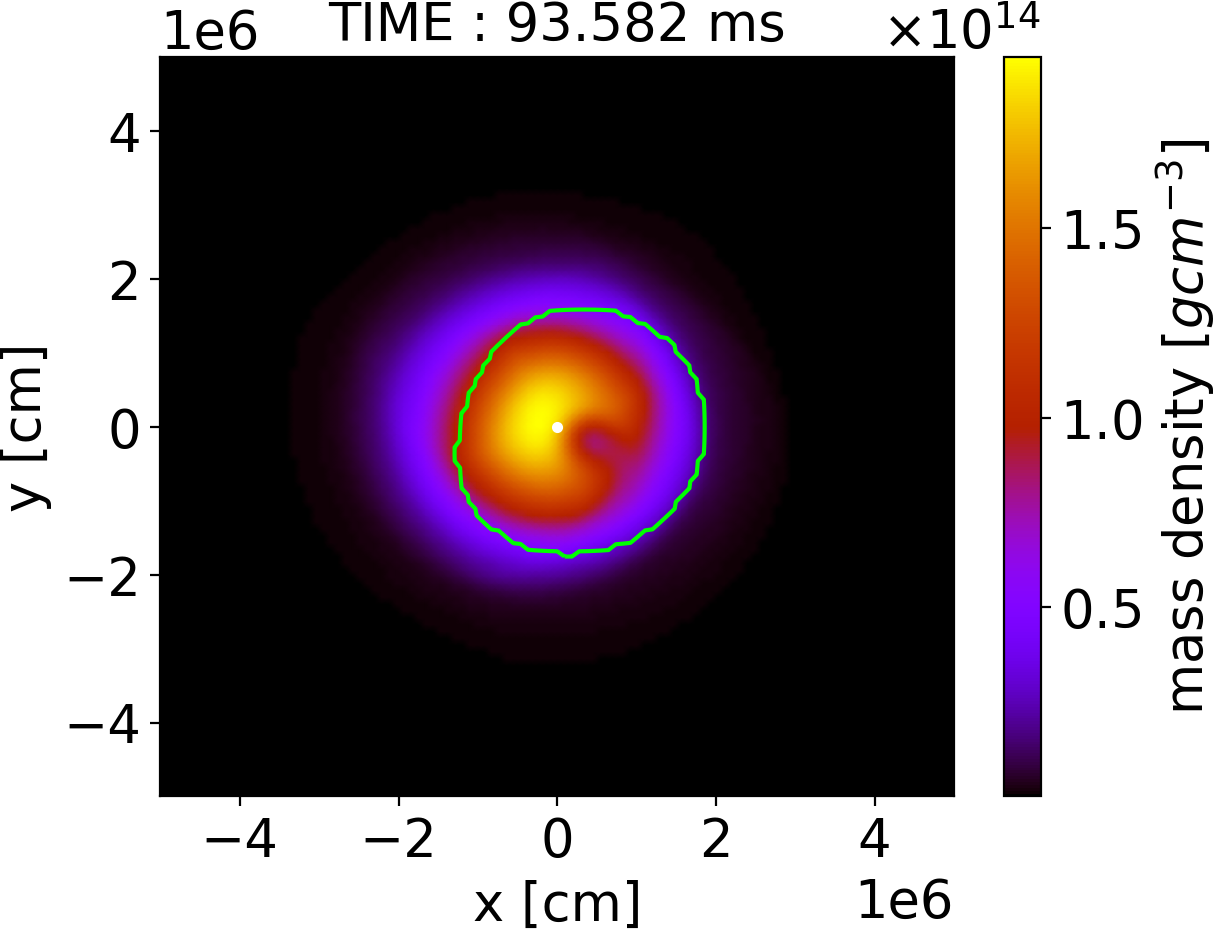}\hspace{-0.0\linewidth}
\includegraphics[width=0.24\linewidth]{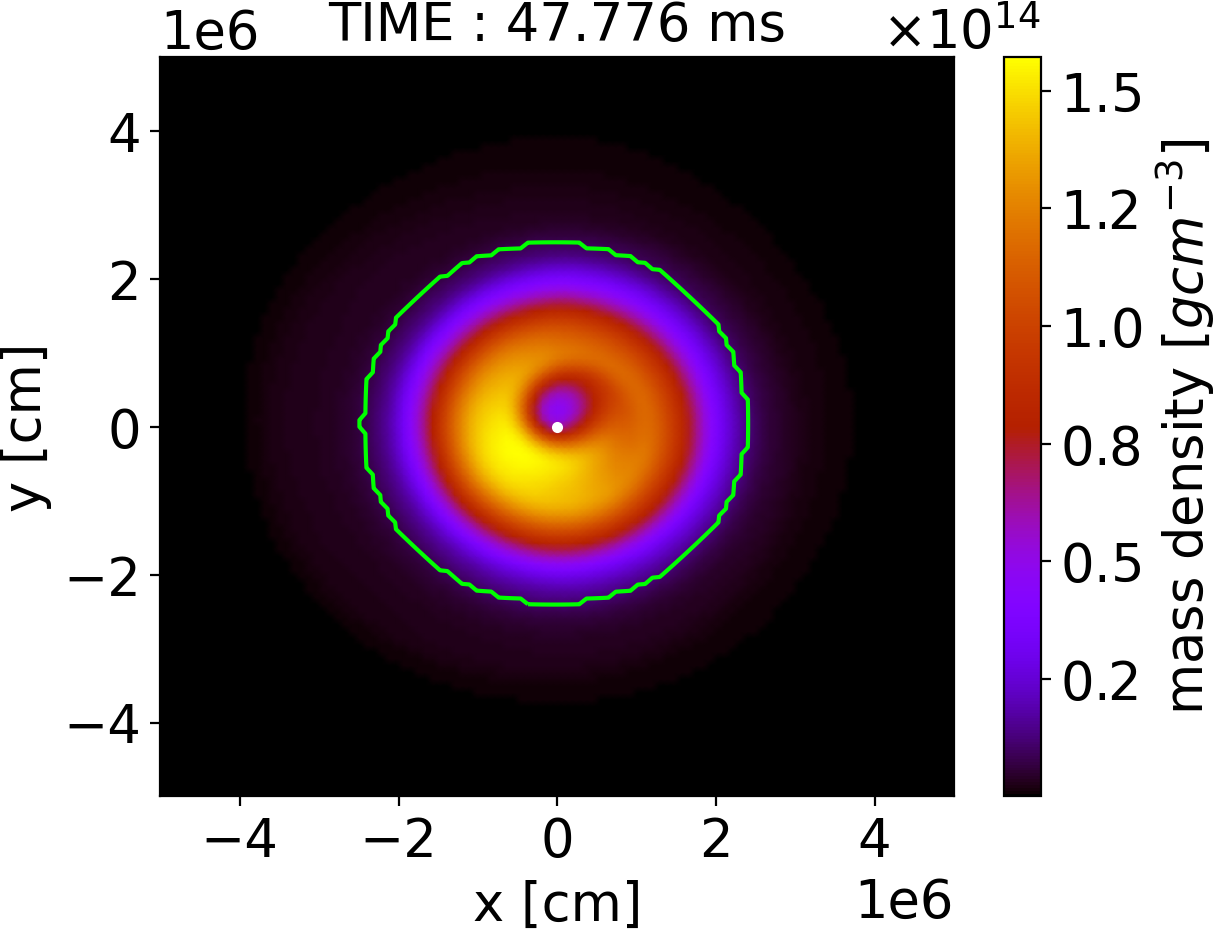}\\
\includegraphics[width=0.24\linewidth]{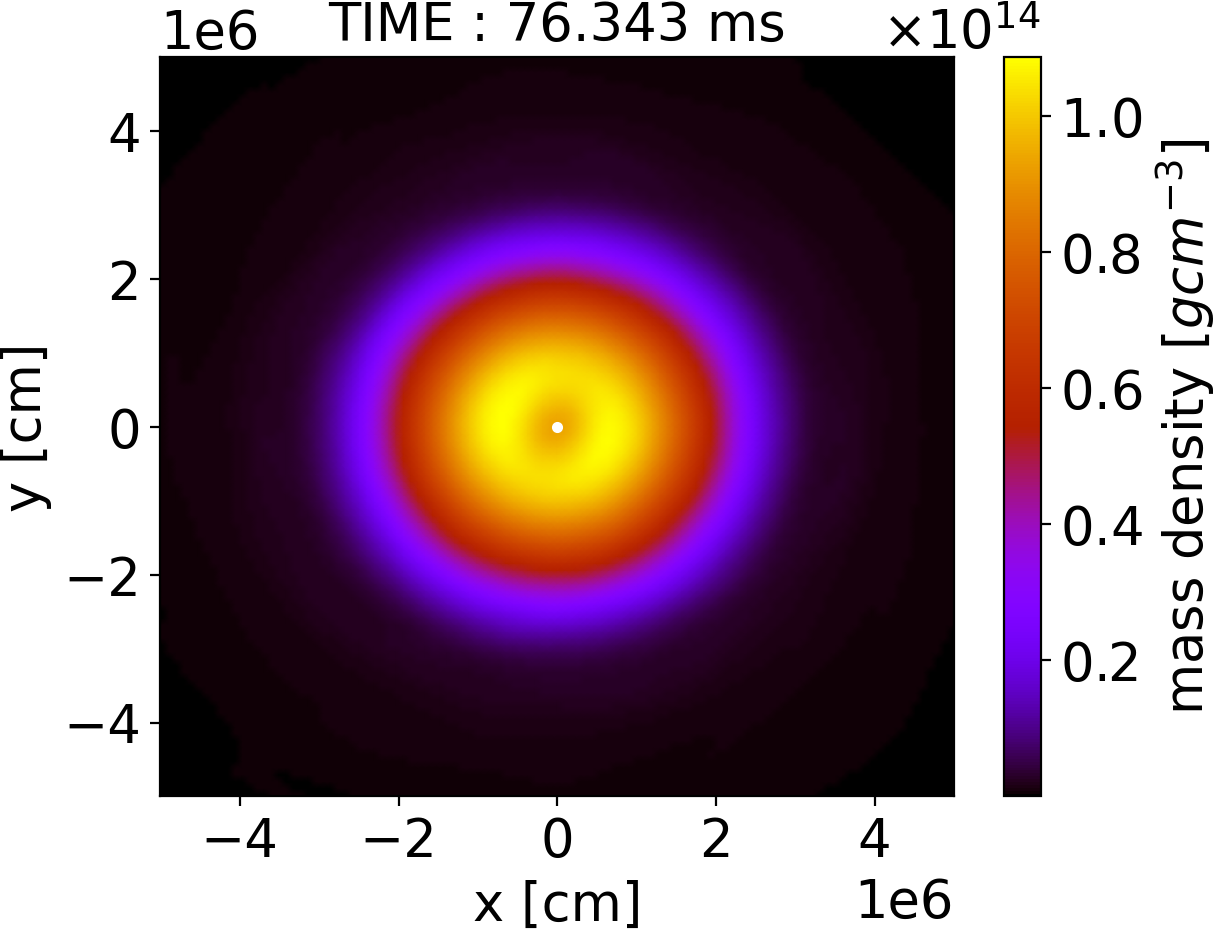}\hspace{-0.0\linewidth}
\includegraphics[width=0.24\linewidth]{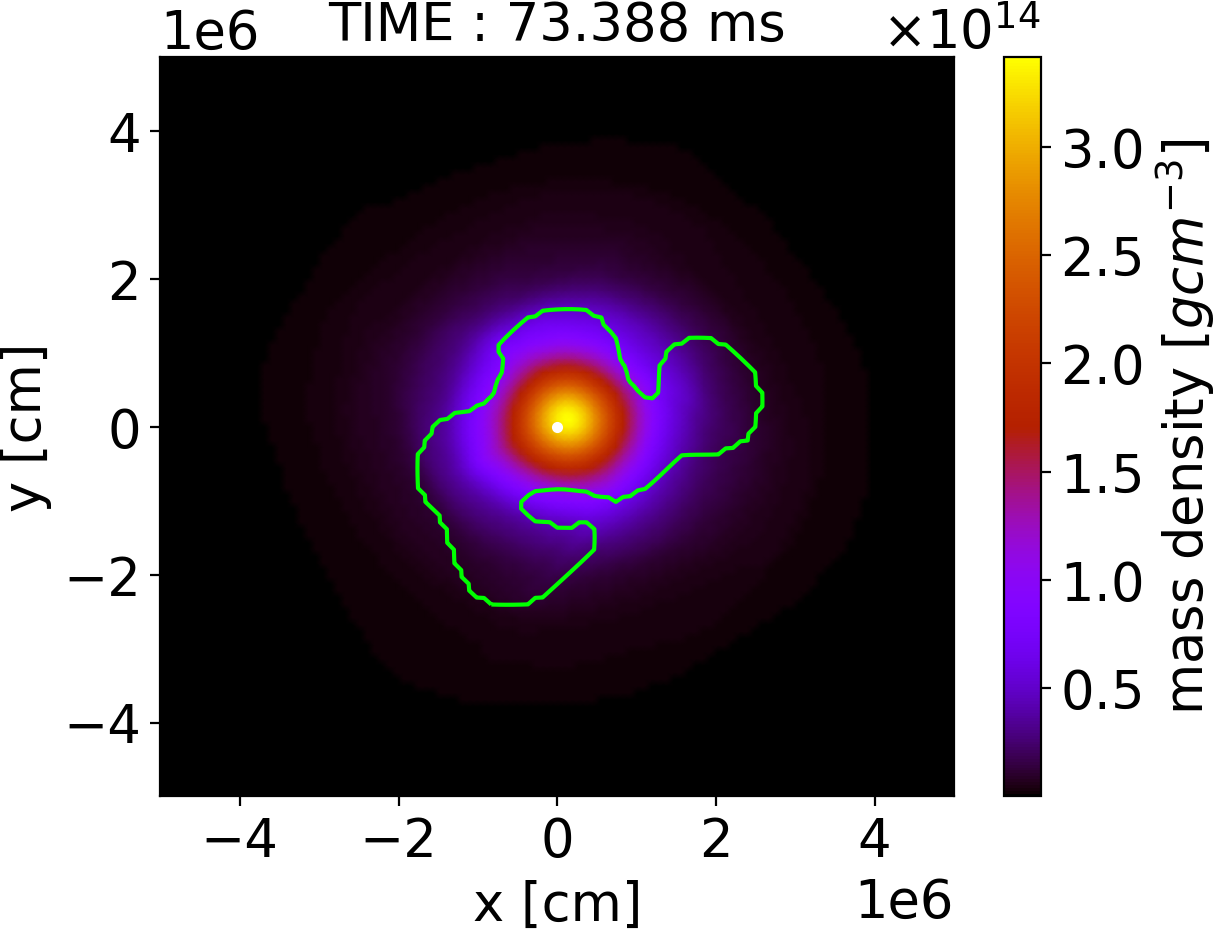}\hspace{-0.0\linewidth}
\includegraphics[width=0.24\linewidth]{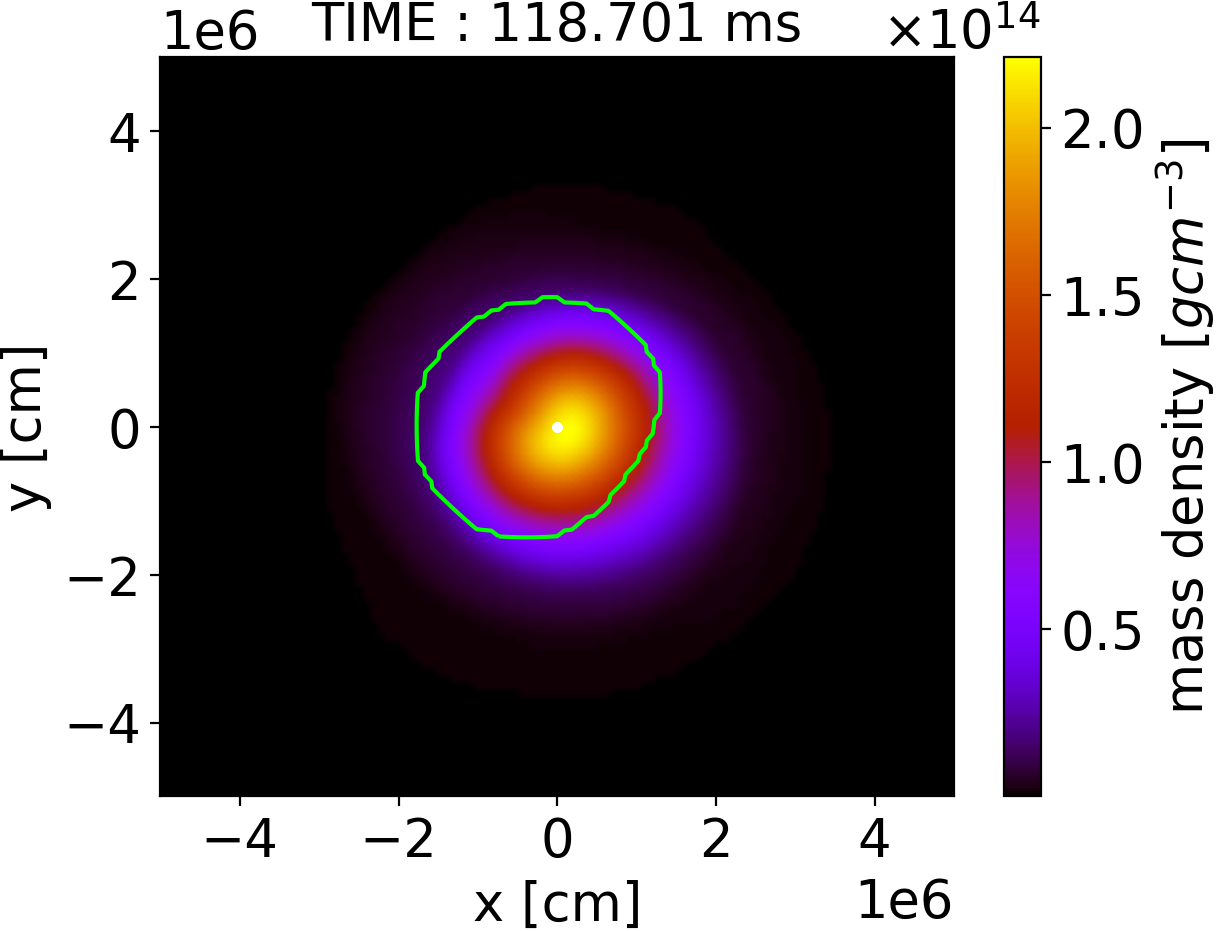}\hspace{-0.0\linewidth}
\includegraphics[width=0.24\linewidth]{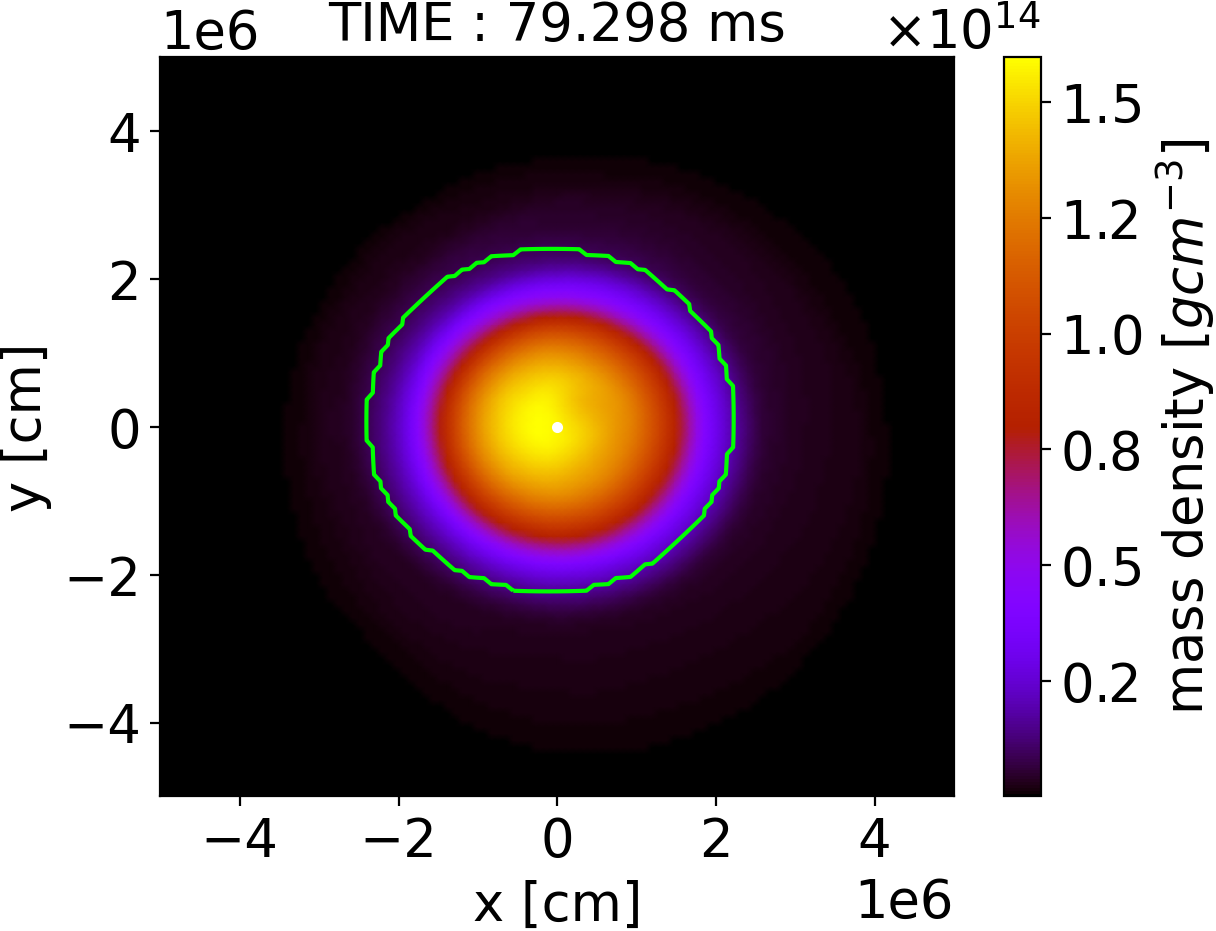}\\
\includegraphics[width=0.24\linewidth]{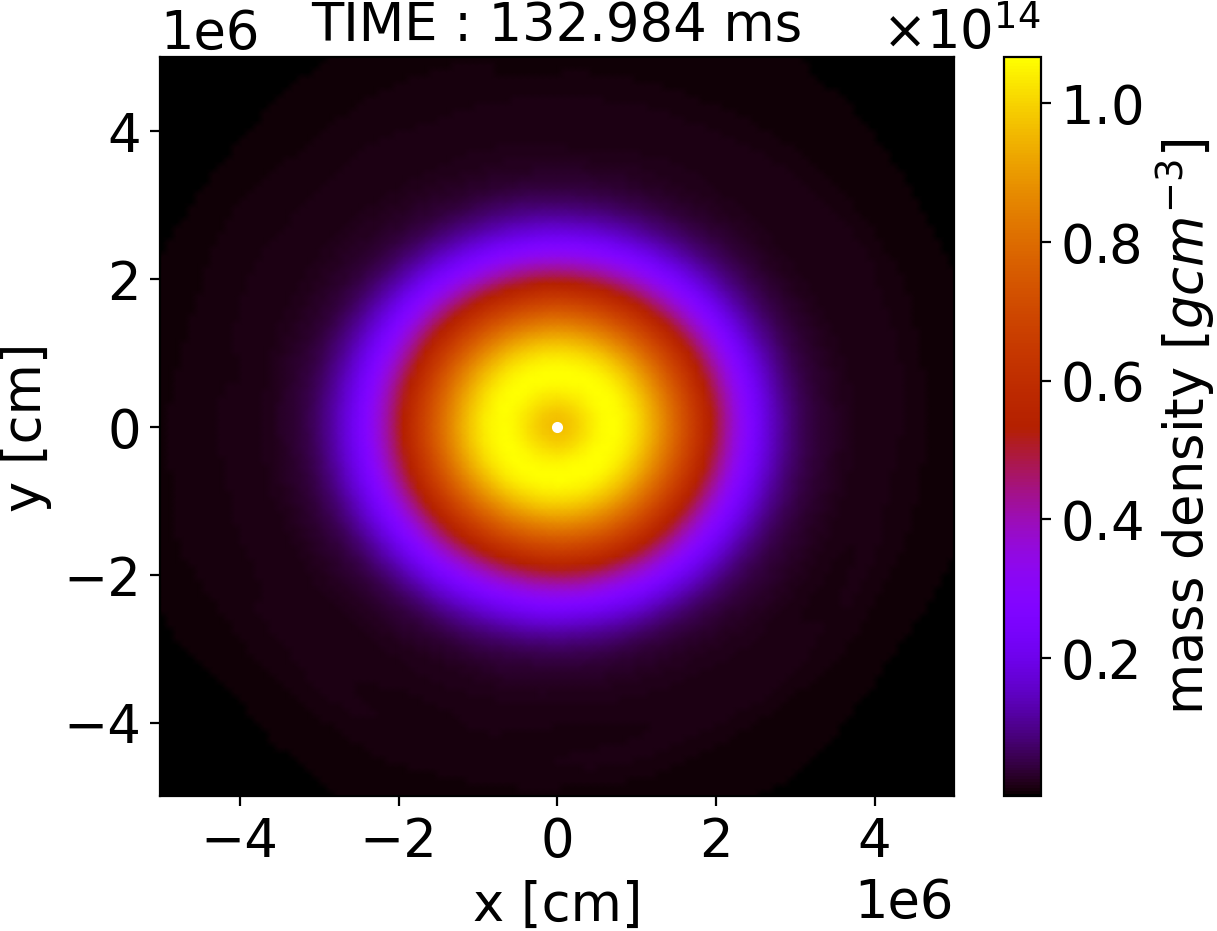}\hspace{-0.0\linewidth}
\includegraphics[width=0.24\linewidth]{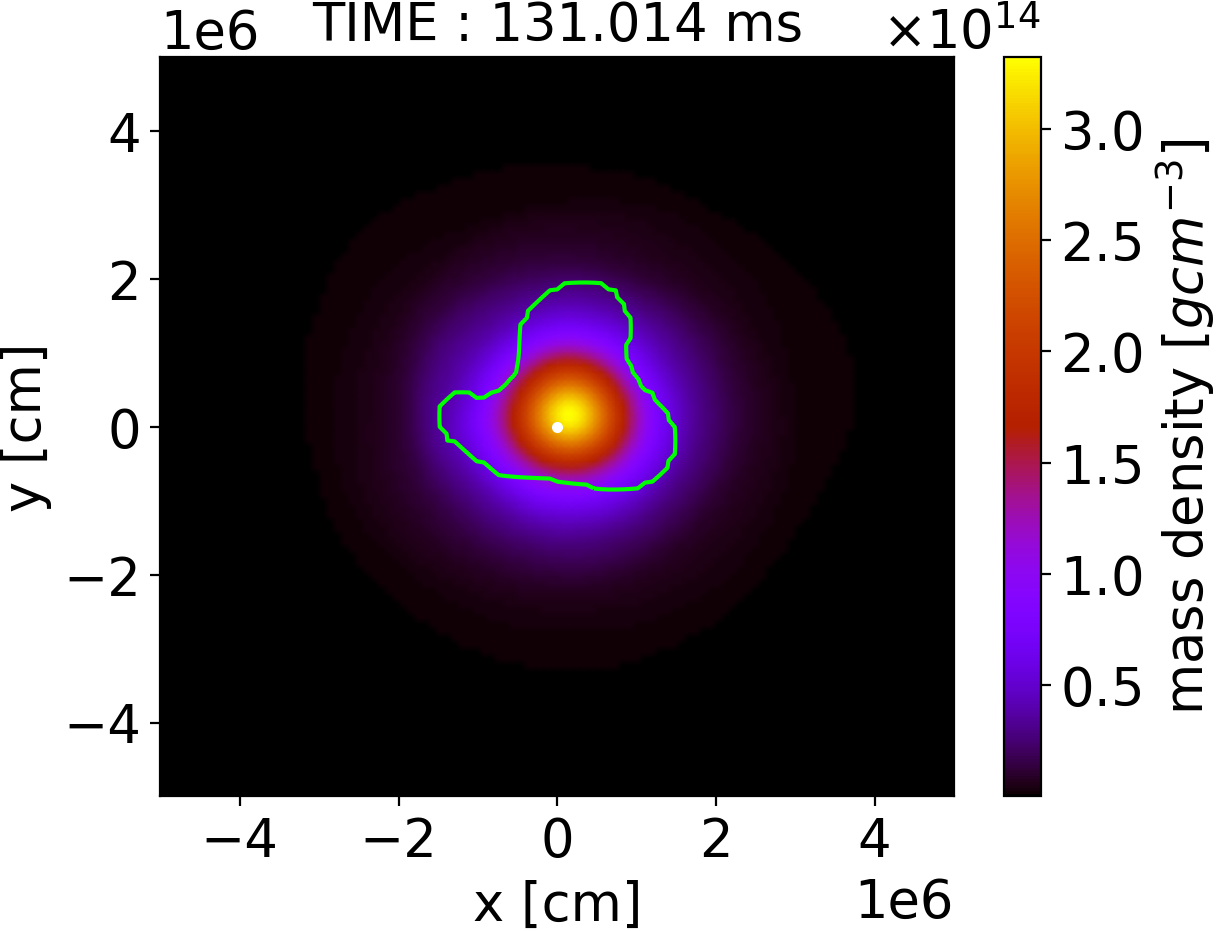}\hspace{-0.0\linewidth}
\includegraphics[width=0.24\linewidth]{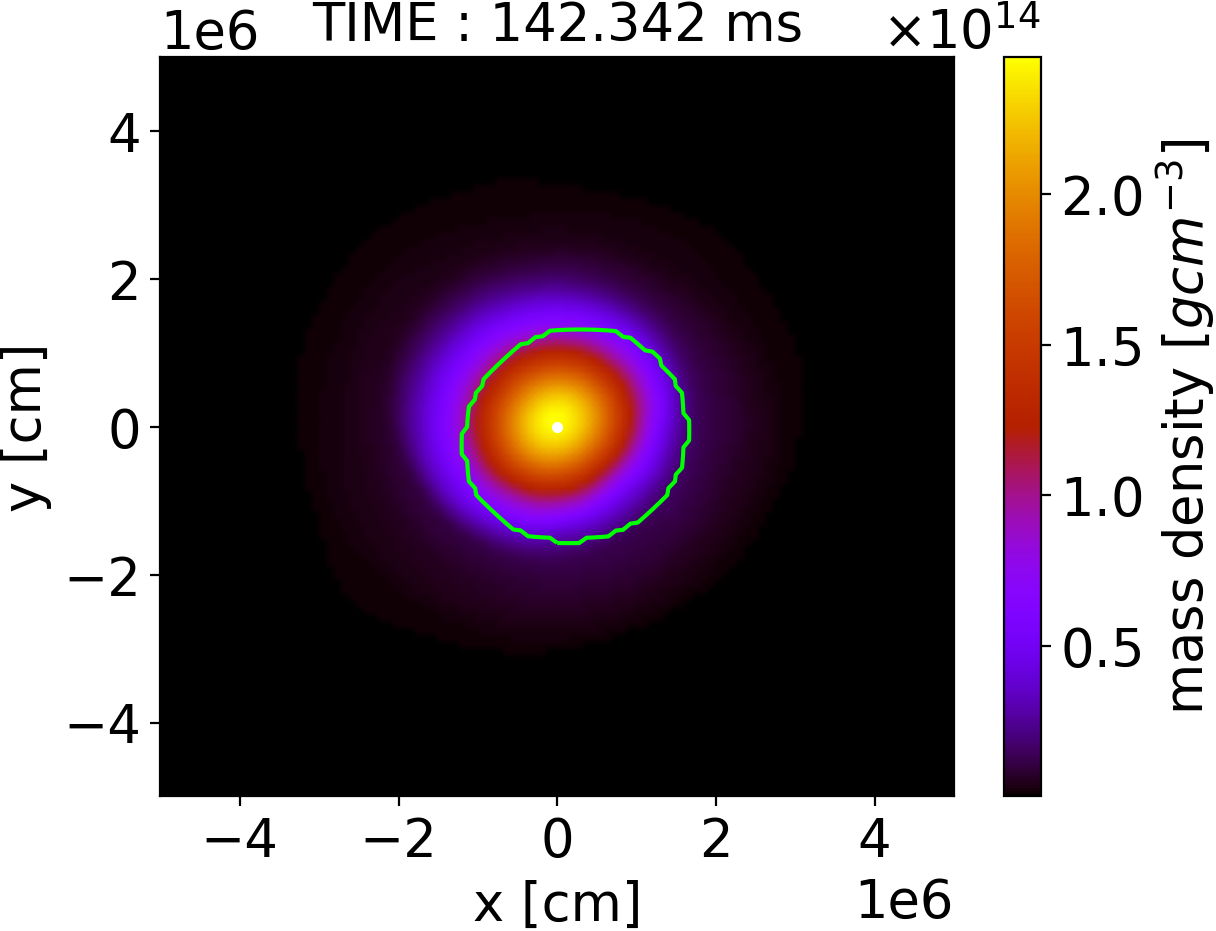}\hspace{-0.0\linewidth}
\includegraphics[width=0.24\linewidth]{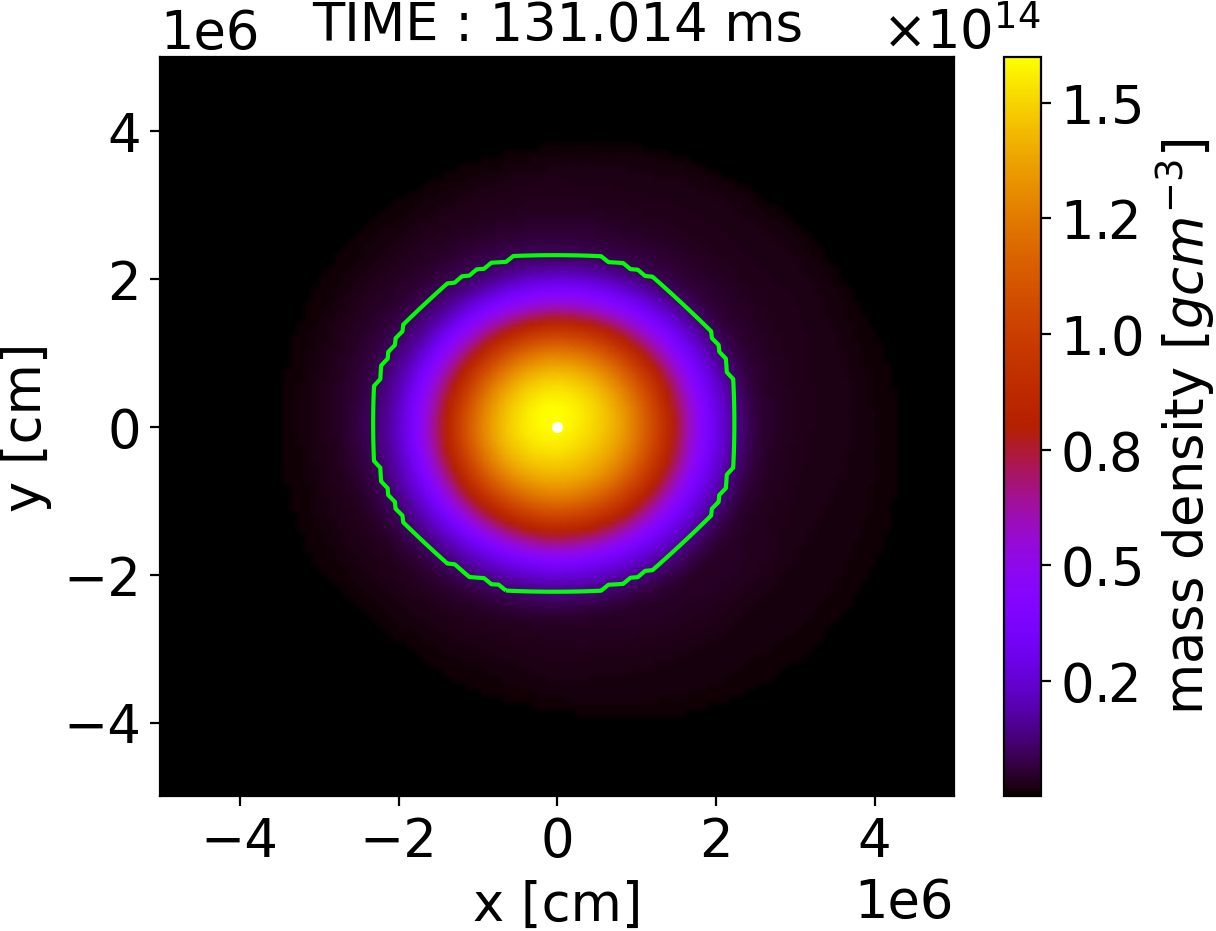}
\caption{Time evolution of the rest-mass density $\rho$ (in cgs units) at the equatorial plane. The four columns correspond, respectively, to the isolated RNS model U13 (left), model U13-a with $\mu=1.0$ (centre-left), model U13-b with $\mu=0.5$ (centre-right), and model U13-c with $\mu=0.33$ (right). The contour in green indicates the level surface of constant bosonic energy density $\E^{(0)} $ which contains 95\% of the total mass of the bosonic cloud. The centre of the computational grid is highlighted with a white dot.
}
\label{fig1}
\end{figure*}

%%%%%%%%%%%%%%%%%%%%%%%%%%%%%%%%%%%%%%%%%%%%%%%%%%%%
\section{Summary of numerical aspects} 
\label{numerics}
%%%%%%%%%%%%%%%%%%%%%%%%%%%%%%%%%%%%%%%%%%%%%%%%%%%%

We employ the community-driven software platform \textsc{EinsteinToolkit}~\cite{EinsteinToolkit:2019_10,Loffler:2011ay,Zilhao:2013hia} for the numerical evolutions, based on the \textsc{Cactus} framework and \textsc{Carpet}~\cite{Schnetter:2003rb,CarpetCode:web} for mesh-refinement
capabilities. We use the \textsc{McLachlan} infrastructure~\cite{reisswig2011gravitational,brown2009turduckening}, which implements the BSSN formulation of Einstein's equations for evolving the spacetime variables. The evolution of the scalar field and the Proca field, along with the computation of their contribution to the stress-energy tensor are managed by a private code that we tested and employed in previous works~\cite{sanchis2019nonlinear,sanchis2019head,DiGiovanni:2020ror,sanchis2021multifield,jaramillo2020dynamical}. The code for the complex Proca field is an extension of the one originally developed in~\cite{Zilhao:2015tya} and currently publicly available in the \textsc{Canuda} repository~\cite{Canuda_2020_3565475} and distributed within each new release of the \textsc{EinsteinToolkit}. We employ \textsc{GRHydro} for the fluid dynamics and \textsc{EOSOmni} for the EoS. The evolutions are carried out using a $\Gamma$-law EoS $ P = (\Gamma-1)\rho\epsilon$.

The Cartesian-coordinate-based numerical grid for our simulations is discretized with five refinement levels, each spanning a different spatial domain with a different resolution. From the outermost to the innermost grid, the spatial domains are $\lbrace 300, 240, 200, 100, 50 \rbrace$ in units of the total mass, and the corresponding ($\Delta x = \Delta y = \Delta z$) resolutions of each level are $\lbrace 10, 5, 2.5, 1.25, 0.65\rbrace $. We choose a Courant factor such that the time step is $\Delta t = 0.25 \Delta x$, where $\Delta x$ is the grid spacing of the innermost grid along the $x$ direction. We assume reflection symmetry with respect to the equatorial plane ($z=0$). We employ radiative (Sommerfeld) outer boundary conditions, which are implemented in the \textsc{NewRad} code.

 %%%%%%%%%%%%%%%%%%%%%%%%%%%%%%%%%%%%%%%%%%%%%%%%%%%%%%%%%%%%%
\section{Results}
\label{results}
%%%%%%%%%%%%%%%%%%%%%%%%%%%%%%%%%%%%%%%%%%%%%%%%%%%%%%%%%%%%%%

As stated before, the aim of this work is to investigate numerically the potential effects of ultralight bosonic dark matter accreting on to differentially RNS on the stability properties of these objects. Our simulations start with a transient phase during which the bosonic cloud accretes on to the neutron star, with a timescale shorter than that of the development of the dynamical bar-mode instability. After the cloud has been accreted its effect on the dynamics of the neutron star are significant, as we discuss next.

%%%%%%%%%%%%%%%%%%%
%\subsection{Rotating neutron stars with a bosonic cloud} \label{ResultsA}
\subsection{Dynamics} \label{ResultsA}
%%%%%%%%%%%%%%%%%%%

\begin{figure*}[t!]
\centering
\includegraphics[width=0.45\linewidth]{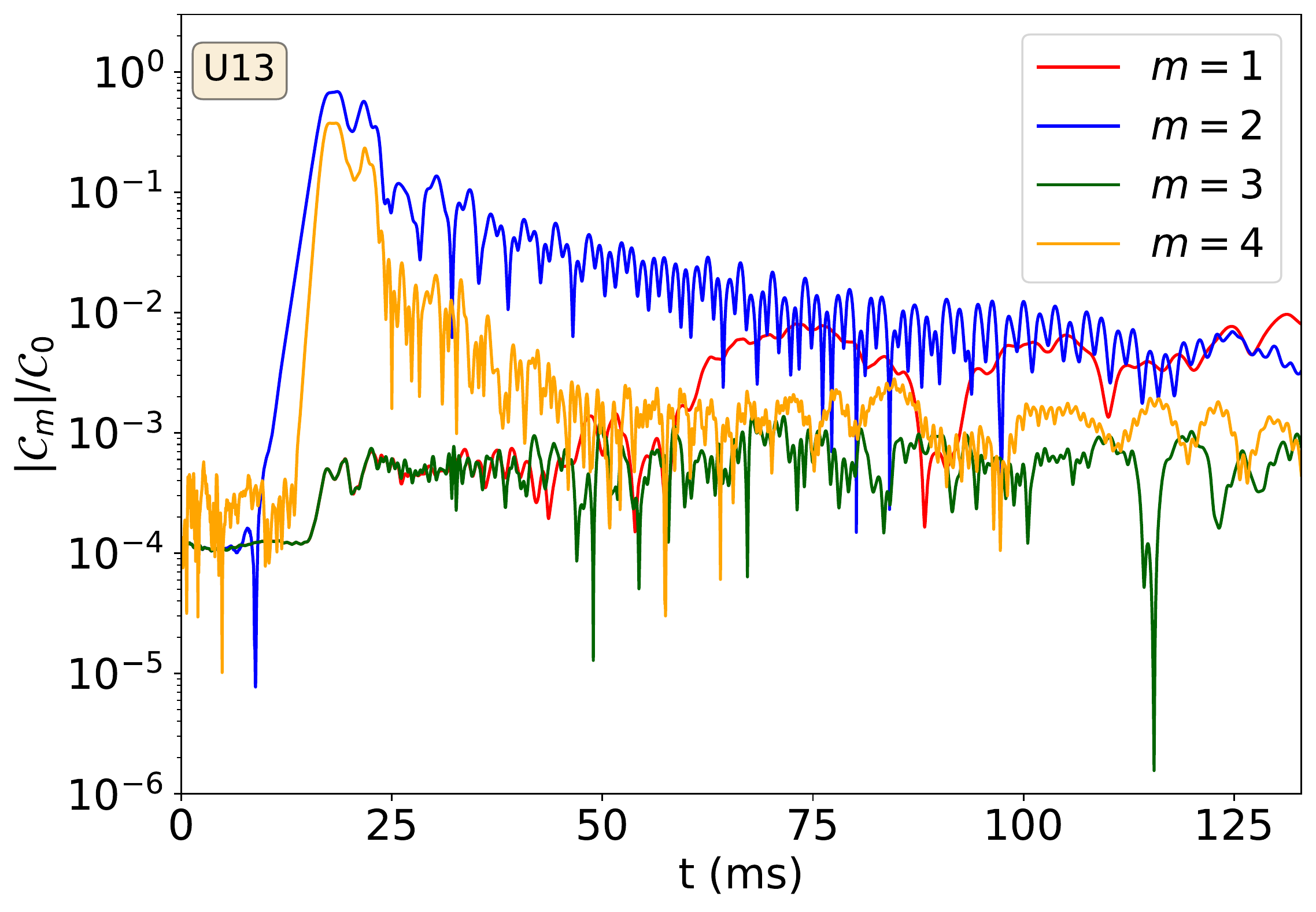} 
\includegraphics[width=0.45\linewidth]{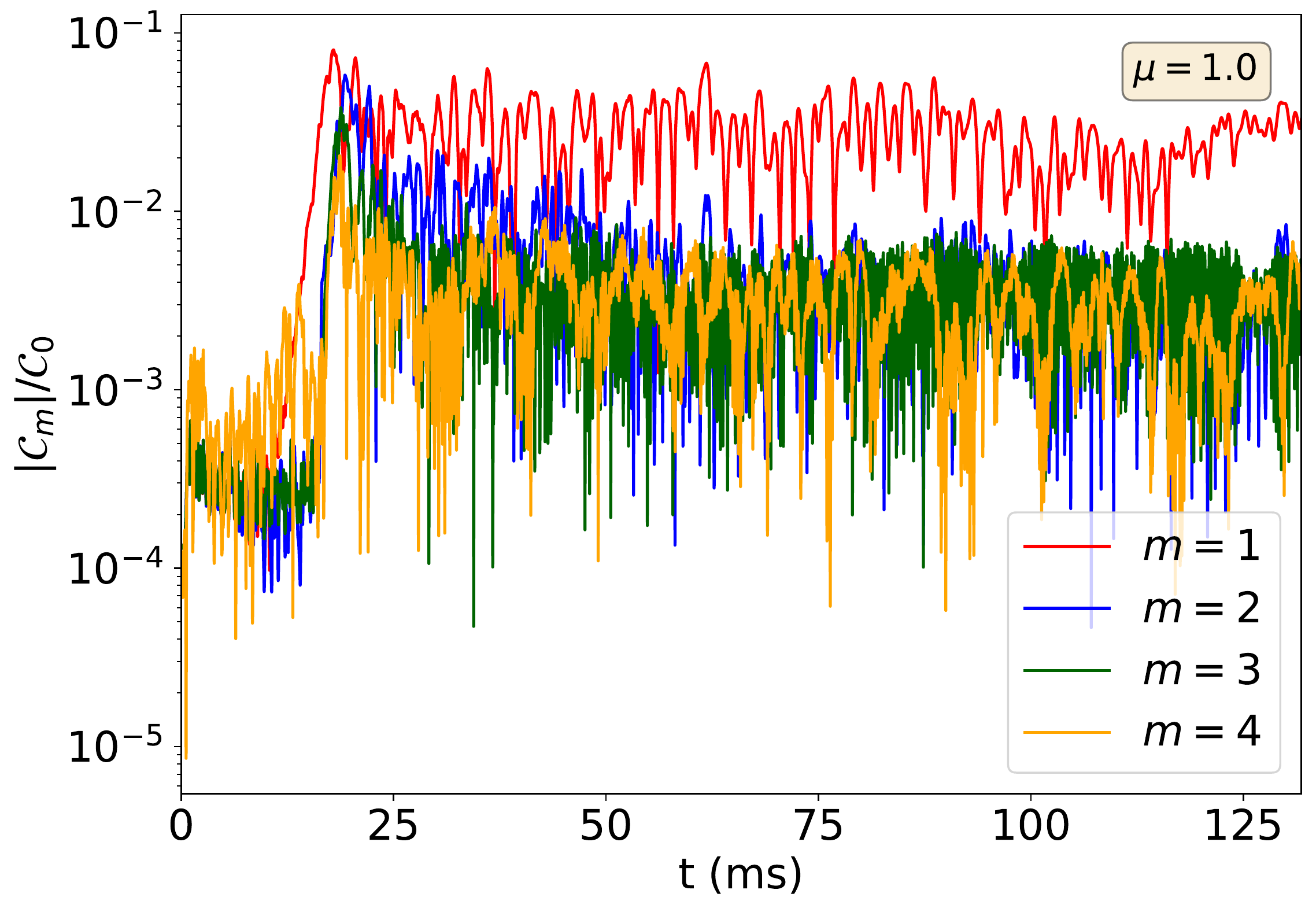} \\
\includegraphics[width=0.45\linewidth]{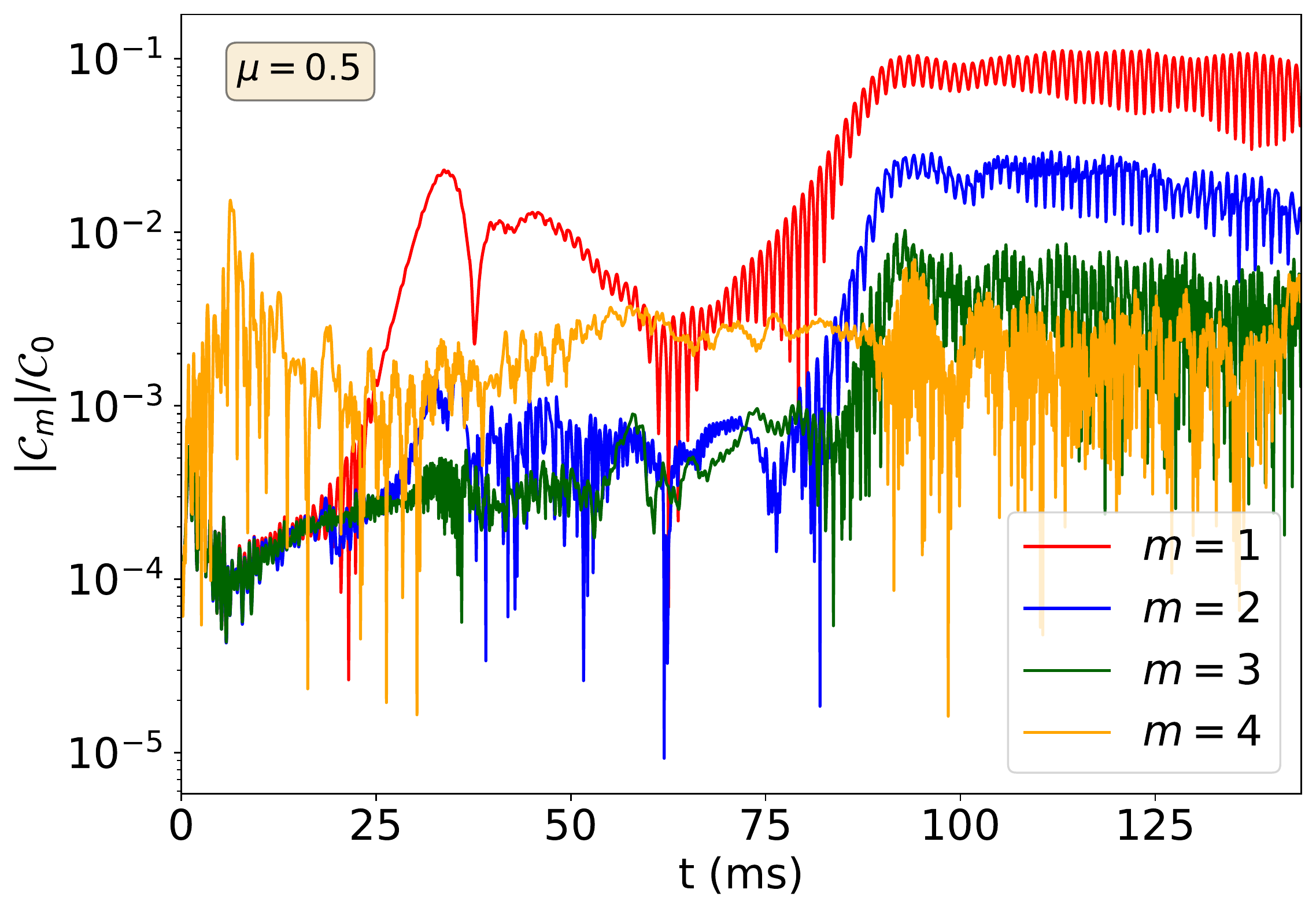} 
\includegraphics[width=0.45\linewidth]{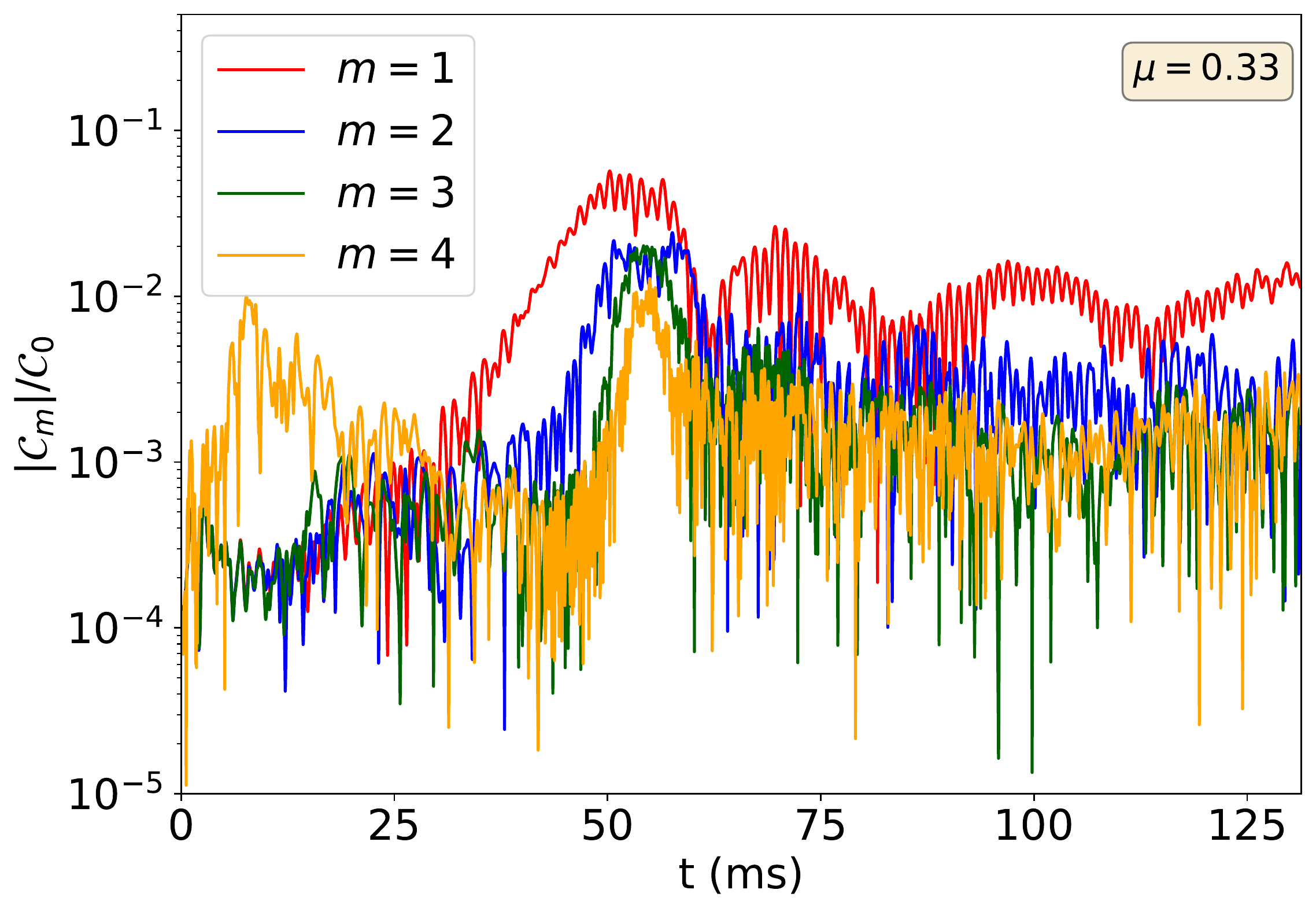} 
\caption{Time evolution of the azimuthal mode decomposition of the fermion energy density, from $m=1$ to $m=4$, for the same models shown in Fig.~\ref{fig1}. Top-left panel: model U13; top-right panel: model U13-a; bottom-left panel: model U13-b; bottom-right panel: model U13-c.  
}
\label{fig2}
\end{figure*}

We perform long-term simulations (${\cal O}(100)$ ms) of the full 16 models of Table~\ref{table:models1}. However, for the sake of clarity in the analysis we present results only for an illustrative subset of models that best display the effects of the bosonic field on the bar-mode instability (and on the gravitational waveforms). We start discussing results for models U13, U13-a, U13-b, and U13-c. For the last three models the mass of the scalar cloud is roughly equal ($M_{\rm cloud}\approx 0.6$) which allows us to isolate the effects of varying the particle mass $\mu$. Moreover, throughout this section we only discuss the scalar-field case, since the conclusions we draw for this case remain unaltered for a Proca-field cloud.

The columns of Figure~\ref{fig1} display snapshots of the rest-mass density $\rho$ at the equatorial plane for those four models. Note that except for the first row (initial data) the snapshots selected in subsequent rows are different for each model. The isolated RNS model U13 is depicted in the left column while the next three columns show the evolution of models U13-a, U13-b, and U13-c for which the scalar-field cloud is built with correspondingly smaller values of the bosonic particle mass, $\mu=1$, $\mu=0.5$, and $\mu=0.33$, respectively. The green contour visible in some of the snapshots of Fig.~\ref{fig1} indicates the level surface of constant bosonic energy density $\E^{(0)} $ which contains 95\% of the total mass of the bosonic cloud. We note that during the accretion process the bosonic cloud loses mass through the mechanism known as gravitational cooling~\cite{seidel1994formation,di2018dynamical}. On the one hand, in models U13-b and U13-c the amount of scalar field expelled is about 5\% of the total stored in the cloud, which means that most of the mass of the cloud accretes on to the RNS in a short timescale, less than $10$ ms . On the other hand, model U13-a undergoes the highest mass loss, losing almost half of the initial bosonic mass by the end of the simulation (at which time the process still continues). The evaluation of the surface containing 95\% of the total mass for model U13-a is affected by the mass released through gravitational cooling during the accretion process, and for this reason we obtain surfaces which are far from being spheroidal. The differences observed in the dynamical evolution of the bosonic cloud in models U13a, U13b, and U13c, can be understand by recalling the stability properties of spherically-symmetric boson stars. Such stars have a maximum mass of $M_{\rm max}=0.633 / \mu$ (see e.g.~\cite{Herdeiro:2017fhv}), separating the stable and unstable branches in the mass-frequency existence plot. We relate the different behaviour of model U13-a with the fact that the mass stored in the cloud ($M_{\rm cloud}=0.628$) is very close to the maximum allowed mass for a boson star with $\mu=1$. As a result, the dynamical process leads to the simultaneous ejection of a large amount of mass from the cloud and the gradual formation of a spherical boson star residing in a more stable region in the parameter space (away from the maximum). On the other hand, models U13-b and U13-c are already initially well inside the corresponding boson-star stable region and, thus, they do not radiate a lot of scalar field to reach stability.

\begin{figure*}[t]
\includegraphics[width=0.34\linewidth]{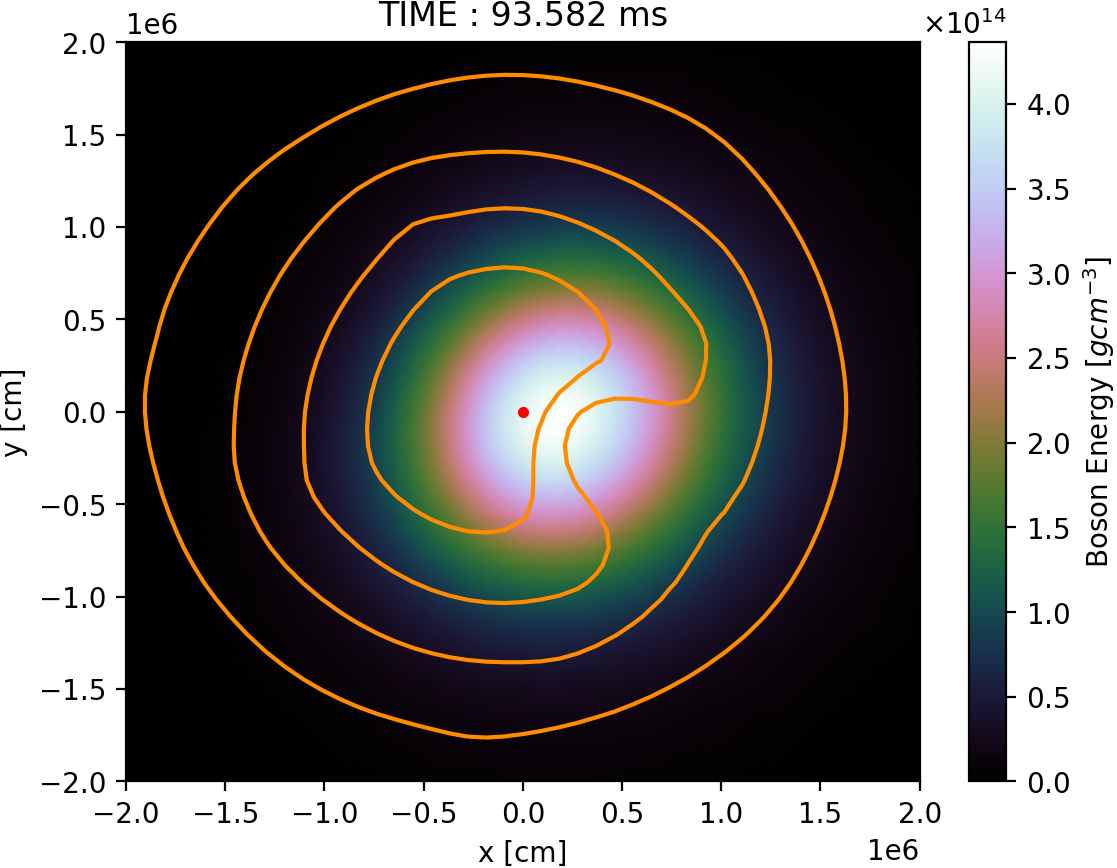}
\hspace{-0.0\linewidth} 
\includegraphics[width=0.34\linewidth]{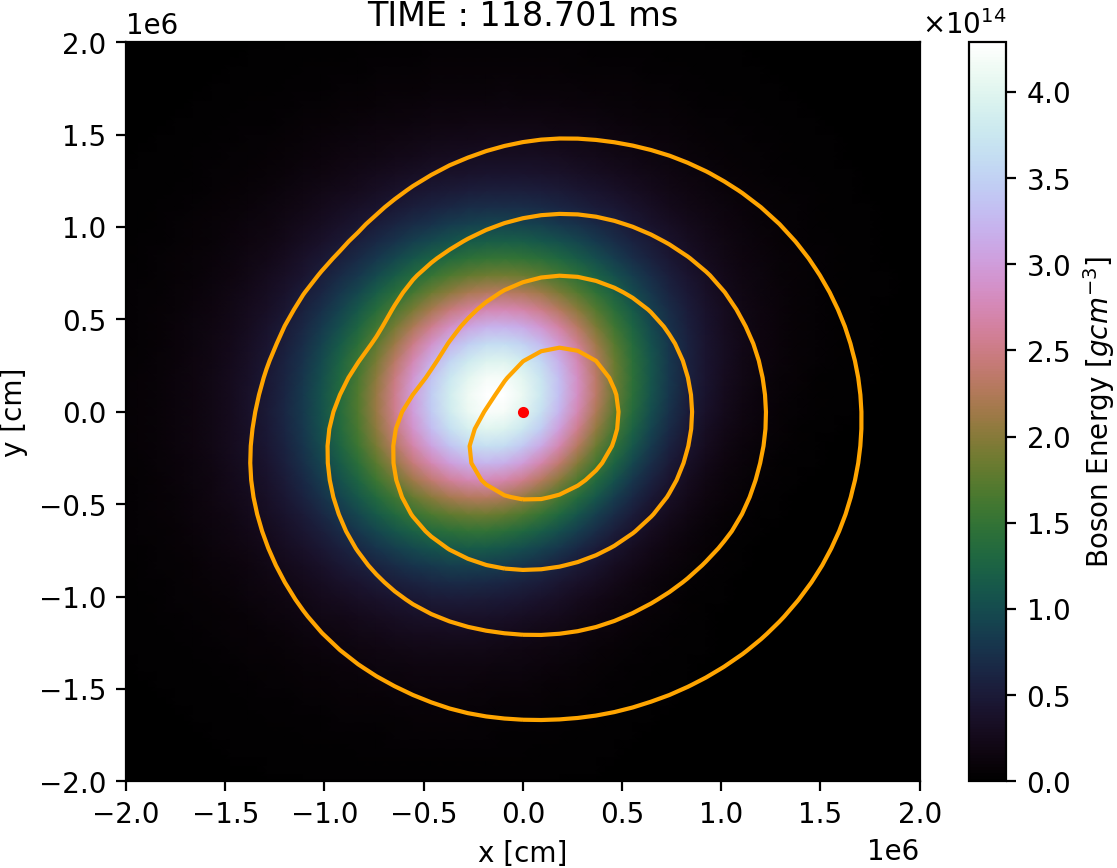}
\hspace{-0.0\linewidth} 
\includegraphics[width=0.27\linewidth]{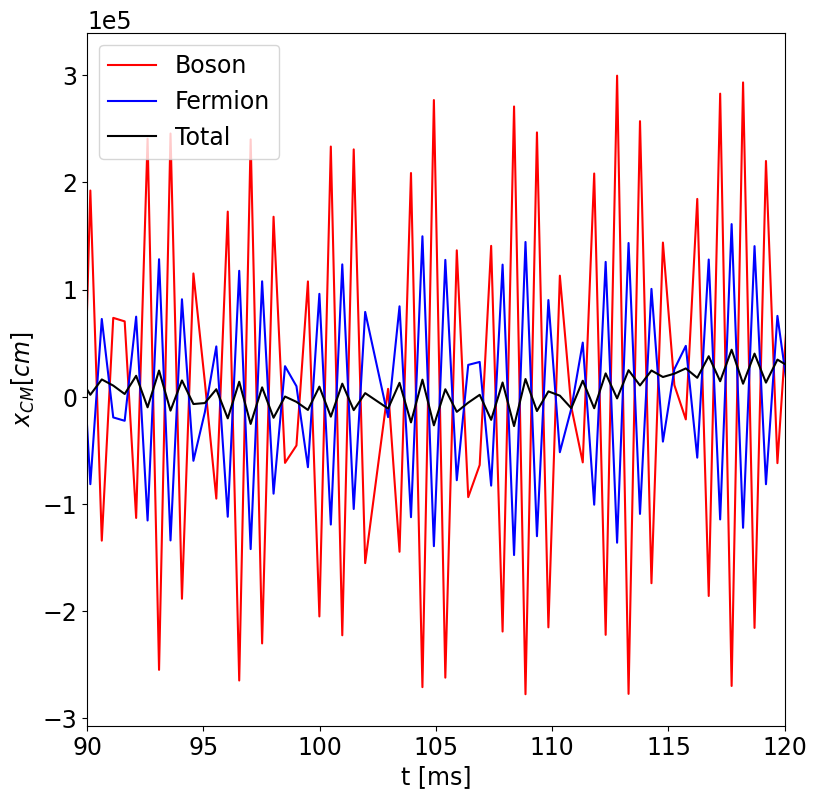}\hspace{-0.0\linewidth} 
\caption{The left and middle panels show two time snapshot of an equatorial cut of $\E^{(0)}$ for model U13-b and of the rest-mass density $\rho$. The latter is indicated by the orange contours. The maximum of the fermionic energy and its centre-of-mass are located inside the smallest contours. These two snapshots reveal the transient formation of a mixed (fermionic-bosonic) bar. In the right panel we show the time evolution (spanning the time interval depicted in the left and centre panels) of the $x$-component of the centre-of-mass evaluated for the bosonic and fermionic contribution, to highlight the $\pi$ phase difference, and for the total object.} 
\label{fig3}
\end{figure*}

All models are subject to non-axisymmetric instabilities throughout the evolution. We note that we do not impose any ad-hoc perturbation on the initial data to trigger those but the only source of perturbation is the discretization error of the finite-difference approximations of the partial derivatives of the equations we solve. The presence of the accreting scalar field leads to different dynamics. For the purely RNS model U13, shown in the leftmost column of Fig.~\ref{fig1}, the development of an $m=2$-dominated instability is apparent at around $t\approx 10$ ms (second row). This dynamical timescale is similar to that reported in~\cite{Baiotti2007} for the same model. This leads to the appearance of a bar-like deformation during which the star sheds mass and angular momentum and finally settles into a perturbed stable configuration. During this process the maximum value of the rest-mass density $\rho$ moves from the end-points of the bar towards the centre of the star, whose morphology changes from toroidal to spheroidal. 

When a scalar field is included, the evolution of the RNS is modified but the star continues to undergo non-axisymmetric instabilities with different timescales and features. The most salient characteristic of all models involving an accreting scalar field is that no bar is formed and the dominant mode of the deformation shifts from $m=2$ to $m=1$, i.e.~those models mostly develop a one-arm instability. This morphological  change can be identified by the appearance of a rotating over-density blob (see, e.g.~the third snapshots from the top in the last two columns of Fig.~\ref{fig1}). Eventually, when angular momentum is radiated away through gravitational waves, this over-density blob collapses into a spheroidal RNS. In addition, the timescale of the $m=1$ instability increases with respect to the isolated RNS case: for U13-a ($\mu=1.0$) it occurs at $t\approx 15$ ms, for U13-b ($\mu=0.5$) at $t\approx 25$ ms, and finally for U13-c ($\mu=0.33$) at $t\approx 40$ ms. We note that the timescale increases as we decrease the value of $\mu$ when keeping the same total initial mass in the cloud. We tentatively identify the reason for this behaviour with the fact that as the cloud becomes more diluted so does the entire configuration, hence the fermionic part gets increasingly less compact and the instability takes longer to set in.

A more quantitative representation of the fundamental properties of the instabilities that develop in our systems can be obtained by monitoring the evolution of the volume-integrated azimuthal Fourier mode decomposition of the fermion energy density, evaluated as
\begin{equation}\label{Am}
\mathcal{C}_{m} = \int d^3\textbf{x} \,\E^{\rm{fluid}}(\textbf{x}) \,e^{i m \varphi}.\\
\end{equation}
We point out that when odd modes (such as $m=1$) start to grow, the centre-of-mass of the object is displaced from the origin of the Cartesian grid. As explained in~\cite{Cerda-Duran2007} we take into account this displacement to properly evaluate $\mathcal{C}_{m}$ and the computation of the angular momentum $J_{\rm{NS}}$ and $J_{\rm{cloud}}$ during the evolution. To this end we evaluate the coordinates of the centre-of-mass of the entire object and we redefine the azimuthal coordinate $\varphi$ with respect to this centre instead of the centre of the numerical grid, as we explain in~\cite{DiGiovanni:2020ror}.

Quantities $\mathcal{C}_{m}$ defined in Eq.~(\ref{Am}) monitor the departure from axisymmetry in the fermion density. In Fig.~\ref{fig2} we show, for the same models discussed in Fig.~\ref{fig1}, the time evolution of the absolute value of the mode decomposition for the first four Fourier modes, $m=\{1,2,3,4\}$, normalized to the total energy $\mathcal{C}_{0}$. In all cases we observe an exponential growth of the different modes. As discussed above we can clearly see that in the case of an isolated RNS (model U13; top-left panel of Fig.~\ref{fig2}) only the even modes $m=2$ and $m=4$ are significantly excited initially, the dominant one being the $m=2$ bar-mode. At later times the amplitude of both modes decay, especially that of the $m=4$ mode which shows a steeper rate, and by the end of the simulation the dominant modes are the $m=2$ and $m=1$. However, their late-time amplitudes are about two orders of magnitude smaller than that of the $m=2$ mode at maximum amplitude (attained around $t=20$ ms). We note that the response of the different modes observed in our simulation of model U13 is in perfect agreement with what was found in~\cite{Baiotti2007} (see, in particular, their Figure 7).

\begin{figure*}[t!]
\includegraphics[width=0.24\linewidth]{U13_rho-0000.png}\hspace{-0.0\linewidth}
\includegraphics[width=0.24\linewidth]{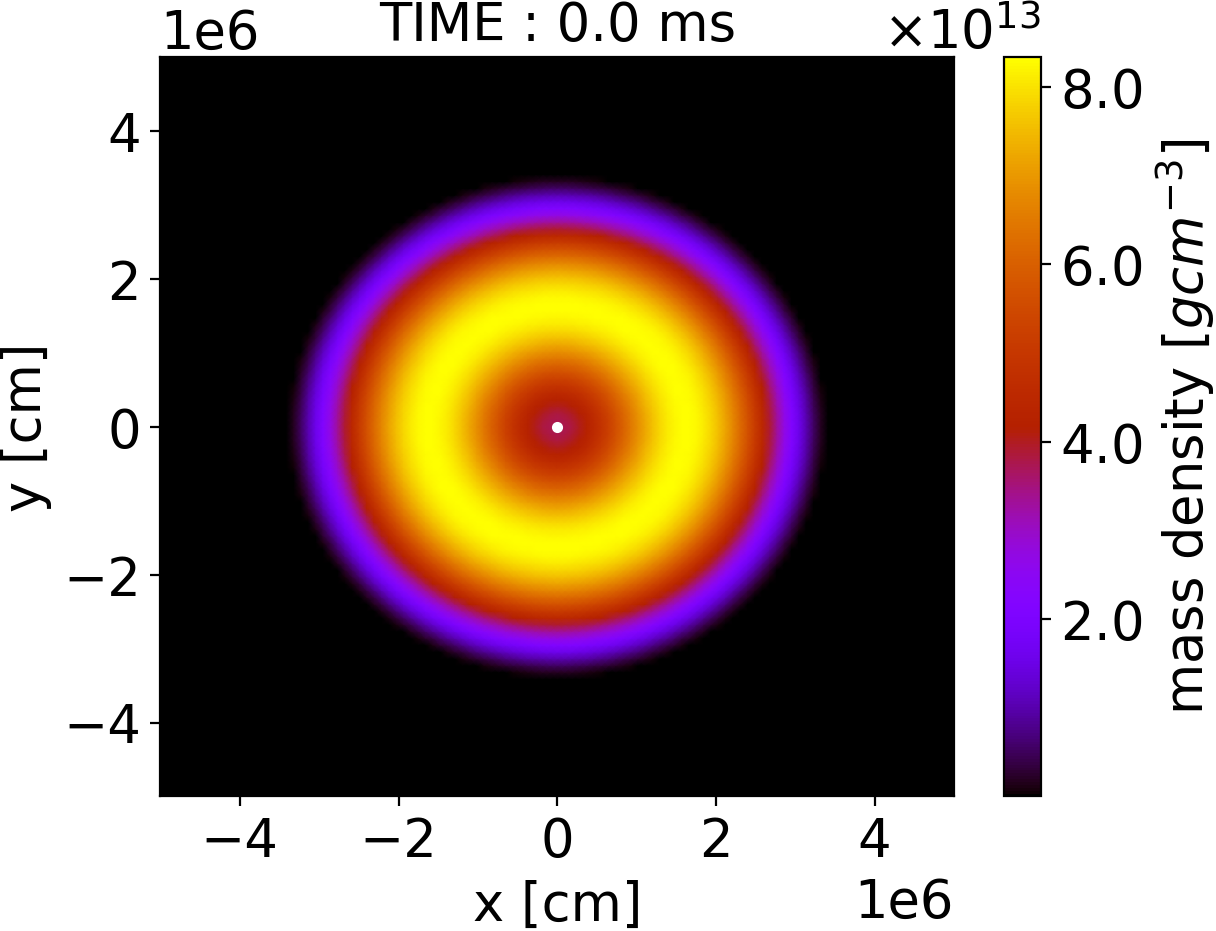}\hspace{-0.0\linewidth}
\includegraphics[width=0.24\linewidth]{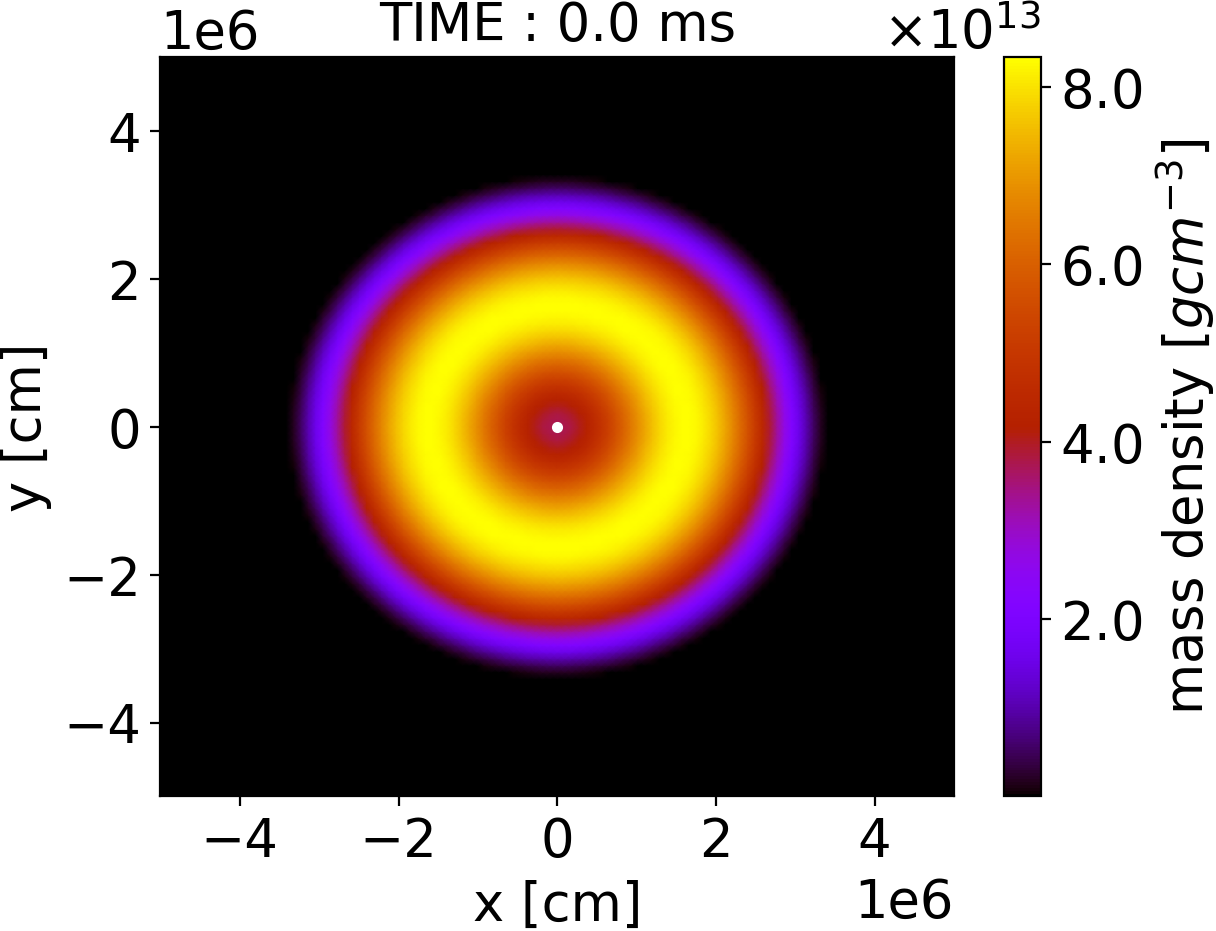}\hspace{-0.0\linewidth}
\includegraphics[width=0.24\linewidth]{U13_mu033_Ap32E-4_rho-0000.png}\\
\includegraphics[width=0.24\linewidth]{U13_rho-0034.png}\hspace{-0.0\linewidth}
\includegraphics[width=0.24\linewidth]{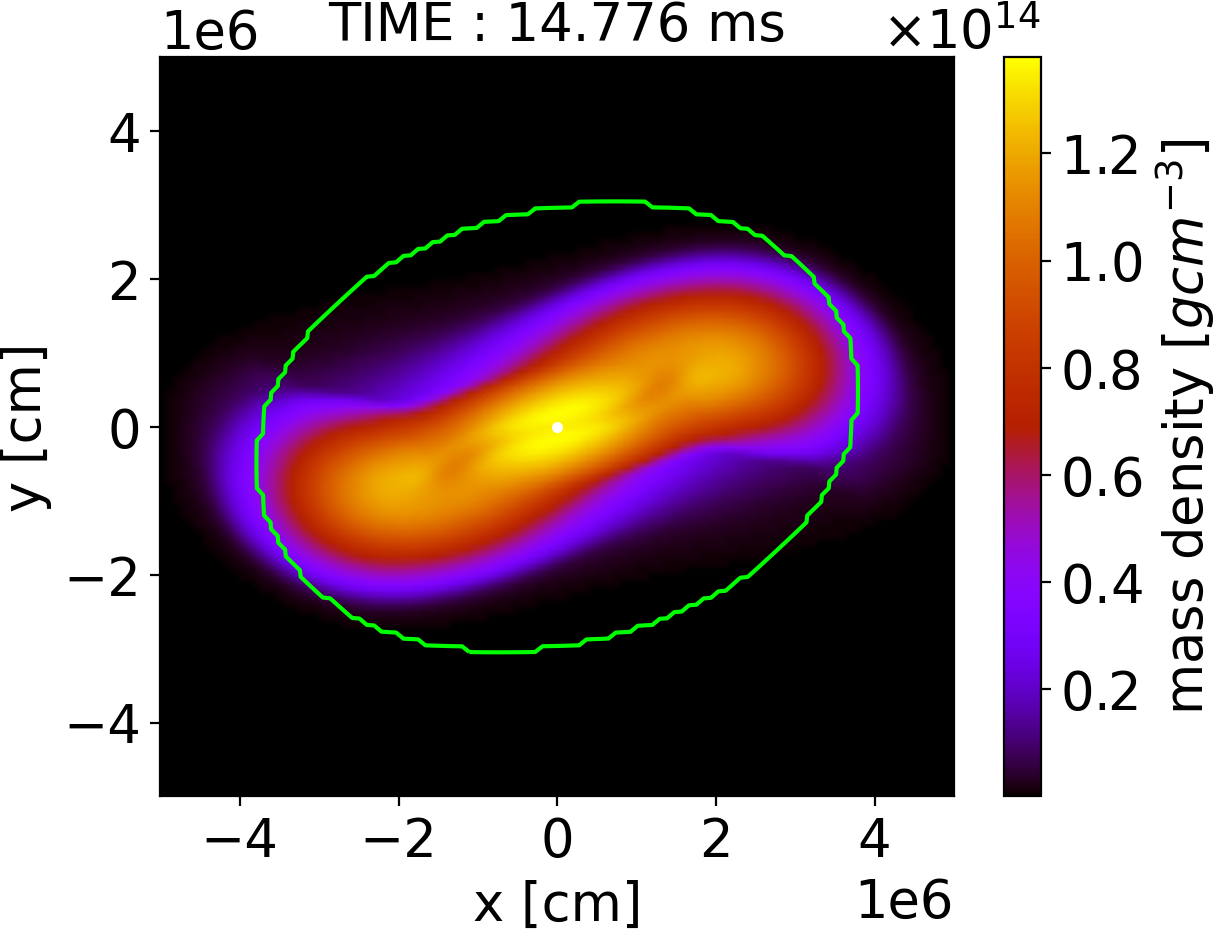}\hspace{-0.0\linewidth}
\includegraphics[width=0.24\linewidth]{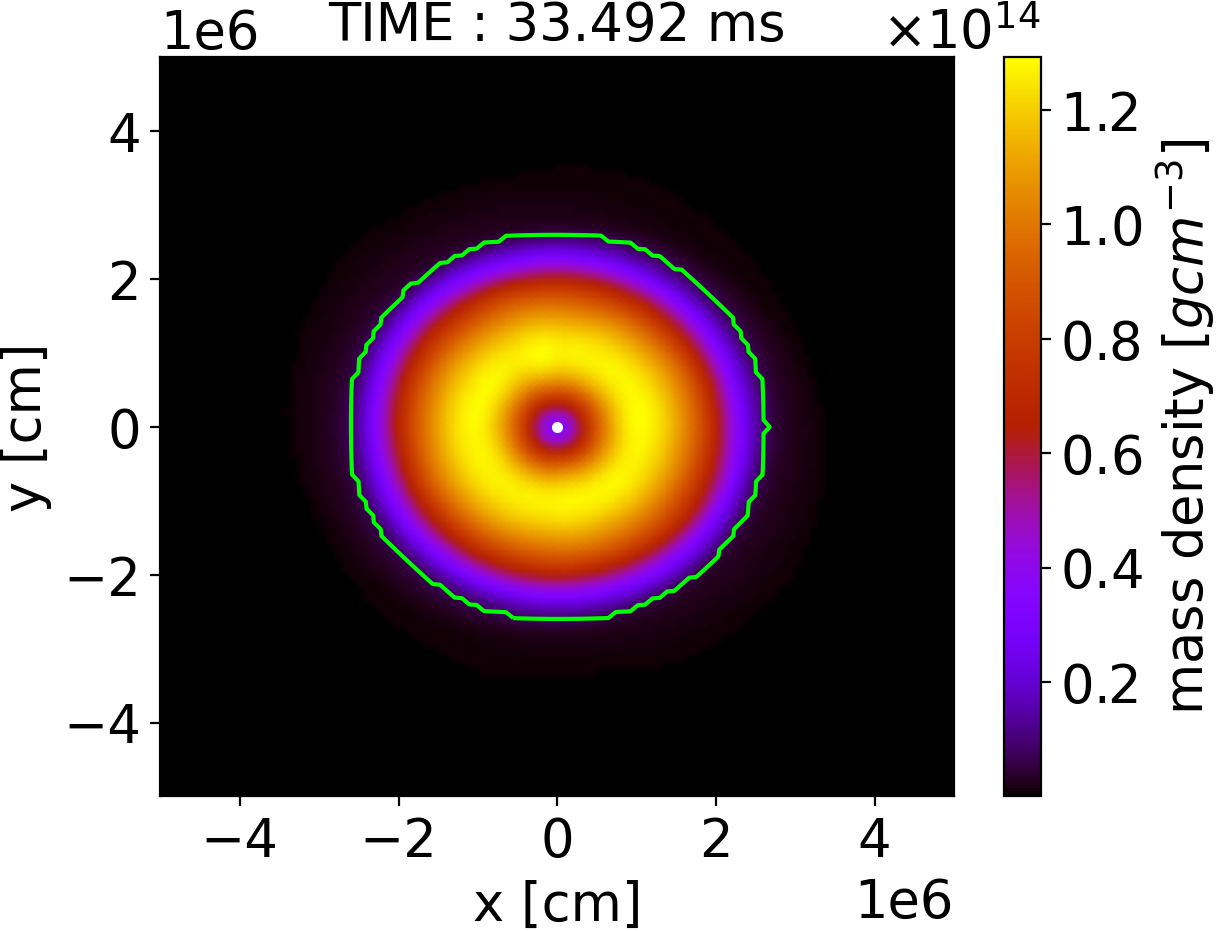}\hspace{-0.0\linewidth}
\includegraphics[width=0.24\linewidth]{U13_mu033_Ap32E-4_rho-0117.png}\\
\includegraphics[width=0.24\linewidth]{U13_rho-0050.png}\hspace{-0.0\linewidth}
\includegraphics[width=0.24\linewidth]{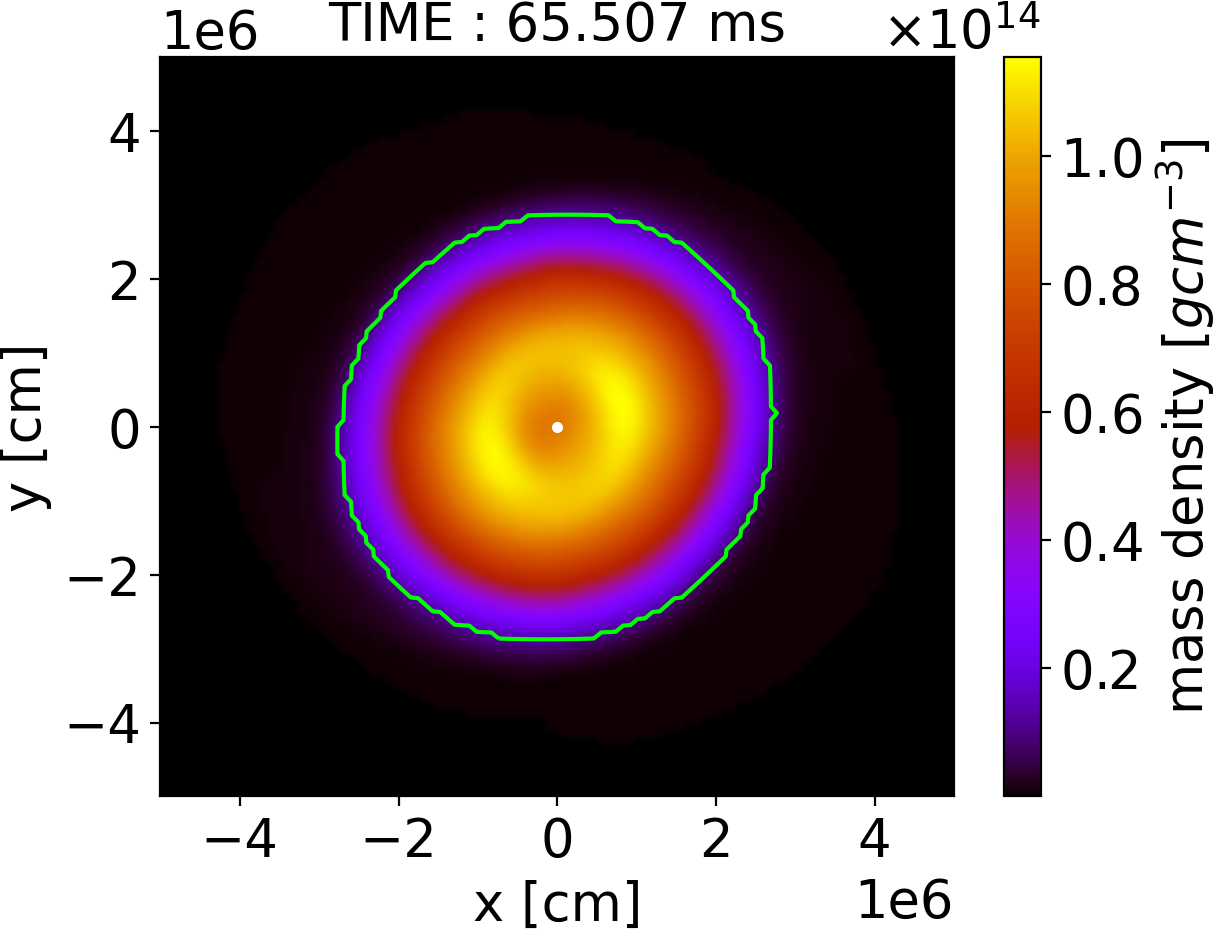}\hspace{-0.0\linewidth}
\includegraphics[width=0.24\linewidth]{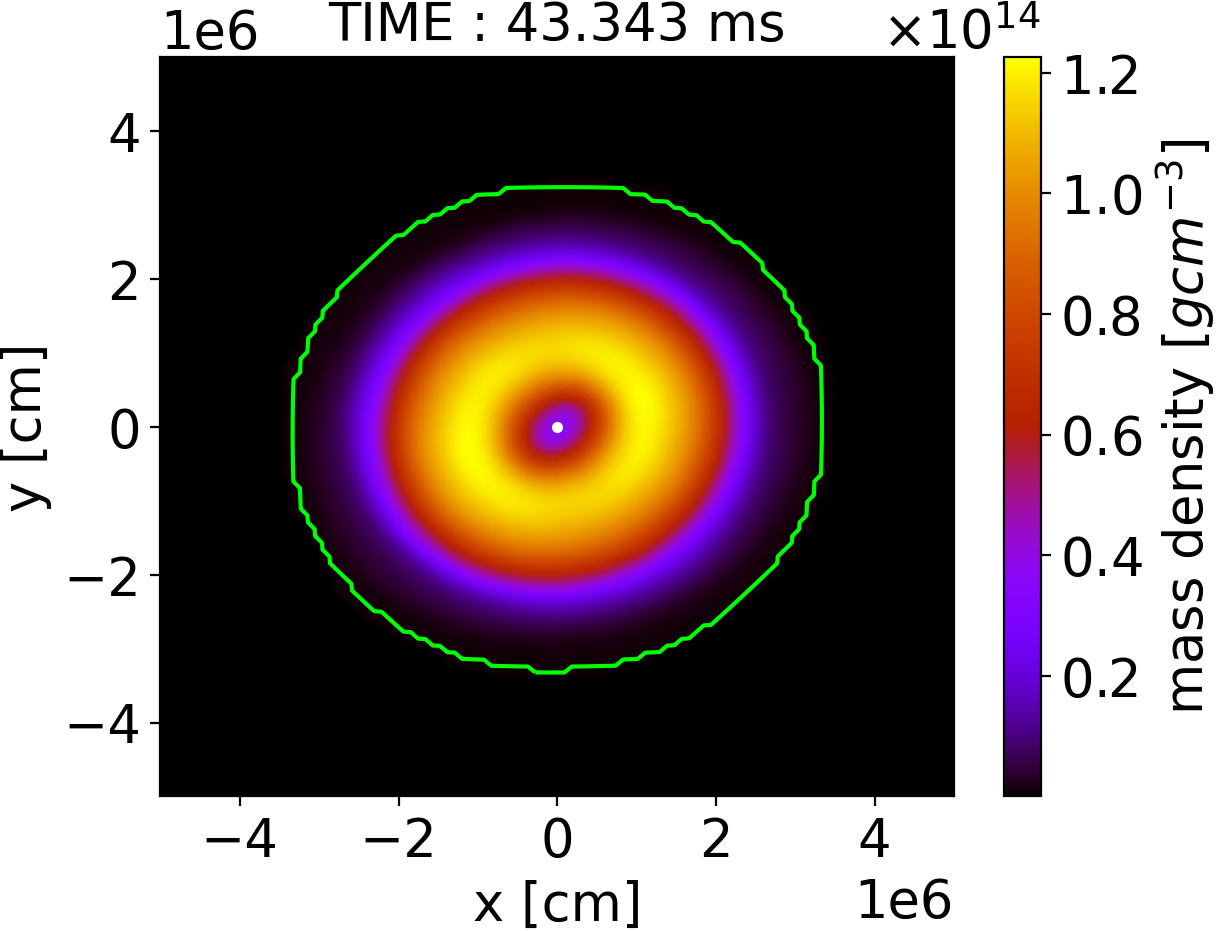}\hspace{-0.0\linewidth}
\includegraphics[width=0.24\linewidth]{U13_mu033_Ap32E-4_rho-0137.png}\\
\includegraphics[width=0.24\linewidth]{U13_rho-0155.png}\hspace{-0.0\linewidth}
\includegraphics[width=0.24\linewidth]{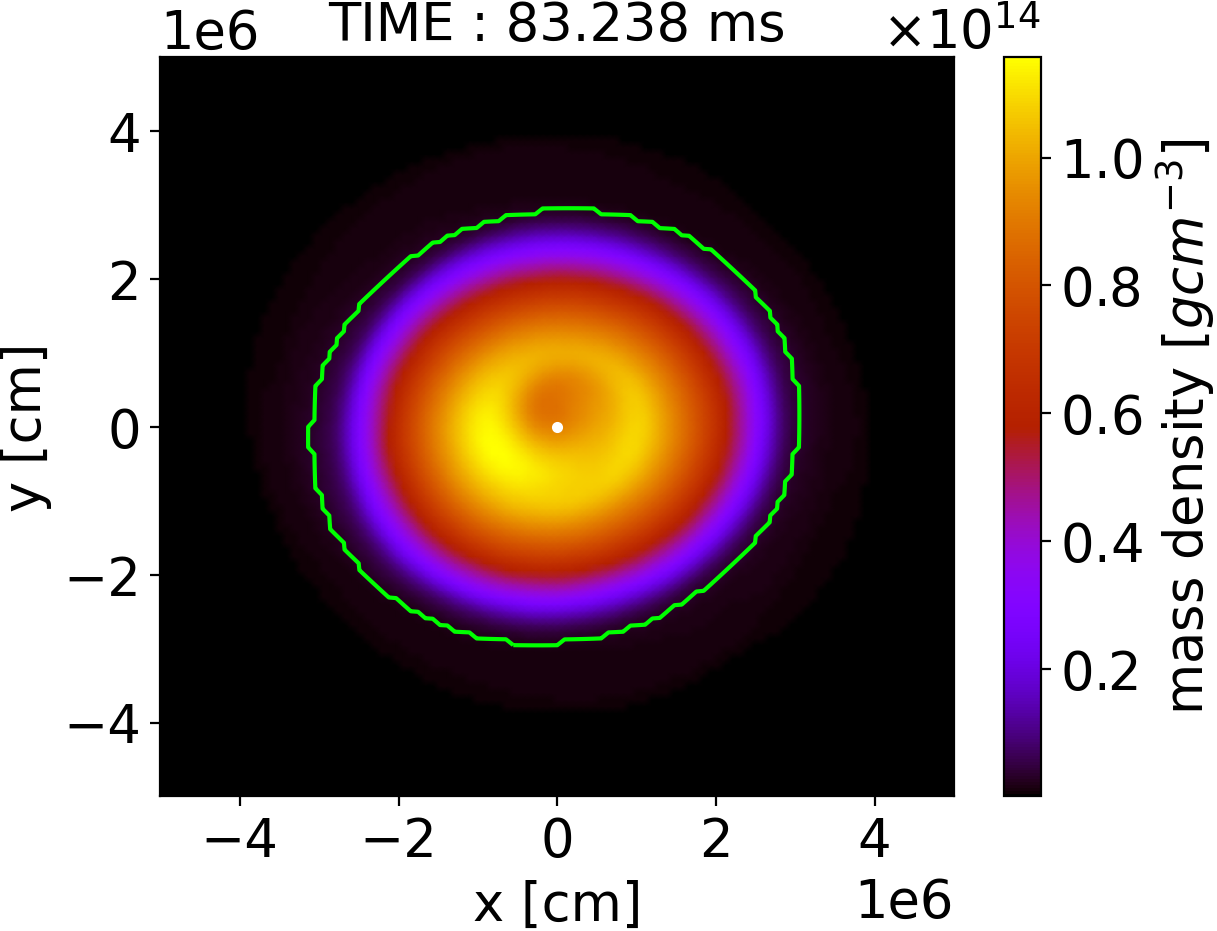}\hspace{-0.0\linewidth}
\includegraphics[width=0.24\linewidth]{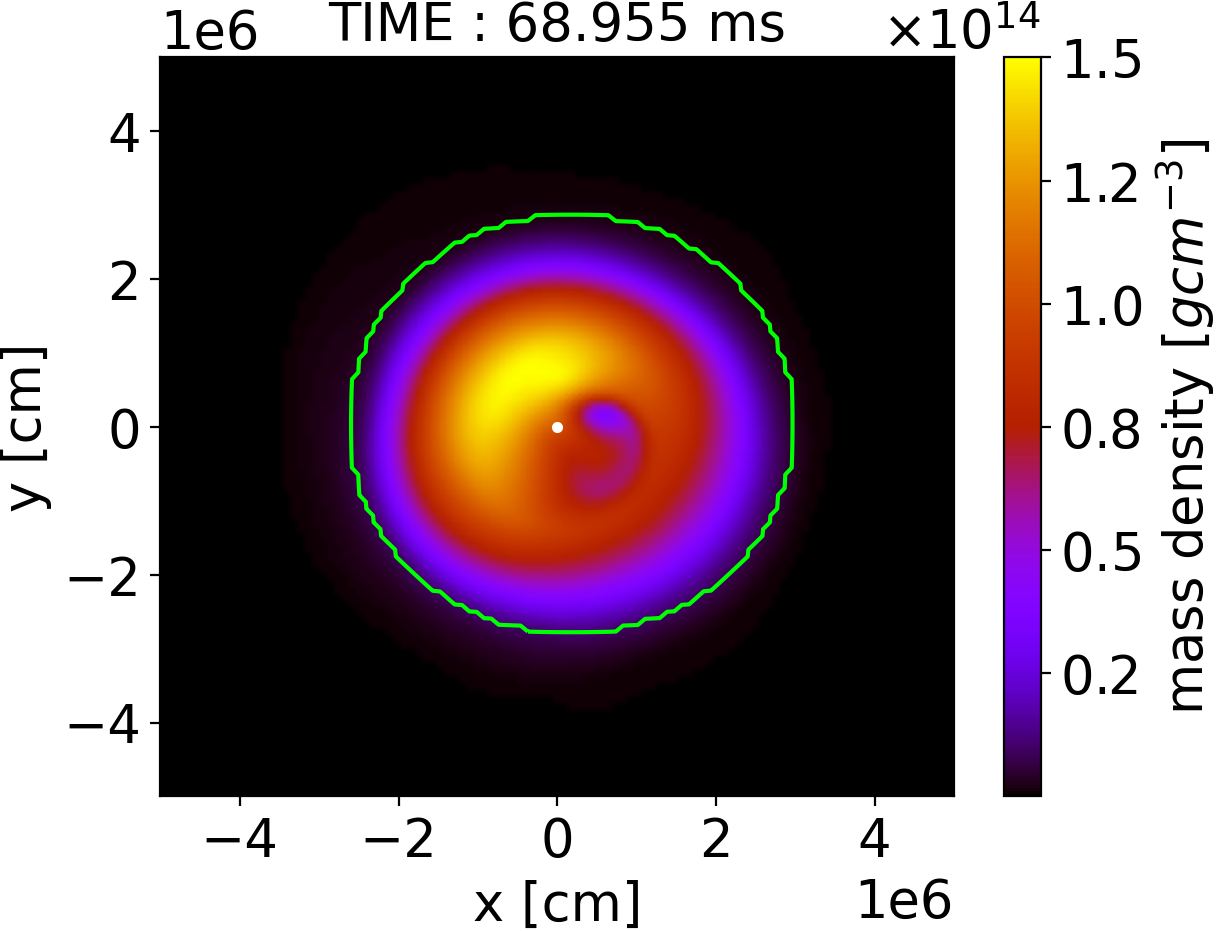}\hspace{-0.0\linewidth}
\includegraphics[width=0.24\linewidth]{U13_mu033_Ap32E-4_rho-0201.png}\\
\includegraphics[width=0.24\linewidth]{U13_rho-0270.png}\hspace{-0.0\linewidth}
\includegraphics[width=0.24\linewidth]{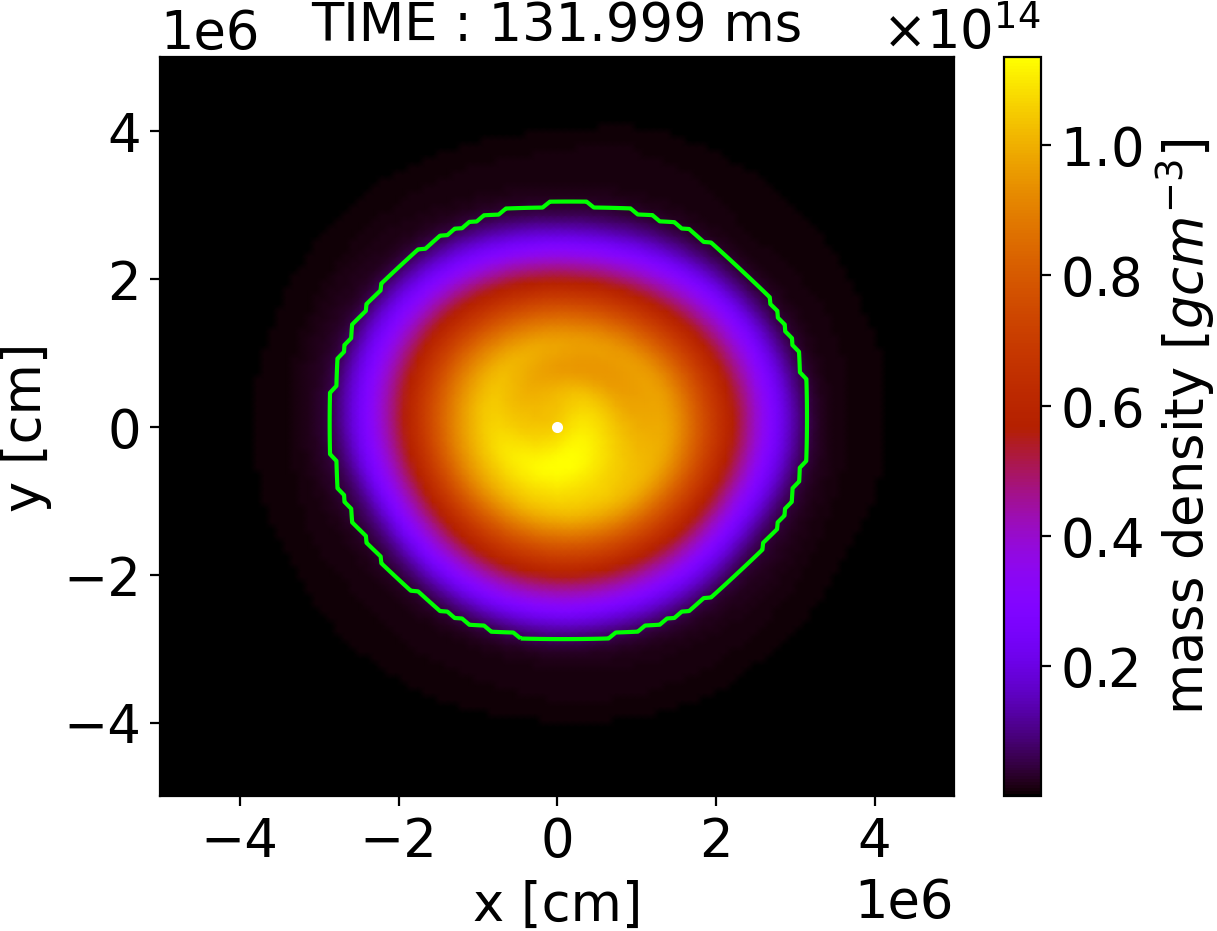}\hspace{-0.0\linewidth}
\includegraphics[width=0.24\linewidth]{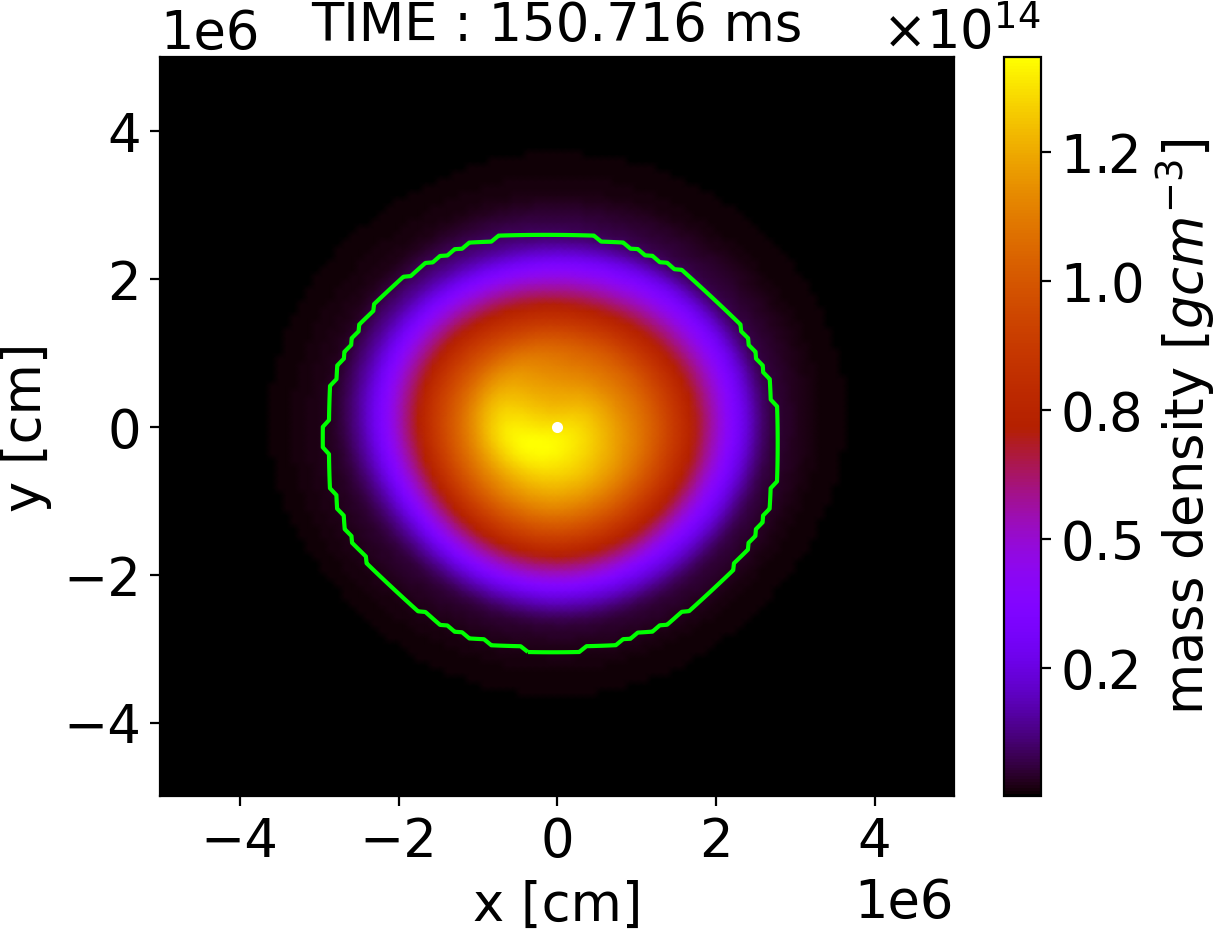}\hspace{-0.0\linewidth}
\includegraphics[width=0.24\linewidth]{U13_mu033_Ap32E-4_rho-0306.png}
\caption{Time evolution of the rest-mass density $\rho$ (in cgs units) at the equatorial plane. The green isocontour indicates the level surface of constant bosonic energy density which contains 95\% of the total mass of the bosonic cloud.  From left to right the columns correspond to model U13, U13-e, U13-d, and U13-c. The last three models have the same value of the bosonic particle mass $(\mu=0.33)$ but the total mass stored in the cloud increases from left to right. 
}
\label{fig4}
\end{figure*}

\begin{figure*}[t!]
\centering
\includegraphics[width=0.45\linewidth]{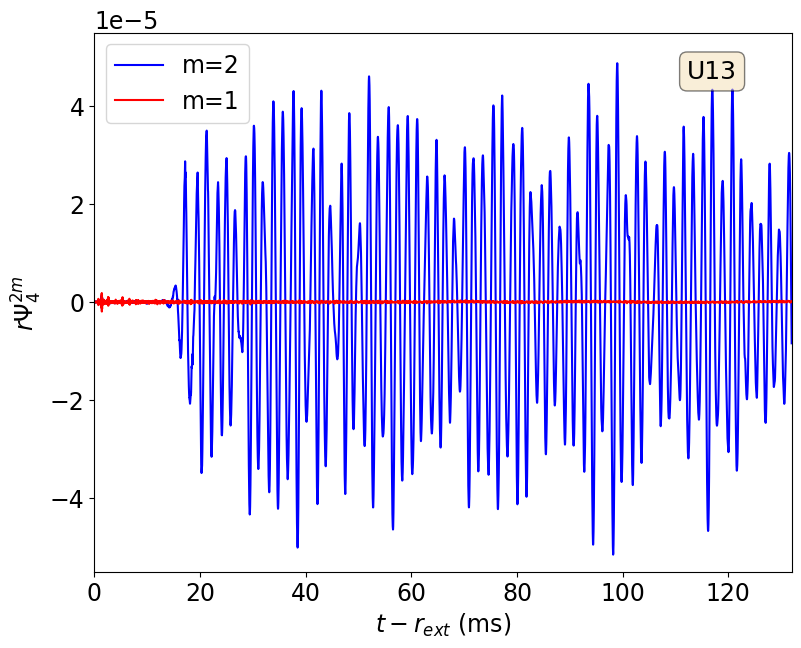} 
\includegraphics[width=0.45\linewidth]{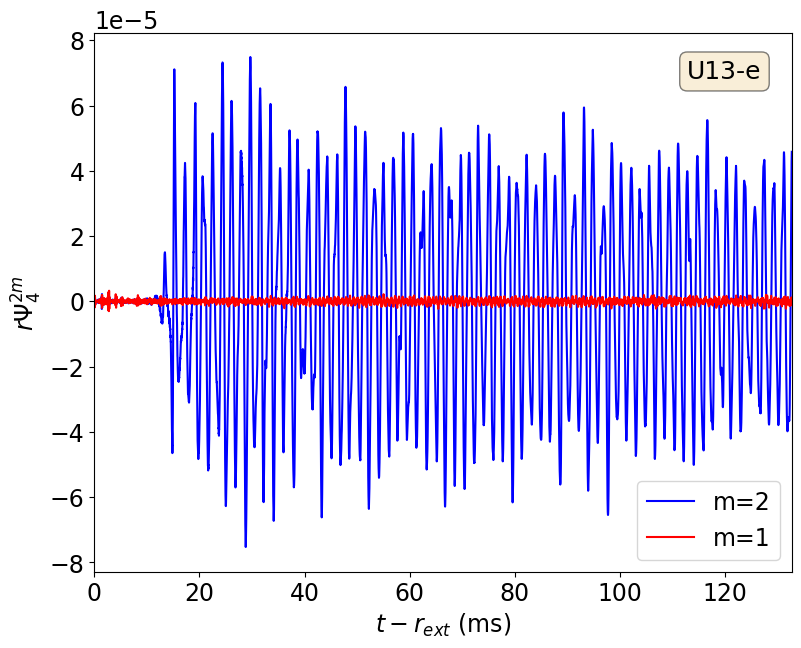} \\
\includegraphics[width=0.45\linewidth]{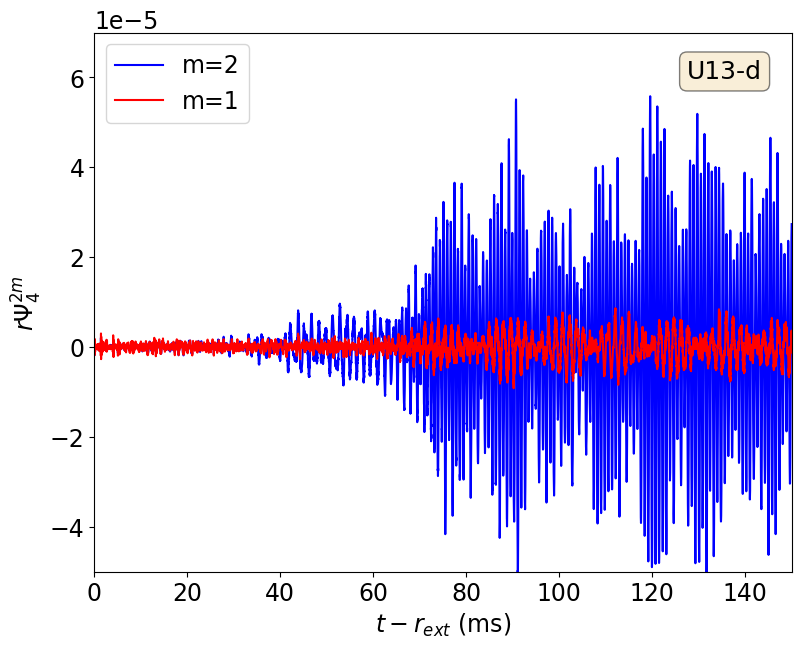} 
\includegraphics[width=0.45\linewidth]{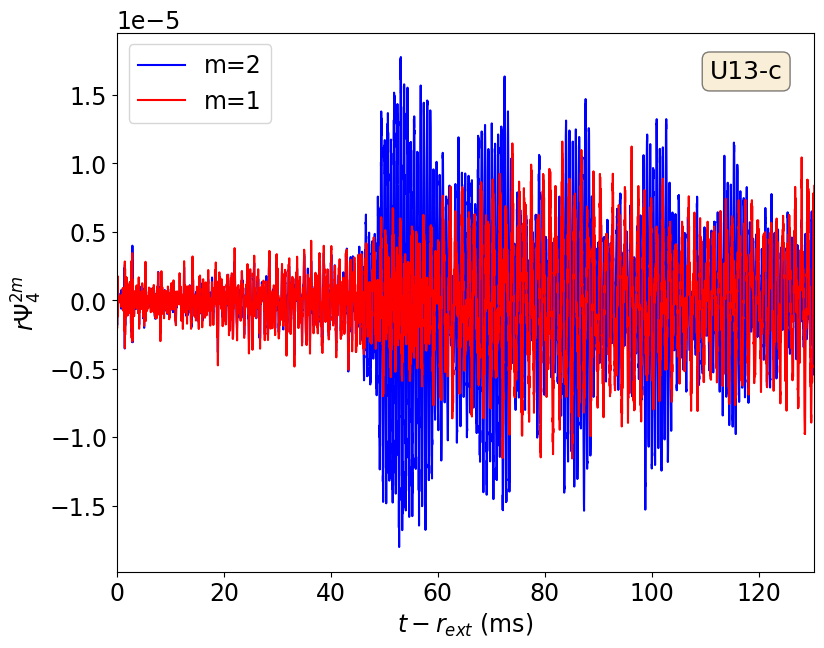} 
\caption{ Real part of $r\Psi_{4}^{2,m}$ with $m=1,2$ as a funtion of the retarded time for models U13 (top-left panel), U13-e (top-right panel), U13-d (bottom-left panel), and U13-c (bottom-right panel). The curves are obtained after a third-order polynomial fit interpolation of the corresponding waveforms from the three extraction radii we select.}
\label{fig5}
\end{figure*}

The remaining panels of Fig.~\ref{fig2} show the time evolution of $|\mathcal{C}_{m}|$ for models U13-a, U13-b, and U13-c.  In the presence of a scalar field all modes are excited to significant levels, with the $m=1$ becoming dominant in all cases. We observe the same excitation also when we depict the Fourier mode decomposition of the bosonic energy density, evaluated in the same fashion as in Eq.~(\ref{Am}). We note that in models U13-b and U13-c the odd modes are excited in both the boson and fermion sectors in such a way that, collectively, they give rise to an {\it even} distribution of the total energy density in the form of a ``mixed bar" (one end of the bar made of bosonic matter, the other of fermionic matter). This morphology guarantees the conservation of the total linear momentum for the case of comparable masses of both sectors. This dynamics is illustrated in the left and centre panels of Figure~\ref{fig3}. The panels exhibit two late-time snapshots of the bosonic energy density $\E^{(0)}$ and the fermionic rest-mass density $\rho$ on the equatorial plane for model U13-b.  The fermionic contribution is shown in orange isocontours. These two panels illustrate how the two different matter components rotate around the Cartesian origin with a $\pi$ phase difference, in such a way that the centre-of-mass of the total object remains close to the centre of the computational grid. This means that the total linear momentum is approximately conserved as the mixed bar compensates the excitation of the dominant $m=1$ modes in both matter constituents. This is further demonstrated in the right panel of Fig.~\ref{fig3} which displays the time evolution of the $x$-component of the centre-of-mass of both the bosonic and fermionic energy density parts (depicted in red and blue, respectively) as well as the total energy density (i.e.~the sum of the two, depicted in black). The evolution shown spans the time interval indicated in the left and centre panels of the figure. A very similar result is observed for the $y$-component of the centre-of-mass. The red and blue curves clearly reveal a constant $\pi$ phase difference between the two matter components as long as the mixed bar persists, while the black curve shows that the total centre-of-mass stays close to the origin. 

A similar behaviour occurs for model U13-c. For both models, U13-b and U13-c, the maximum displacement of the centre-of-mass from the origin is about 3 times smaller than the resolution of our finest grid. On the other hand, for model U13-a we observe a small displacement of the centre-of-mass which starts to be significant at $t\approx 30$ ms. We tentatively associate the different behaviour of model U13-a with respect to models U13-b and U13-c with the larger ejection of scalar field during the accretion process, the formation of a more compact bosonic star core, and the transfer of angular momentum to the scalar component due the dragging of the neutron star. A larger amount of angular momentum is then expected to be emitted in the form of gravitational waves for model U13-a, as we discuss below.

We turn now to briefly discuss the dynamics of models with constant boson particle mass $\mu$ and varying initial cloud mass $M_{\rm{cloud}}$. Those models are U13-c, U13-d, and U13-e in Table~\ref{table:models1}, all with $\mu=0.33$ and correspondingly decreasing $M_{\rm{cloud}}$. Time snapshots on the equatorial plane of the rest-mass density $\rho$ for these models, also including the purely RNS model U13, are plotted in Figure~\ref{fig4}. As in Fig.~\ref{fig1} the green contour visible in most snapshots corresponds to the surface containing 95\% of the bosonic energy density which allows to better evaluate the effects of the scalar field on the dynamics of the neutron stars. Model U13-e, plotted in the second column from the left, is the one with less initial bosonic mass $M_{\rm{cloud}}$. During its early evolution the neutron star develops the bar-mode instability, as in the no-scalar-field model U13 plotted in the first column, and in a very similar timescale of ${\cal O}$(10 ms). However, at late times an $m=1$ spiral mode develops in the energy profile (see the last two snapshots of the second column) which is not present in model U13. This transition from an initial $m=2$-dominated neutron star to a final $m=1$-dominated one is still in effect as   the initial mass of the bosonic cloud increases, as shown in models U13-d and U13-c plotted in the third and fourth columns of Fig.~\ref{fig4}, respectively. As $M_{\rm{cloud}}$ increases the transition accelerates - the $m=2$ bar-like deformation quickly disappears while the $m=1$ blob-like deformation becomes dominant.

%%%%%%%%%%%%%%%%%%%%%
\subsection{Gravitational-wave emission} 
\label{ResultsB}
%%%%%%%%%%%%%%%%%%%%%

We characterize the gravitational-wave emission by computing the mode decomposition of the Newman-Penrose scalar $\Psi_{4}$ in spin-weighted spherical harmonics with spin $-2$. We extract the coefficients $\Psi_{4}^{\ell,m}$  for $\ell=2$ and $m=1,2$ at three different radii, namely $r=\{100,150,200\}$. These extraction radii are both far enough from the source (to be in the wave zone) and not too close to the outer boundary of our numerical grid (to avoid unphysical effects from spurious numerical reflections). We interpolate with a third-order polynomial fit the values from the three different extraction radii to obtain $r\Psi_{4}^{2,m}$. The waveforms for models U13, U13-e, U13-d, and U13-c, the last three having the same bosonic particle mass $(\mu=0.33)$ and increasing initial boson cloud mass, are shown in Fig.~\ref{fig5}. We display the (retarded) time evolution of the real part of $r\Psi_{4}^{2,m}$ for $m=1,2$. The waveforms shown in the top-left panel of Fig.~\ref{fig5}  correspond to model U13, which has no accreting bosonic field. As expected, the dominant contribution to the waveform is the $\ell=m=2$ mode, reflecting  the distinct bar-mode deformation this model undergoes. The $m=1$ mode  is hardly excited for this model, its amplitude being a few orders of magnitude smaller than that of the $m=2$ mode. By the end of the simulation the amplitude of the $m=2$ waveform has not yet decreased substantially. 

\begin{figure*}[t!]
\centering \hspace{-0.15cm}
\includegraphics[width=0.436\linewidth]{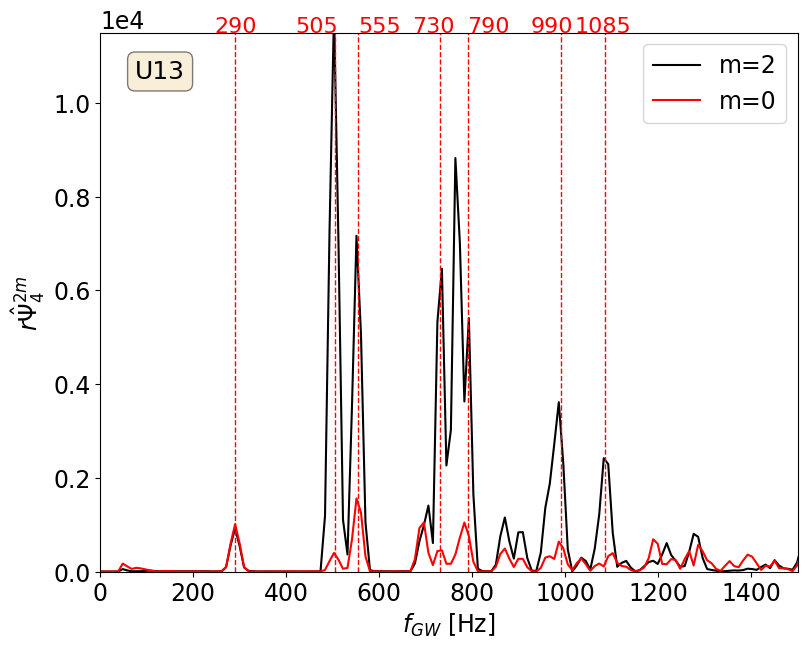} 
\centering \hspace{0.15cm}
\includegraphics[width=0.45\linewidth]{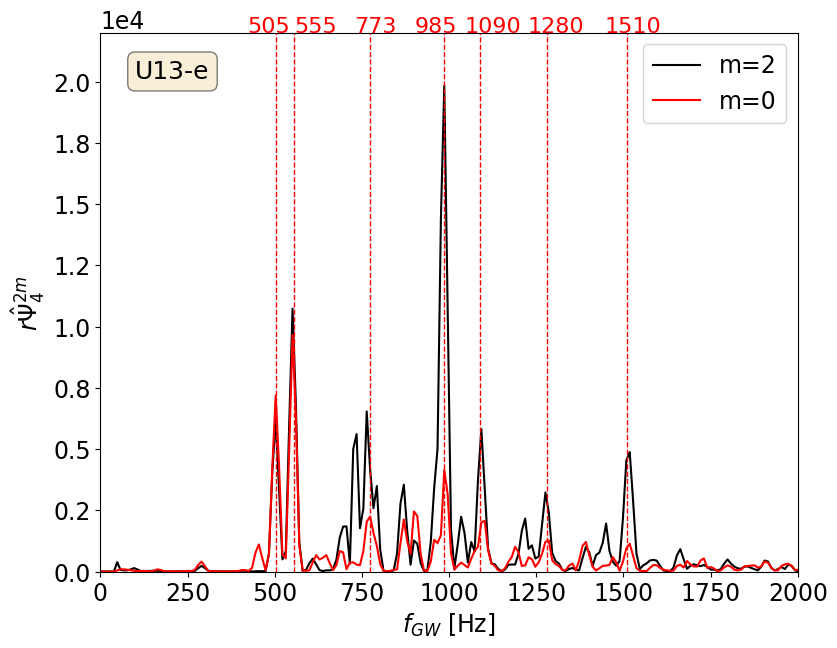} \\
\includegraphics[width=0.45\linewidth]{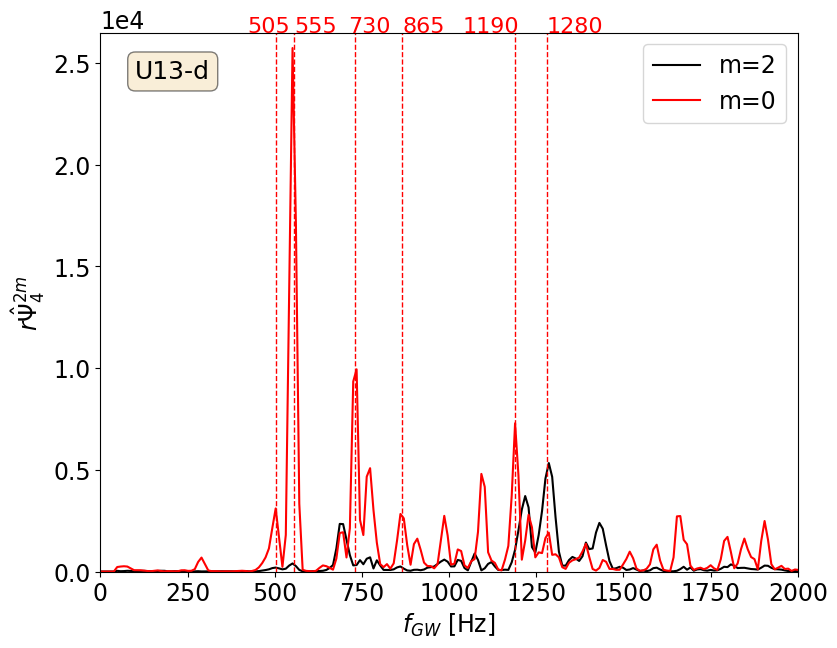} 
\includegraphics[width=0.45\linewidth]{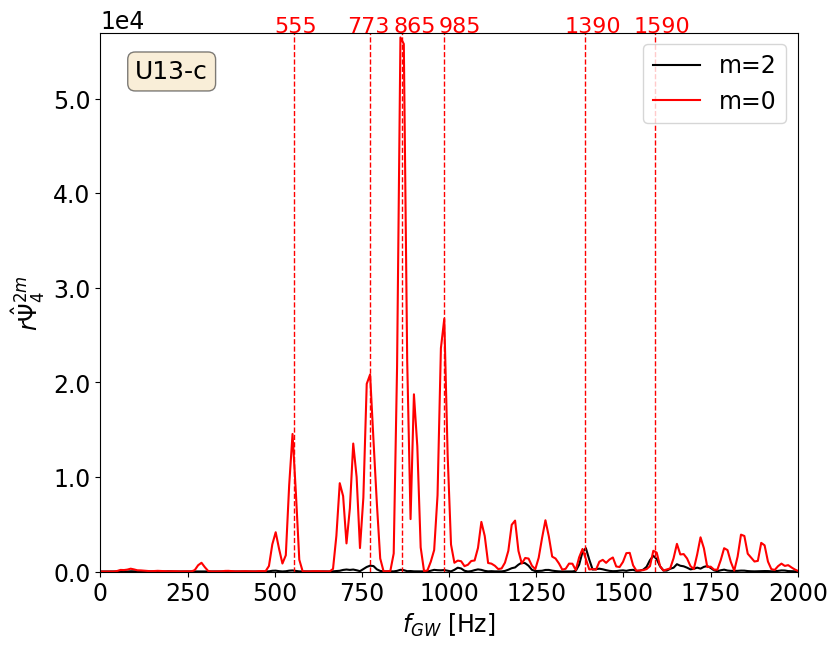} \\
\caption{Fourier transform of $r\Psi_{4}^{2,m}$ for $m=0,2$ for models U13 (top-left panel), U13-e (top-right panel), U13-d (bottom-left panel) and U13-c (bottom-right panel).
}
\label{fig6}
\end{figure*}

As we discussed in the previous section, the presence of an accreting bosonic cloud has a major impact on the dynamics of the stars. This is also imprinted on the waveforms. From Fig.~\ref{fig5} we observe that the more massive the scalar-field cloud, the larger (smaller) the amplitude of the $m=1$ $(m=2)$ gravitational-wave mode. Indeed, by the end of our simulations the $m=1$ amplitude of the most massive case, U13-c, becomes comparable to the amplitude of the $m=2$ mode (see bottom-right panel). For model U13-e (top-right) the $m=1$ mode is still largely suppressed. Therefore, the fact that by increasing $M_{\rm cloud}$ the appearance of the bar-mode instability becomes less clear as the spiral-mode instability becomes more prominent, has a recognizable manifestation on the gravitational-wave signals as well. 

By direct inspection of Fig.~\ref{fig5} we can also observe that the frequency of the $m=2$ mode seems to significantly increase as the mass stored in the cloud is larger. Moreover, a noticeable feature of the waveforms is the presence of beating patterns in the oscillations. This is a clear sign of the existence and superposition of more than one significant oscillation frequency of comparable values. We compute those frequencies by performing fast Fourier transforms of the gravitational-wave time series $r\Psi_{4}^{2, m}$. The associated magnitudes are depicted in Fig.~\ref{fig6} for the same four models of Fig.~\ref{fig5}. We show two types of modes in this figure, namely the $\ell=m=2$ and the $\ell=2$, $m=0$ modes, to emphasize the possible contribution of quasi-radial oscillations ($\ell=0$) to the frequency pattern. The top-left panel of Fig.~\ref{fig6} shows the main frequencies that are excited during the evolution of model U13. The first thing to notice is that the spectrum for this model, and those of the other models, present the same essential features, with a fundamental mode and a series of overtones. We note the presence of a double-peaked feature at $f_{\rm GW}\approx 505$ Hz and $f_{\rm GW}\approx 555$ Hz. We identify the former with the fundamental bar-mode frequency (see below). The proximity of the two frequencies could explain the beating pattern shown in the blue curve of Fig.~\ref{fig5}. For the same model~\cite{Baiotti2007} did not observe such beating and only reported a single frequency of 457 Hz, in broad agreement with our value, given the different resolutions employed in the two simulations and the length of the time series (much shorter in the case of~\cite{Baiotti2007}) which limits the accuracy of the computation of the frequency. As a consistency check we have verified that the same frequency pattern is obtained when evolving the same U13 model but constructing the initial data with the \textsc{Hydro\_RNSID} numerical code. Details on this comparison are provided in Appendix~\ref{AppendixA}.

While the value of the $\beta$ parameter of the U13 model is high enough for the model to develop the nonaxisymmetric bar-mode instability, the star is also subjected to axisymmetric pulsating modes during its evolution. 
The frequency spectrum of non-linear axisymmetric pulsations of rotating relativistic stars was studied in detail by~\cite{Dimmelmeier2006}. Their sequence of differentially rotating models with a fixed rest mass of $M_0=1.506$ (same as that of U13) extends from the non-rotating model to a model with $\beta=0.223$ (model A10 in~\cite{Dimmelmeier2006}). Hence, those models are stable against the dynamical bar-mode deformation. Their frequency spectrum is dominated by the fundamental quasi-radial ($\ell=0$) $F$ mode (and its first overtone), the fundamental quadrupole ($\ell=2$) mode (and its first two overtones), and three inertial modes (see Table 2 and Figure 1 in~\cite{Dimmelmeier2006}). Along their sequence, the frequency of the $F$ mode  decreases fairly linearly with $\beta$. Extrapolating that trend to our U13 model, with $\beta=0.29$, would yield a value of the $F$ mode frequency of $\approx 400$ Hz (and of $\approx 450$ Hz for $\beta=0.28$ as used in~\cite{Baiotti2007}). To infer the actual frequency of the $F$ mode we monitor the time evolution of the rest-mass density $\rho$ at the centre of the star for model U13. This particular choice is motivated by the fact that as $\rho$ at the centre is unaffected by even mode deformations (such as the bar) we can isolate the effects of the quasi-radial oscillations. By evaluating the Fourier transform of $\rho$ in the first 35 ms we observe a wide peak for the $F$ mode at around 407 Hz. The limited time window does not allow us to better resolve the frequency but our result is in broad agreement with the value we extrapolated from~\cite{Dimmelmeier2006}. In addition, we repeat the same procedure for a fixed point on the equatorial plane, namely at $r\approx 9$ km, in order to obtain the spectrum of frequencies of {\it both} the quasi-radial oscillations and the bar-mode instability. By subtracting the magnitude of the Fourier transform at $r\approx 9$ km and at the centre of the star, we eliminate from the former the contribution of the quasi-radial oscillations and we isolate the frequency of the bar. This yields a frequency peak at $\approx 500$ Hz which is in close agreement with what we infer from Fig.~\ref{fig6}.

After the bar has mostly dissipated, we also observe the appearance of a well defined frequency at $\approx 785$ Hz. We interpret this frequency as associated with the actual $\ell=0$ quasi-radial $F$ mode oscillation of the {\it new} equilibrium configuration reached by model U13 once the bar deformation has disappeared. In addition, we speculate that the secondary peak of $\approx 555$ Hz depicted in the top-left panel of Fig.~\ref{fig6} may have been originated by the coupling with the frequency of the $\ell=2,m=0$ mode. However, the presence of the fundamental quasi-radial $F$ mode may have also helped triggering the double-peaked structure seen in the figure. 

\begin{figure*}[t!]
\centering
\includegraphics[width=0.95\linewidth]{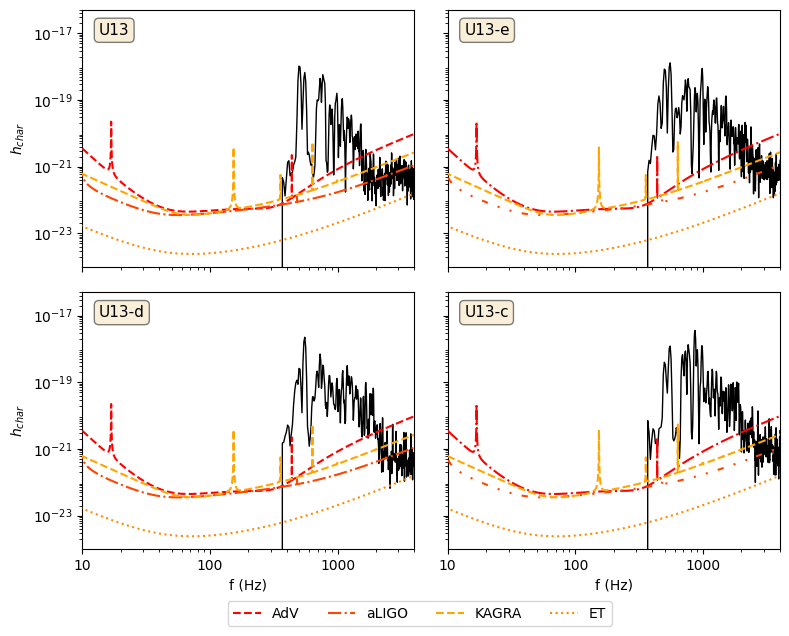} 
\caption{Characteristic gravitational-wave strain against frequency for model U13 (top-left panel), U13-e (top-right panel), U13-d (bottom-left panel) and U13-c (bottom-right panel), compared with the sensitivity curves of current second-generation detectors and the planned Einstein Telescope. A source distance of $D = 10$ kpc is assumed.}
\label{fig7}
\end{figure*}

The discrete modes we observe in the PSD are non-linear harmonics of linear pulsation modes, which is a general property of non-linear systems~\cite{Landau1976}. To lowest order, these arise as linear sums and differences of different linear modes, including self-couplings. For a system with eigenfrequencies $f_i$, the non-linearity of the equations of motion excites modes at frequencies $f_i \pm f_j$. Such non-linear harmonics have been noted in other types of oscillating compact objects, as e.g.~thick disks around black holes~\cite{Zanotti2005,Montero2007} and pulsating relativistic stars~\cite{Dimmelmeier2006}. In our case one such harmonic appears at $f_{\rm GW}\approx 1$ kHz, where we observe the same double-peaked structure at a frequency which corresponds, roughly, to twice the frequency of the fundamental mode (a self-coupling). In between those two modes the spectrum depicts two further combinations of intermediate frequencies which may correspond to  other non-linear harmonics arising as linear sums or differences of the bar-mode frequency and other modes. In  the top-left panel of Fig.~\ref{fig6} we highlight the peaks at $f_{\rm GW}\approx 730$ Hz and $f_{\rm GW}\approx 790$ Hz which can be identified with linear combinations of the peak at $505$ Hz and the one at $290$ Hz, namely $2\times 505 - 290 = 720$ Hz and $505 + 290 = 795$ Hz.

The presence of the scalar field which interacts gravitationally with the baryonic matter and modifies the evolution of the whole system makes the gravitational-wave emission more complex, as we saw in Fig.~\ref{fig5}. In general, as we increase the scalar field contribution, the $m=2$ amplitude becomes smaller, due to the fact that the bar-mode instability tends to disappear. Moreover the $m=0$ spectra in Fig.~\ref{fig6} become increasingly prominent, due to the higher perturbation the neutron star undergoes and the gravitational cooling process of the scalar cloud which leads to a radially perturbed stationary configuration (see also~\cite{seidel1994formation,di2018dynamical}). The model with the lightest scalar cloud (top-right panel of Fig.~\ref{fig6}) still displays a similar frequency pattern than model U13, associated with the formation of the bar. The dominant peak is now, however, at $\approx 1$ kHz, and an additional overtone is present at $\approx 1510$ Hz. For the last 2 models (U13-d and U13-c; bottom-left and bottom-right panels of Fig.~\ref{fig6}) the dominant contributions come from the $m=0$ mode, where we observe the double peak structure at around 500-550 Hz  and at 730-790 Hz and the appearance of a new peak at $\approx 865$ Hz, which could be roughly identified as the sum of the frequencies at 555 Hz and 290 Hz. In model U13-c, in particular, the dominant peak appears at this new overtone at $\approx 865$ Hz.

In Fig.~\ref{fig7} we show the characteristic gravitational-wave strain $h_{\rm char}$ at a distance $D=10$ kpc for models U13, U13-e, U13-d, and U13-c compared with the designed sensitivity curves of ground-base detectors Advanced LIGO  (aLIGO)~\cite{aLIGO2015}, Advanced Virgo (AdV)~\cite{aVirgo2015}, KAGRA\cite{Somiya:2012}, and the future Einstein Telescope (ET)~\cite{Hild:2011}. For burst-like sources the characteristic GW strain is (see e.g.~\cite{Flanagan:1998a})
\begin{equation}
h_{\rm char} (f) = \frac{1}{\pi D}\sqrt{ 2 \frac{dE}{df}[f]}\,, \label{eq:hchar}
\end{equation}
where $D$ is the distance of the source and $dE/df$ is the energy spectrum of the gravitational waves. The interested reader is addressed to~\cite{DiGiovanni:2020ror} for further details on the definition of the energy spectrum. We note that for the reasons explained in Appendix~\ref{AppendixA} we cut the contribution at high frequencies of the $m=0$ mode in the evaluation of $h_{\rm char}$.

The spectra shown in Fig.~\ref{fig7} closely parallel the Fourier transforms depicted in Fig.~\ref{fig6}. For the U13 model (no scalar field cloud) the maximum of $h_{\rm char}$ is at $f_{\rm GW}\approx 505$ Hz which is the main peak of the $m=2$ mode shown in Fig.~\ref{fig6}, linked to the bar-mode instability. For models U13-e and U13-d, the maxima in the spectra are at around 555 Hz which we associate with the $m=0$ mode that is excited by quasi-radial oscillations of the neutron star. Moreover the overtones at higher frequencies become more relevant. Finally for model U13-c, which is the one with the largest amount of scalar field, the $m=0$ overtone at frequency $f_{\rm GW}\approx 860$ Hz becomes the maximum of the characteristic strain. 

We evaluate the matched-filtering signal-to-noise ratio (SNR) squared for an optimally oriented detector, averaged over all possible source directions as~\cite{Flanagan:1998a}
\begin{equation}
\rho^2_{\rm optimal} = \int_0^\infty  d (\ln f) \frac{h_{\rm char}(f)^2}{f S_{\rm n}(f)}, \label{eq:SNR}
\end{equation}
where $S_{\rm n}(f)$ is the power spectral density (PSD) of the detector noise. We consider the SNR averaged over all possible detector orientations and sky localizations, which is simply $\langle\rho^2\rangle = \rho^2_{\rm optimal} / 5$.
We define the horizon distance as the distance at which SNR=8. We report in Table~\ref{table:SNR} this quantity for the four models shown in figure~\ref{fig7} and for the four gravitational-wave detectors considered. For second-generation detectors, the signal studied in this paper could be detectable up to distances of about 1 Mpc while ET could observe it up to a distance of a few tens of Mpc. Interestingly, Table~\ref{table:SNR} shows that the horizon distance of the signal increases, in most cases, as the amount of scalar field in the models becomes larger, to almost reach a factor two in model U13-c with respect to model U13.

\begin{table}
\caption{Horizon distances of the gravitational-wave signal studied in this work for models U13, U13-e, U13-d, and U13-c with increasing bosonic contribution, evaluated for the ground-based detectors Advanced Virgo (AdV), Advanced LIGO (aLIGO), KAGRA, and Einstein Telescope (ET).}
\centering
\begin{tabular}{c | c  c  c  c }
\hline
 & \multicolumn{4}{c}{ Horizon distance (Mpc) }\\
\hline
Model & AdV  & aLIGO & KAGRA &  ET  \\
\hline

U13   & 0.747 & 1.460 & 0.872 & 14.791 \\
U13-e & 0.813 & 1.690 & 0.994 & 16.880 \\
U13-d & 1.037 & 2.245 & 1.318 & 22.371 \\
U13-c & 0.860 & 2.695 & 1.425 & 24.403 \\
\hline
\end{tabular}
\label{table:SNR}
\end{table}

The LIGO-Virgo-KAGRA (LVK) Collaboration has conducted various searches of continuous signals generated by nonaxisymmetric neutron stars, including r-modes and other types of instabilities (se e.g.~\cite{PhysRevD.104.082004} for the most recent search employing O3 data). The results reported in our work might be relevant for those studies. Taking our findings at face value the potential detection of such continuous signals  in an unexpected range of frequencies could hint at the possible presence of dark matter in neutron stars. On the other hand, a lack of detections could also convey information about the composition and dynamics of such composite stars, since the frequencies of the gravitational-wave emission could be outside the LVK sensitivity range. 

%%%%%%%%%%%%%%%%%%%%%%%%%%%%%%%%%%%%%%%%%%%%%%%%%%%%%%%%%%%%%
\section{Conclusions}
\label{conclusions}
%%%%%%%%%%%%%%%%%%%%%%%%%%%%%%%%%%%%%%%%%%%%%%%%%%%%%%%%%%%%%%

We have investigated the effects ultralight bosonic field dark matter may have on the dynamics of unstable differentially-rotating neutron stars prone to the bar-mode instability. We have found that the presence of the bosonic field can critically modify the development of the bar-mode instability of neutron stars, depending on the total mass of the bosonic field and on the boson particle mass. This, in turn, implies that dark-matter accretion in neutron stars could change the frequency of the expected gravitational-wave emission from the bar-mode instability, which would have an impact on ongoing searches for continuous gravitational waves. In this paper we have focused on ultralight bosonic dark matter but our results could be extrapolated to other dark matter models.

The kind of composite (fermion-boson) stars we have studied in this work remain an intriguing possibility. Dark matter can pile up in neutron stars, either by accretion during the life of the supernova progenitor star, by capture during the evolution of the neutron star itself, or both. A number of theoretical works have explored such scenarios in the context of fermion-boson stars (see e.g.~\cite{brito2015accretion,brito2016interaction,DiGiovanni:2021ejn}). In the case dark matter is accreted before the formation of the neutron star, a similar ratio between the bosonic and fermionic components in all composite stars should be expected. On the other hand, if dark matter is captured during the neutron star evolution, older stars might have a higher bosonic contribution than younger ones. In this situation, one could expect that in BNS mergers the contribution of the bosonic field could be large enough to have an impact in the dynamics. Concerning rotation, highly differentially rotating composite stars might form as a result of the merger of two such fermion-boson stars~\cite{Bezares:2019}, or of one neutron star with a boson star. Current simulations are, however, still unable to prove this as the latter are restricted to head-on collisions~\cite{Clough2018,Dietrich2019}).

Our results have been obtained from a large set of numerical simulations in general relativity of rotating neutron stars accreting an initial spherically symmetric bosonic field cloud, solving the Einstein-(complex, massive) Klein-Gordon-Euler and the Einstein-(complex) Proca-Euler systems. For our purely neutron star models (no bosonic field) a bar-like deformation appears and we observe, as expected, the exponential growth of the Fourier density modes of the star, with the $m=2$ mode being the dominant one. Incorporating the bosonic field leads to different dynamics and mode excitation, with the $m=1$ becoming now the dominant mode. In some of our models, the accreting bosonic field can effectively quench the dominant $\ell=m=2$ mode of the bar-deformation by dynamically forming a mixed (fermion-boson) star that retains part of the angular momentum of the original neutron star. Interestingly, the mixed star undergoes the development of a mixed bar that leads to significant gravitational-wave emission, substantially different to that of the isolated neutron star. The timescale of the instability is also affected by the presence of dark matter, being significantly delayed as the amount of bosonic field increases. We note, however, that our setup is such that the unstable neutron star accretes a large amount of bosonic field in a short period of time. It might be possible that in another region of the parameter space of the problem the bar-mode instability could actually be quenched without triggering the $m=1$ deformation, e.g.~for different neutron star  models or further exploring different bosonic masses. It would also be interesting to perform evolutions of equilibrium sequences of stationary, rotating fermion-boson star models to address their stability in a more controlled system. Given the absence of such models presently, this is a task we defer for the future.

We have also found that the differences in the evolution of the composite stars due to the presence of the bosonic field are imprinted in the gravitational-wave emission. This was studied by computing the Newman-Penrose scalar $\Psi_{4}$ to evaluate the gravitational-wave frequency for our models. Those quantities are affected by the presence of the bosonic field, yielding complex gravitational-wave signals in which different modes contribute and leading, in particular, to a remarkable increase of the dominant frequency. The signals studied in this work are within reach of current ground-base detectors up to distances of about 1 Mpc. This increases to a few tens of Mpc for third-generation detectors as the ET. Therefore, the results reported might be of some interest for searches of continuous signals from neutron stars, routinely carried out by the LVK Collaboration in every scientific run. The potential detection of such continuous signals  in an unexpected range of frequencies could hint at the possible presence of dark matter in neutron stars.
 
%%%%%%%%%%%%%%%%%%%%%%%%%%%%%%%%%%%%%%%%%%%%%%%
\begin{acknowledgments}
Work  supported by the Spanish Agencia Estatal de Investigaci\'on (PGC2018-095984-B-I00), by the Generalitat Valenciana (PROMETEO/2019/071  and  GRISOLIAP/2019/029), by the Center for Research and Development in Mathematics and Applications (CIDMA) through the Portuguese Foundation for Science and Technology (FCT - Funda\c c\~ao para a Ci\^encia e a Tecnologia), references UIDB/04106/2020 and UIDP/04106/2020, by national funds (OE), through FCT, I.P., in the scope of the framework contract foreseen in the numbers 4, 5 and 6 of the article 23, of the Decree-Law 57/2016, of August 29, changed by Law 57/2017, of July 19 and by the projects PTDC/FIS-OUT/28407/2017,  CERN/FIS-PAR/0027/2019 and PTDC/FIS-AST/3041/2020, and by  the  European  Union's  Horizon  2020  research  and  innovation  (RISE) programme H2020-MSCA-RISE-2017 Grant No.~FunFiCO-777740. NSG acknowledges support by the Spanish Ministerio de Universidades, reference UP2021-044, within the European Union-Next Generation EU.  MMT acknowledges support by the Spanish Ministry of Science, Innovation and Universities (Ministerio de Ciencia, Innovaci\'on y Universidades del Gobierno de Espa\~na) through the FPU Ph.D.~grant No.~FPU19/01750. DG acknowledges support by the Spanish Ministry of Science and Innovation and the Nacional Agency of Research (Ministerio de Ciencia y Innovaci\'on, Agencia Estatal de Investigaci\'on) through the grant PRE2019-087617. Computations have been performed at the Servei d'Inform\`atica de la Universitat de Val\`encia. The authors gratefully acknowledge the Italian Instituto Nazionale di Fisica Nucleare (INFN), the French Centre National de la Recherche Scientifique (CNRS) and the Netherlands Organization for Scientific Research, for the construction and operation of the Virgo detector and the creation and support of the EGO consortium. This manuscript has ET preprint number ET-0105A-22.
\end{acknowledgments}

% \bigskip

%\newpage

\appendix

\section{Comparison between XCFC and \textsc{Hydro\_RNSID} Initial Data} \label{AppendixA}
\begin{figure*}[t!]
\centering \hspace{-0.18cm}
\includegraphics[width=0.43\linewidth]{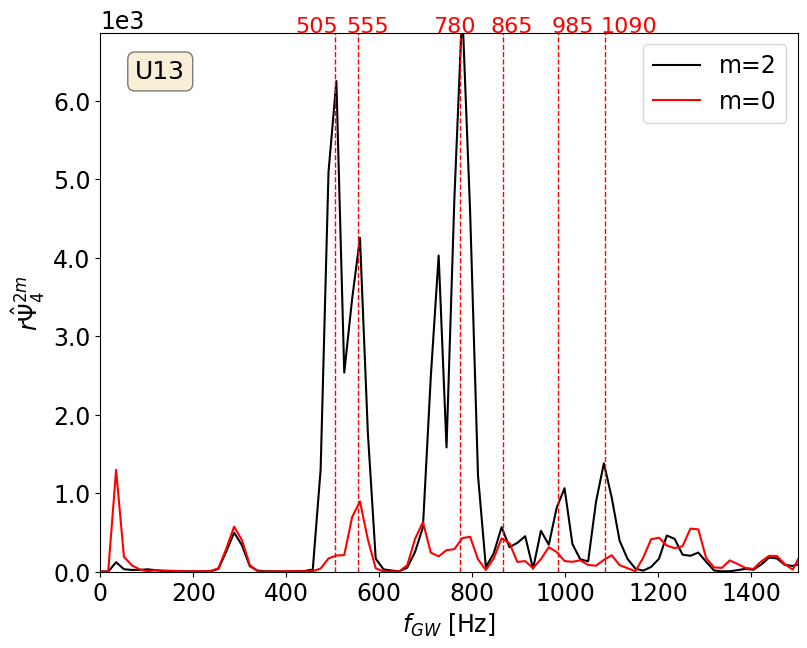} 
%\centering 
\hspace{0.15cm}
\includegraphics[width=0.43\linewidth]{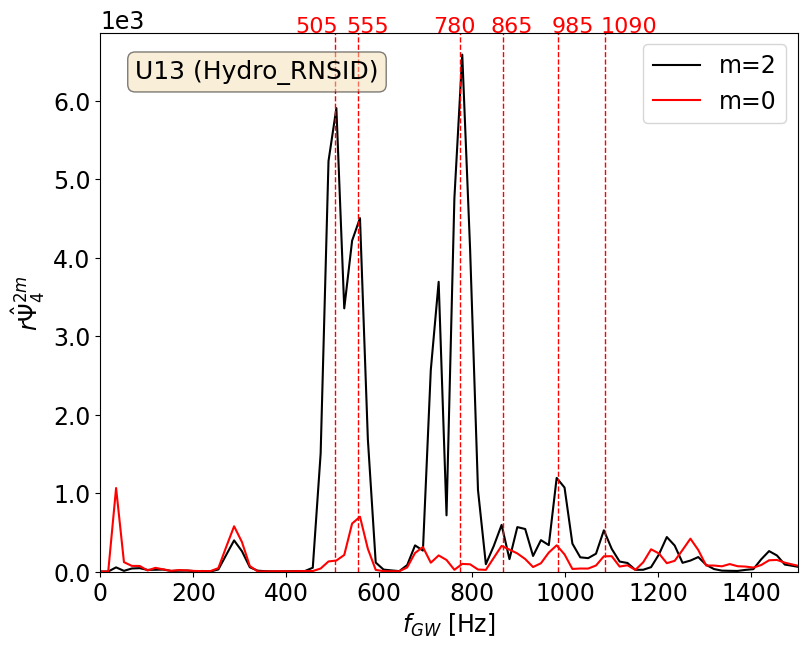} \\
\centering
\includegraphics[width=0.45\linewidth]{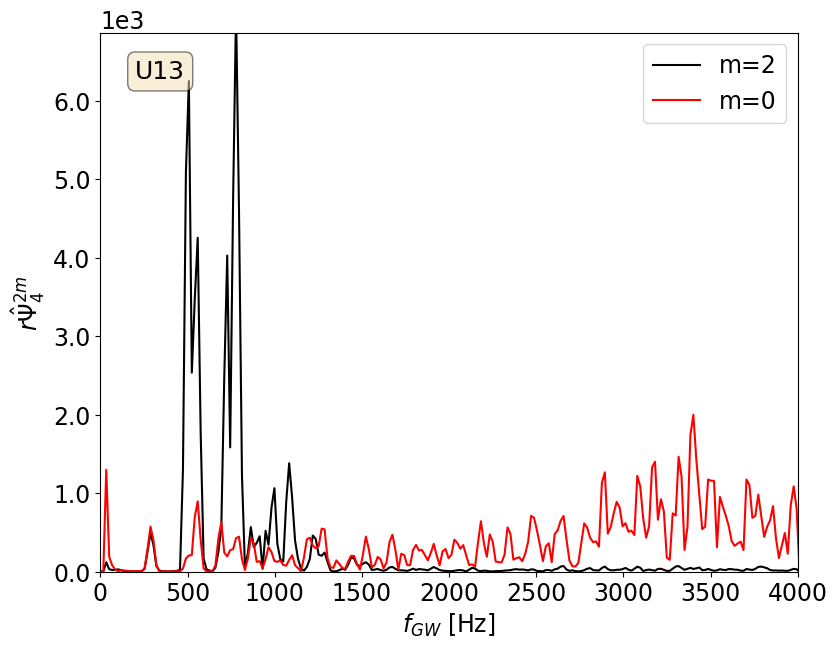} 
\includegraphics[width=0.45\linewidth]{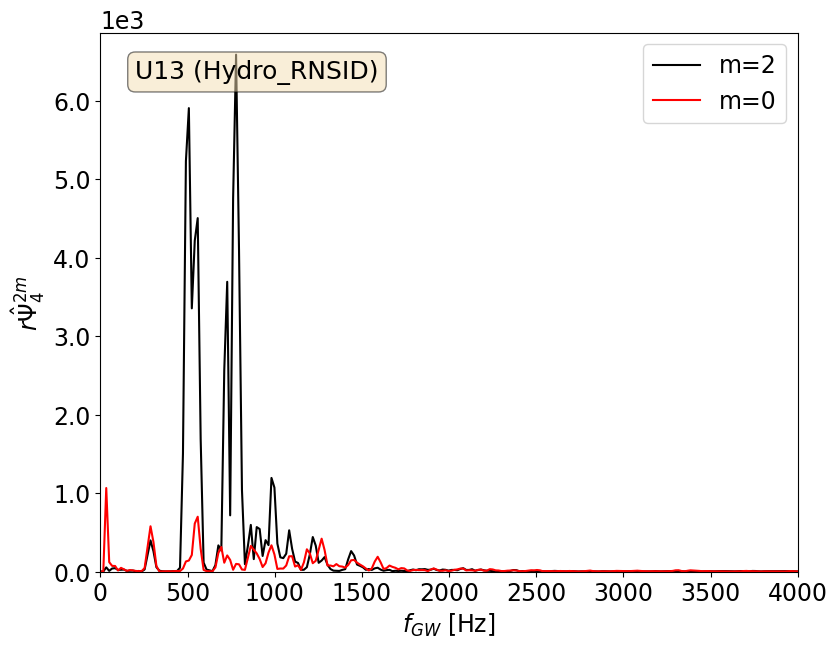} \\
\caption{Fourier transform of $r\Psi_{4}^{2,m}$ for $m=0$ and $m=2$ for model U13. The left panels correspond the evolutions performed using our CFC initial data code while the right panels are the ones using \textsc{Hydro\_RNSID}. The frequency range in the top row extends up to 1500 Hz while in the bottom row it goes up to 4 kHz. The agreement between the two appraches, as seen in the top panels, is remarkable.
}
\label{fig8}
\end{figure*}

The initial data for the evolutions reported in this work have been constructed using the numerical code introduced in~\cite{CorderoCarrion:2008nf} in polar spherical coordinates and assuming the conformal flatness condition for the Einstein equations. We developed a private \textit{thorn}, which is a component of the \textsc{EinsteinToolkit} software, to read and linearly interpolate the initial data into the Cartesian grid used for the evolutions.

In this Appendix we show a brief comparison of the results obtained evolving the neutron star model U13 up to $t\approx 80$ ms, making use of our initial data thorn and \textsc{Hydro\_RNSID} which is part of the official release of the \textsc{EinsteinToolkit}. The main dynamical features of the two evolutions are essentially identical, with the appearence of the bar-mode instability at around $t\approx 10$ ms. The main difference we observe is in the gravitational-wave emission, specifically in the $\ell=2$, $m=0$ component of $\Psi_{4}$.

 In Fig.~\ref{fig8} we compare the frequency spectrum obtained for model U13 using our thorn (left panels) and \textsc{Hydro\_RNSID} (right panels). In the top plots, where we show only the frequencies up to $f_{\rm GW}=1500$, we can appreciate that the main peaks connected to the bar-mode instability and to the quasi-radial oscillations are essentially the same with the two approaches, validating the results obtained in this work. The frequency range of the bottom plots extends to 4 kHz. This is to highlight the presence of high-frequency noise in the $m=0$ component of $\Psi_{4}$ when using the CFC initial data (left panel). This feature is not present if we use the \textsc{Hydro\_RNSID} initial data (right panel). We suspect that the reasons behind this difference might be the poor interpolation into the Cartesian grid and the low resolution used in the angular coordinate of our initial data models, which is 5 times coarser than the one employed in \textsc{Hydro\_RNSID}. These two factors do not influence the evolutions in a significant way but they do induce a stronger initial perturbation on the CFC initial data which triggers stronger quasi-radial oscillations in the star from the beginning of the simulation. This effect is visible in the gravitational-wave emission but was not evident from the time snapshots of the energy density on the equatorial plane. For this reason, as we write in the main text, we depict in Figure~\ref{fig7} the evaluation of $h_{\rm char}$ without the contribution of the $m=0$ mode at high frequencies.

%\newpage

\bibliography{num-rel2}

\end{document}